\documentclass[prd,aps,floats,twocolumn]{revtex4}

\usepackage[dvips]{graphicx}
\usepackage{amssymb} 
\usepackage{amsmath}
\usepackage{ctable}
\usepackage{caption}

\begin{document}

\title{Coxeter Pairs, Ammann Patterns and Penrose-like Tilings}

\author{Latham Boyle$^1$ and Paul J. Steinhardt$^1$}

\affiliation{$^1$Perimeter Institute for Theoretical Physics, \\ 
Waterloo, Ontario N2L 2Y5, Canada \\
$^2$Princeton Center for Theoretical Science, Princeton University \\
Princeton, NJ, 08544 USA \\
$^3$Department of Physics, Princeton University\\Princeton, NJ, 08544 USA}  

\begin{abstract}
We identify a precise geometric relationship between: (i) certain natural pairs of irreducible reflection groups (``Coxeter pairs"); (ii) self-similar quasicrystalline patterns formed by superposing sets of 1D quasi-periodically-spaced lines, planes or hyper-planes (``Ammann patterns"); and (iii) the tilings dual to these patterns (``Penrose-like tilings").  We use this relationship to obtain all irreducible Ammann patterns and their dual Penrose-like tilings, along with their key properties in a simple, systematic and unified way, expanding the number of known examples from four to infinity.  For each symmetry, we identify the minimal Ammann patterns (those composed of the fewest 1d quasiperiodic sets) and construct the associated Penrose-like tilings: 11 in 2D, 9 in 3D and one in 4D.  These include the original Penrose tiling, the four other previously known Penrose-like tilings, and sixteen that are new.  We also complete the enumeration of the quasicrystallographic space groups corresponding to the irreducible non-crystallographic reflection groups, by showing that there is a unique such space group in 4D (with nothing beyond 4D).
\end{abstract}

\maketitle


\section{Introduction}

The Penrose tilings \cite{Penrose74, Gardner77, Penrose78} (see Fig.~\ref{2D_10fold_AmmannLines}) have been a source of fascination for mathematicians and physicists ever since their discovery in the 1970s (see {\it e.g.}\ \cite{GrunbaumShephard, Janot, Senechal, Baake2002, SteurerDeloudi, BaakeGrimme}).  
Here we present a new perspective on these important objects (and others like them).   In particular, we point out a precise geometric relationship between: (i) certain natural pairs of irreducible reflection groups (which we call ``Coxeter pairs"); (ii) self-similar quasicrystalline patterns formed by superposing sets of 1D quasi-periodicially-spaced lines, planes or hyper-planes (which we call ``Ammann patterns"); and (iii) the Penrose-like tilings that correspond to these patterns.  More specifically, the perspective we propose is that a Penrose-like tiling should be regarded as the dual of a more fundamental object: an Ammann pattern; and this Ammann pattern, in turn, can be derived from the relationship between the two members of a Coxeter pair, in a way that we will make precise.  We hope to convey some of the advantages of this perspective.  

Let us start by briefly summarizing our results.  We first introduce the notion of a Coxeter pair, and enumerate all such pairs (there are two infinite families plus four exceptional cases).  We then provide an explicit, simple, systematic procedure for using these pairs to construct all irreducible Ammann patterns.  The construction relies upon the list of reflection quasilattices enumerated in Ref.~\cite{ReflectionQuasilattices} and the list of the ten special self-similar 1D quasilattices enumerated in Ref.~\cite{BoyleSteinhardt1D}.  However, more is required.  In order for an Ammann pattern to be self-similar, a precise choice of relative ``phases" of the 1D quasilattices must be adopted.  An important feature of our construction is that it automatically generates the correct phases.  Consequently, all such Ammann patterns and their self-similarity transformations turn out to be described explicitly by a single, closed-form analytic expression, which is one of our main results.  

Our next step is to derive a precise ``dualization formula" (\ref{dualization_formula}).  We explain how this formula may be used to convert each Ammann pattern into a dual Penrose-like tiling, and to systematically derive key properties of that tiling including its Ammann decoration, its inflation transformation, and its matching rules.  We emphasize that our approach is rather different from the previous generalized dual \cite{deBruijn81b, SocolarLevineSteinhardt85} and cut-and-project \cite{deBruijn81b, KramerNeri84, DuneauKatz85, GahlerRhyner86, MoodyModelSetsSurvey, BaakeGrimme} methods in that these other approaches generate a wide range of tilings for any given symmetry that are not generally self-similar and do not possess inflation rules.

Using these techniques, we then present the complete set of {\it minimal} Ammann patterns and Penrose-like tilings.   These have the minimal set of 1D Ammann directions compatible with their orientational symmetry.  There are only a handful of such minimal patterns/tilings: eleven in 2D, nine in 3D, one in 4D, and none in higher dimensions.  These include all of the previously-obtained Ammann patterns/Penrose-like tilings: the original Penrose tiling \cite{Penrose74, Gardner77, Penrose78, deBruijn81b, SocolarSteinhardt86, GrunbaumShephard} (in 2D, with 10-fold symmetry), the Ammann-Beenker pattern \cite{GrunbaumShephard, Beenker, AmmannGrunbaumShephard, Socolar89} (in 2D, with 8-fold symmetry), the Ammann-Socolar tiling \cite{Socolar89} (in 2D, with 12-fold symmetry), a previously-unpublished tiling found by Socolar (in 2D, with 12-fold symmetry) \cite{SocolarPrivate}, and the Socolar-Steinhardt tiling \cite{SocolarSteinhardt86} (in 3D, with icosahedral symmetry).  The remaining minimal patterns/tilings are new: seven of the eleven 2D patterns, eight of the nine 3D patterns, and the 4D pattern.   We provide figures displaying each of the eleven minimal 2D patterns/tilings along with their Ammann decorations and inflation rules.  

Our construction of the minimal Ammann patterns also yields, as a by-product, a number of interesting ``Ammann cycles" -- sets of minimal Ammann-like patterns that are {\it different} from one another ({\it i.e.}\ in different local isomorphism classes), but that cycle into one another under inflation.  These are not strictly Ammann patterns, as they are not self-similar after a single inflation (but only after two or three inflations).  However, they are closely related, and of comparable interest.  We find that there are four such minimal Ammann cycles, all in 2D: (i) a trio of three different patterns with 10-fold symmetry (previously found by Socolar \cite{SocolarPrivate}) that cycle into one another under inflation; and (ii) three separate duos of 12-fold symmetric patterns (all new), where each duo consists of two different 12-fold symmetric patterns which cycle into one another in alternating fashion under inflation.

Finally, we complete the enumeration of the quasicrystallographic space groups \cite{Mermin2D1988, Mermin3D1988, Mermin1992} corresponding to the irreducible non-crystallographic reflection point groups, by showing that there is a unique such space group in 4D.  Since there are none in higher dimensions, this 4D space group is the maximal one in terms of both its dimension and point symmetry.

As part of this investigation, we wish to promote the view that the Ammann pattern is an important entity in and of itself, not simply the decoration of a certain tesselation.  First, the Ammann pattern is a quasicrystal tiling in its own right, since the Ammann lines/planes/hyperplanes divide up space into a finite number of polytopes arranged quasiperiodically in a crystallographically forbidden pattern.  Its diffraction pattern consists of Bragg peaks arranged with the same crystallographically forbidden symmetry.  (This is in contrast to de Bruijn's periodic pentagrid \cite{deBruijn81b}, which contains an infinite number of different ``tiles", including tiles of arbitrarily small size.)  While a Penrose-like tiling has the simplifying property that all the edge lengths of all the tiles are the same, an Ammann pattern (regarded as a tiling) has the simplifying property that all the tile edges join up to form infinite unbroken straight lines (or, more correctly, all the codimension-one tile ``faces" join up to form infinite unbroken codimension-one affine spaces).  

In fact, the Ammann pattern with orientational symmetry $G$ is in many ways the {\it simplest} type of quasicrystal with orientational symmetry $G$.  In particular, as far as we are aware, the Ammann pattern is the only type of quasicrystal (with orientational order $G$) that can be explicitly described by a closed-form analytic expression.  The same is true for its diffraction pattern \cite{LevineSteinhardt86}.  By contrast, we cannot explicitly describe the corresponding Penrose-like tiling by a closed-form analytic expression, and must instead content ourselves with an algorithm for constructing it ({\it e.g.}\ by the cut-and-project method, by one of the two dualization methods described above, or by simply piecing together tiles according to the matching rules).  This point becomes increasingly significant as we move from the original (2D, 10-fold) case to other analogues in higher dimension and/or with more complicated orientational symmetries: in these more complicated cases, the number of Penrose-like tiles proliferates, and constructing and analyzing the tiling becomes unwieldy.  By contrast, all of the different Ammann patterns (regardless of their symmetry or dimension) are described by essentially the same formula, so that the higher-dimension or higher-symmetry cases are no more complicated than the original one.  Since one of the main purposes of these tilings is to provide a simple and useful model for investigating quasicrystalline order, this is an important point.  

We also note that the Ammann pattern brings out most directly a deep fact about quasicrystalline order with orientational symmetry $G$: namely, that it may be built up from (or decomposed into) 1D quasiperiodic constituents.  Of course, in many physics problems, separation of a higher-dimensional problem into 1D problems is an important step.  In this regard, we note that in quasicrystals in particular, some striking analytical results have been obtained in 1D (see {\it e.g.}\ \cite{KohmotoKadanoffTang, KaluginKitaevLevitov, KohmotoSutherlandTang}), while analogous problems in higher dimension have often resisted solution.  The fact that Ammann patterns are described by a closed-form analytical expression, and are already decomposed into their 1D constituents, suggests that they are likely to be a particularly fruitful starting point for future investigations.  

The outline of the paper (in more detail) is as follows.

In Section \ref{Coxeter_Pairs}, we introduce the notion of a Coxeter pair.  We begin, in Subsection \ref{root_systems}, by briefly reviewing Coxeter's classification of the finite reflection groups in terms of root systems and Coxeter-Dynkin diagrams, and some basic facts about non-crystallographic root systems.  Then, in Subsection \ref{coxeter_pairs}, we explain that, in some cases, a non-crystallographic reflection group (of lower rank) has a natural crystallographic partner (of higher rank).  We find all such ``Coxeter pairs" in Appendix \ref{Finding_all_Coxeter_pairs} and collect them in Table \ref{CoxeterPairs}: they are organized into two infinite families plus four exceptional cases.  

In Section \ref{general_construction}, we show how to construct all irreducible Ammann patterns.  For starters, in Subsection \ref{Introducing_Ammann_patterns}, we introduce the idea of an irreducible Ammann pattern -- the natural generalization of the original Ammann pattern shown in Fig.~\ref{2D_10fold_AmmannLines}.  Then, in Subsection \ref{constructing_ammann_patterns} we explain how to obtain all irreducible Ammann patterns via a geometrical construction based on the Coxeter pairs introduced previously; and we obtain a useful closed-form analytic expression that describes all such Ammann patterns in a simple and unified way.  

In Section \ref{Penrose_Tilings}, we show how to construct all of the Penrose-like tilings dual to these  Ammann patterns.  First, in Subsection \ref{dualization}, we derive a dualization formula that allows us to convert any irreducible Ammann pattern into a dual Penrose tiling, and to scale and shift this Penrose tiling by the right amount (so that, if we superpose it on the original Ammann pattern, the Ammann lines decorate the Penrose prototiles in only a finite number of different ways).  Then, in Subsections \ref{AmmannDecoration} and \ref{InflationDecoration}, we show to use this formula to derive the Ammann decorations and inflation rules for the prototiles in the Penrose-like tilings.

Applying these techniques, in Section \ref{Minimal_Penrose_Tilings} we present the complete set of {\it minimal} Ammann patterns and Penrose-like tilings (eleven in 2D, nine in 3D, one in 4D, and none in higher dimensions).  At the end of the section, we present figures explicitly displaying the eleven 2D patterns/tilings, along with their Ammann decorations and inflation rules, as well as the four minimal ``Ammann cycles" (one 10-fold-symmetric "Ammann trio" and three "12-fold-symmetric "Ammann duos") obtained as a by-product.

In Section \ref{Discussion}, we discuss matching rules: in particular, we show how the Ammann-pattern perspective makes the existence of perfect local matching rules particularly transparent; and we point out that the Coxeter-pair perspective suggests a beautiful implementation of these rules.

Finally, in Appendix B, we complete the enumeration of the quasicrystallographic space groups corresponding to the irreducible non-crystallographic reflection point groups, by showing that there is a unique such space group in 4D (with nothing in higher dimensions, so that this 4D space group is maximal in terms of both its dimension and point symmetry).

Throughout the paper, to illustrate our formalism, we will use the following two examples:
\begin{itemize}
\item Example 1: the Penrose tiling \cite{Penrose74, Gardner77},
\item Example 2: the Ammann-Beenker tiling \cite{GrunbaumShephard, Beenker, AmmannGrunbaumShephard}.
\end{itemize}

In what follows $\mathbb{R}$ are the real numbers, $\mathbb{Q}$ are the rationals, and $\mathbb{H}$ are the quaternions.

\section{Root systems and Coxeter Pairs}
\label{Coxeter_Pairs}

\subsection{Root systems}
\label{root_systems}

The modern classification of the finite reflection groups (finite Coxeter groups) in terms of irreducible root systems and Coxeter-Dynkin diagrams is due to Coxeter  \cite{Cox1, Cox2, Cox3, RegularPolytopes, CoxeterMoser}.  For an introduction to these topics, see Chapter 4, Section 2 in \cite{ConwaySloane} (for a brief introduction) and Part 1 ({\it i.e.}\ Chs. 1-4) in \cite{Humphreys} (for more detail).  Here we review a few relevant points.

\begin{figure}
  \begin{center}
    \includegraphics[width=2.4in]{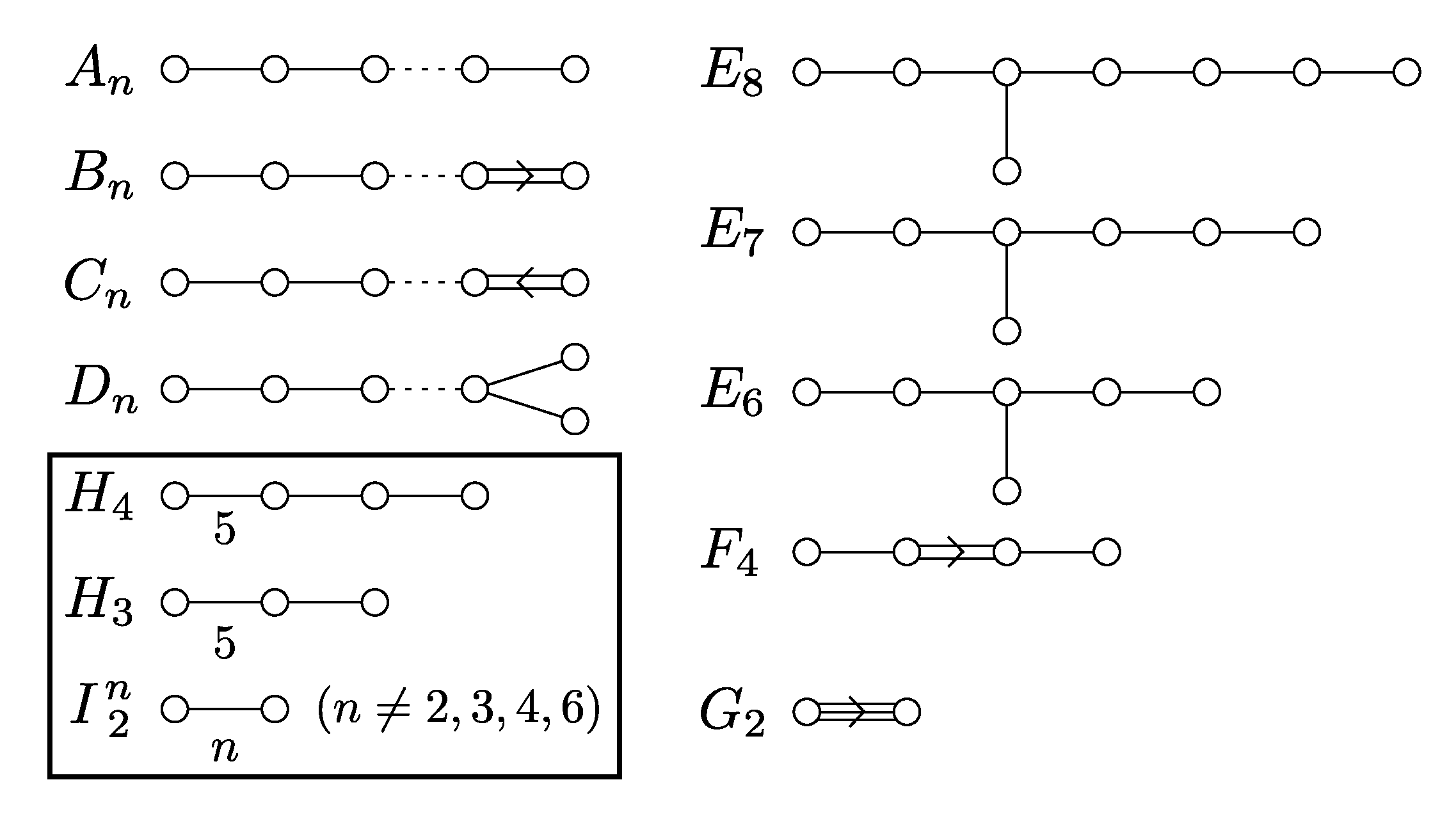}
  \end{center}
  \caption{The Coxeter-Dynkin diagrams for the finite irreducible root systems (or, equivalently, the finite irreducible reflection groups).  The non-crystallographic cases are boxed.}
 \label{CoxeterDynkinDiagrams}
\end{figure}

The irreducible finite reflection groups and their corresponding root systems may be neatly described by Coxeter-Dynkin diagrams (see \cite{ConwaySloane, Humphreys} and Fig. \ref{CoxeterDynkinDiagrams}).  These come in two varieties: crystallographic and non-crystallographic.  The crystallographic cases are familiar from the theory of Lie groups and Lie algebras, and are summarized in Table 4.1 in \cite{ConwaySloane}: they come in four infinite families ($A_{n}$ with $n\geq1$, $B_{n}$ with $n\geq2$, $C_{n}$ with $n\geq3$ and $D_{n}$ with $n\geq4$) and five exceptional cases ($G_{2}$, $F_{4}$, $E_{6}$, $E_{7}$ and $E_{8}$).  The remaining roots systems are non-crystallographic: almost all of these are in 2D ($I_{2}^{n}$, $n=5,7,8,9,\ldots$), with just one in 3D ($H_{3}$), one in 4D ($H_{4}$), and none in higher dimensions.

Let us briefly describe the non-crystallographic roots systems, since they will be of particular interest to us in this paper.

First consider $I_{2}^{n}$.  In geometric terms, the $2n$ roots of $I_{2}^{n}$ are perpendicular to the $n$ mirror planes of a regular $n$-sided polygon; note that when $n$ is odd, these mirror planes are all equivalent (each intersects a vertex and its opposite edge), but when $n$ is even the mirror planes split into two sets (those that intersect two opposite vertices, and those that intersect two opposite edges).  In algebraic terms, we can think of the $2n$ roots as $2n$ complex numbers.  When $n$ is odd, these are the $(2n)$th roots of unity: $\zeta_{2n}^{k}$ ($k=1,\ldots,2n$), where $\zeta_{n}\equiv{\rm exp}(2\pi i/n)$.  When $n$ is even, the $2n$ roots break into two rings: (i) a first ring of roots $\zeta_{n}^{k}$ ($k=1,\ldots,n$) which point to the vertices of the regular $n$-gon; and (ii) a second ring of roots that point to the edge midpoints of the regular $n$-gon (and may be expressed as integer linear combinations of the roots in the first ring).  The $I_{2}^{n}$ reflections generate the symmetry group of the regular $n$-gon, of order $2n$.  

Next consider $H_{3}$.  If $\tau$ and $\sigma$ are the golden ratio and its Galois conjugate, respectively:
\begin{equation}
  \tau\equiv \frac{1}{2}(1+\sqrt{5}),\qquad\qquad \sigma\equiv\frac{1}{2}(1-\sqrt{5}),
\end{equation}
then the $H_{3}$ roots are the 30 vectors obtained from
\begin{subequations}
  \label{H3_roots}
  \begin{eqnarray}
    &\{\pm1,0,0\}& \\
    &\frac{1}{2}\{\pm\tau,\pm1,\pm\sigma\}&
  \end{eqnarray}
\end{subequations}
by taking all combinations of $\pm$ signs, and all even permutations of the three coordinates.  These point to the 30 edge midpoints of a regular icosahedron \cite{RegularPolytopes}, and the corresponding reflections generate the full symmetry group of the icosahedron (of order 120).

Finally consider $H_{4}$.  The $H_{4}$ roots are the 120 vectors obtained from
\begin{subequations}
  \label{H4_roots}
  \begin{eqnarray}
    &\{\pm1,0,0,0\}& \\
    &\frac{1}{2}\{\pm1,\pm1,\pm1,\pm1\}& \\
    &\frac{1}{2}\{0,\pm\tau,\pm1,\pm\sigma\}&
  \end{eqnarray}
\end{subequations}
by taking all combinations of $\pm$ signs, and all even permutations of the four coordinates.  From a geometric standpoint, these are the 120 vertices of a 4D regular polytope called the 600 cell \cite{RegularPolytopes}.  From an algebraic standpoint, each 4-vector $v\in\mathbb{R}^{4}$ corresponds to a quaternion $q\in\mathbb{H}$: that is, the 4-vector $v=\{w,x,y,z\}\in\mathbb{R}^{4}$ corresponds to the quaternion $q=w+x{\bf i}+y{\bf j}+z{\bf k}\in\mathbb{H}$ (where the three imaginary quaternion units $\{{\bf i},{\bf j},{\bf k}\}$ satisfy Hamilton's celebrated relations ${\bf i}^{2}={\bf j}^{2}={\bf k}^{2}={\bf i}{\bf j}{\bf k}=-1$ \cite{ConwaySmith}).  In this way, the $H_{4}$ roots are mapped to a special set of 120 quaternions known as the unit icosians \cite{ConwaySloane, ElserSloane}.  The $H_{4}$ reflections generate the full symmetry group of the 600 cell: this group has $120^{2}=14400$ elements, corresponding to all maps from $\mathbb{H}\to\mathbb{H}$ of the form $q\to\bar{q}_{1}q q_{2}$ or $q\to\bar{q}_{1}\bar{q}q_{2}$, where $q_{1}$ and $q_{2}$ are unit icosians \cite{ConwaySmith, ElserSloane}.

\begin{table}
\begin{center}
\begin{tabular}{c|c|c}
non-crystallographic root system & rank $d$ & rational rank $d_{\mathbb{Q}}$ \\
\hline
$I_{2}^{(n)}$ & $2$ & $\phi(n)$ \\
$H_{3}$ & $3$ & $6$ \\
$H_{4}$ & $4$ & $8$ 
\end{tabular}
\end{center}
\caption{The non-crystallographic roots systems, with their ordinary and rational ranks.  Here Euler's totient function, $\phi(n)$, is the number of natural numbers $< n$ that are relatively prime to $n$.}
\label{indexing_dimension}
\end{table} 

Given a root system $\theta$: 
\begin{itemize}
\item its ``rank" $d$ is the dimension of the vector space (over $\mathbb{R}$) generated by taking all {\it real} linear combinations of the roots; and
\item its ``{\it rational} rank" $d_{\mathbb{Q}}$ is the dimension of the vector space (over $\mathbb{Q}$) generated by taking all {\it rational} linear combinations of the roots. 
\end{itemize}
When $\theta$ is crystallographic, the rational rank is the same as the ordinary rank ($d_{\mathbb{Q}}=d$); but in the non-crystallographic case, the rational rank is larger than the ordinary rank ($d_{\mathbb{Q}}>d$).  Thus, a non-crystallographic root system lives a double life: in one sense, its roots live in the lower-dimensional space $\mathbb{R}^{d}$; but in another sense, they live in the higher-dimensional space $\mathbb{Q}^{d_{\mathbb{Q}}}$.  The ordinary and rational ranks of the non-crystallographic root systems are summarized in Table \ref{indexing_dimension}.

\subsection{Coxeter Pairs}
\label{coxeter_pairs}

Now consider two irreducible root systems, $\theta$ and $\theta^{\parallel}$:
\begin{enumerate}
\item $\theta$ is a crystallographic root system of rank $d$, whose $j$th root (denoted ${\bf r}_{j}$) corresponds to a reflection $R_{j}$ that acts on the $d$-dimensional coordinates ${\bf x}$ as:
\begin{subequations}
\begin{equation}
  R_{j}:{\bf x}\to{\bf x}-2\frac{{\bf x}\cdot{\bf r}_{j}}{{\bf r}_{j}\cdot{\bf r}_{j}}{\bf r}_{j}.
\end{equation} 
\item $\theta^{\parallel}$ is a non-crystallographic root system of rank $d^{\parallel}$, whose $j$th root (denoted ${\bf r}_{j}^{\parallel}$) corresponds to a reflection $R_{j}^{\parallel}$ that acts on the $d^{\parallel}$-dimensional coordinate ${\bf x}^{\parallel}$ as:
\begin{equation}
  R_{j}^{\parallel}:{\bf x}^{\parallel}\!\to{\bf x}^{\parallel}\!-2\frac{{\bf x}_{}^{\parallel}\!\cdot{\bf r}_{j}^{\parallel}}
  {{\bf r}_{j}^{\parallel}\!\cdot{\bf r}_{j}^{\parallel}}{\bf r}_{j}^{\parallel}.
\end{equation} 
\end{subequations}
\end{enumerate}
We say $\theta$ and $\theta^{\parallel}$ form a ``Coxeter pair" (of degree $N$) if:
\begin{enumerate}
\item they have the same rational rank ({\it i.e.}\ they both live in $\mathbb{Q}^{d}$); and 
\item from the maximally symmetric orthogonal projection of the $\theta$ roots onto a $d^{\parallel}$-dimensional plane (the ``parallel space") we obtain ($N$ copies of) the $\theta^{\parallel}$ roots. 
\end{enumerate}
Let us illustrate with our two basic examples:
\begin{itemize}
\item Example 1: The root systems $\theta=A_{4}$ (with $d=4$) and $\theta^{\parallel}=I_{2}^{5}$ (with $d^{\parallel}=2$) form a Coxeter pair (of degree $N=d/d^{\parallel}=2$).  In both cases, the roots live in $\mathbb{Q}^{4}$; and if we take the maximally-symmetric 2D projection of the 20 roots of $A_{4}$, we obtain ($N=2$ copies of) the 10 roots of $I_{2}^{5}$ (see Fig.~\ref{A4rootfig}).  This is the Coxeter pair relevant to the Penrose tiling (Fig.~\ref{2D_10fold_PurePenrose}).
\item Example 2: The root systems $\theta=B_{4}$ (with $d=4$) and $\theta^{\parallel}=I_{2}^{8}$ (with $d^{\parallel}=2$) form a Coxeter pair (of degree $N=d/d^{\parallel}=2$).  In both cases, the roots live in $\mathbb{Q}^{4}$; and if we take the maximally-symmetric 2D projection of the 32 roots of $B_{4}$, we obtain ($N=2$ copies of) the 16 roots of $I_{2}^{8}$ (see Fig.~\ref{B4rootfig}).  This is the Coxeter pair relevant to the Ammann-Beenker tiling (Fig.~\ref{2D_8foldA1_PurePenrose}).
\end{itemize}
In Appendix \ref{Finding_all_Coxeter_pairs}, we obtain the complete list of Coxeter pairs. The results are summarized in Table \ref{CoxeterPairs}: the Coxeter pairs fall into two infinite families and four exceptional cases.  Note that most non-crystallographic root systems do {\it not} belong to a Coxeter pair; and if the non-crystallographic root system $\theta^{\parallel}$ {\it does} belong to a Coxeter pair, it belongs to a {\it unique} Coxeter pair ({\it i.e.}\ it has a unique crystallographic partner $\theta$).

\begin{table}
\begin{center}
\begin{tabular}{c|c|c}
$\theta^{\parallel}$ & $\theta$ & $N=d/d^{\parallel}$ \\
\hline
$I_{2}^{p}$ ($p$ any prime $\geq5$) & $A_{p-1}$ & $(p-1)/2$ \\
$I_{2}^{2^{m}}$ ($m$ any integer $\geq3$) & $B_{2^{m-1}}/C_{2^{m-1}}$ & $2^{m-2}$ \\
$I_{2}^{12}$ & $F_{4}$ & $2$ \\
$I_{2}^{30}$ & $E_{8}$ & $4$ \\
$H_{3}$ & $D_{6}$ & $2$ \\
$H_{4}$ & $E_{8}$ & $2$
\end{tabular}
\end{center}
\caption{The complete list of Coxeter pairs.  Here $\theta^{\parallel}$ is the non-crystallographic root system, $\theta$ is the crystallographic partner, and $N=d/d^{\parallel}$ is the degree.}
\label{CoxeterPairs}
\end{table}

\begin{figure}
  \begin{center}
    \includegraphics[width=2.4in]{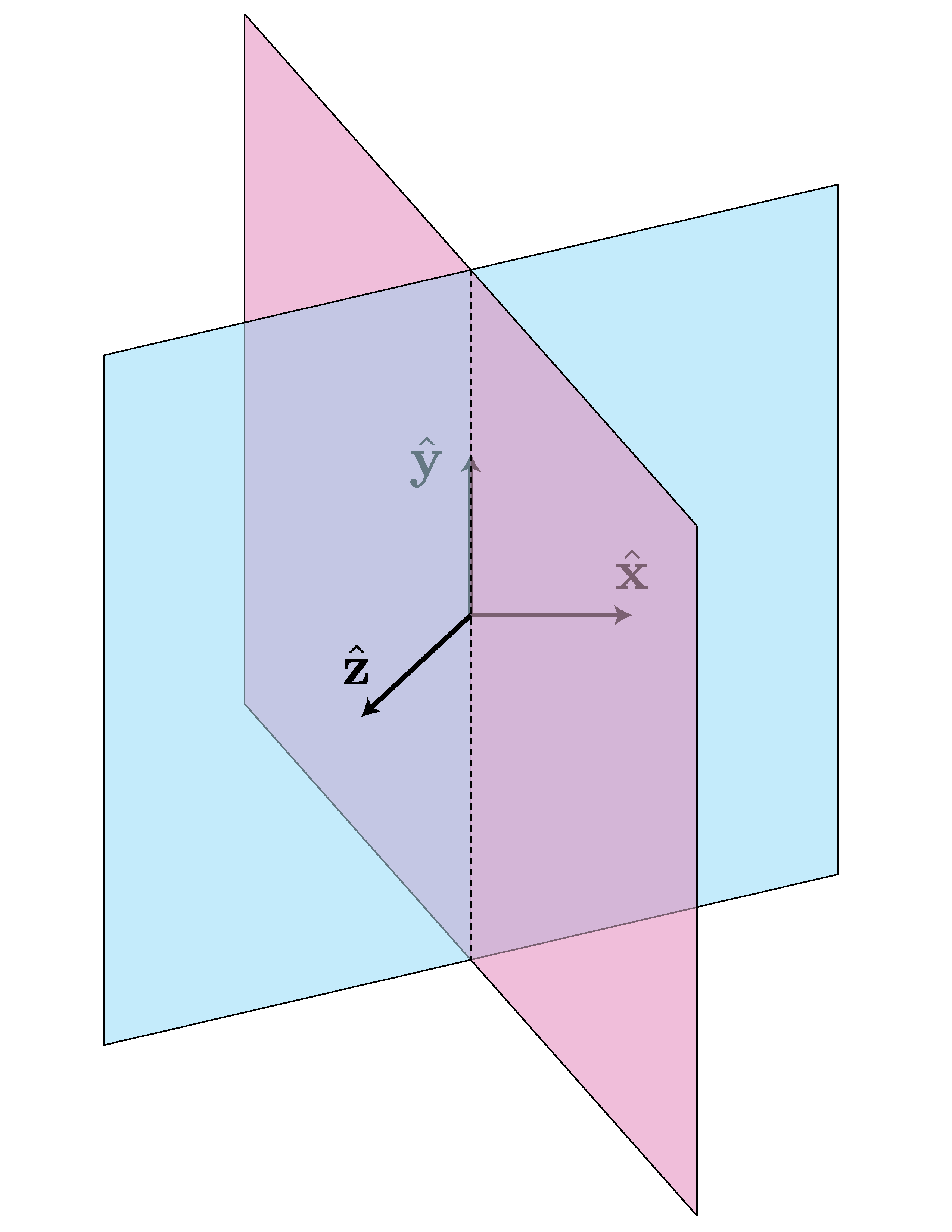}
  \end{center}
  \caption{In this figure, the pink and blue planes represent two different $\theta$ mirrors that live in the higher-dimensional space ({\it i.e.}\ the $d$-dimensional embedding space in which the $\theta$ mirrors act), while the $xy$-plane represents the lower-dimensional space ({\it i.e.}\ the $d^{\parallel}$-dimensional parallel space in which the $\theta^{\parallel}$ mirrors act).  The intersection of the blue $\theta$ mirror with the parallel space (the $xy$-plane) defines a $\theta^{\parallel}$ mirror (which, in this case, is the $y$-axis); and the pink $\theta$ mirror intersects the parallel space in exactly the {\it same} place, and thus defines exactly the {\it same} $\theta^{\parallel}$ mirror.  In a Coxeter pair of degree $N$, there are $N$ such $\theta$ mirrors that all have the same intersection with the parallel space, and thus all define the same $\theta^{\parallel}$ mirror; all the $\theta^{\parallel}$ mirrors arise in this way, and the $\theta$ mirrors are naturally grouped into $N$-fold multiplets in the process.}
  \label{mirrorfig}
\end{figure}

There is another way to think about the relationship between the higher-dimensional root system $\theta$ and the lower-dimensional root system $\theta^{\parallel}$: the intersection of any $\theta$ mirror (a codimension-one plane in $d$ dimensions) with the parallel space yields a $\theta^{\parallel}$ mirror (a codimension-one plane in $d^{\parallel}$ dimensions); and, in fact, each $\theta^{\parallel}$ mirror arises in this manner in $N$ distinct ways ({\it i.e.}\ from $N$ distinct $\theta$ mirrors that differ in $d$ dimensions, but all degenerate with one another in their intersection with the $d^{\parallel}$-dimensional parallel space -- see Fig.~\ref{mirrorfig}).

Now let ${\bf f}_{k}$ ($k=1,\ldots,d$) be the fundamental roots of $\theta$ (the $d$ roots perpendicular to one of the fundamental regions bounded by the mirror planes of $\theta$ -- see Section 4.2 in \cite{ConwaySloane}); let ${\bf f}_{k}^{\parallel}$ ($k=1,\ldots,d^{\parallel}$) be the fundamental roots of $\theta^{\parallel}$; and let $P^{\parallel}$ denote the orthogonal projection operator from $d$ dimensions onto the $d^{\parallel}$-dimensional parallel space: ${\bf x}^{\parallel}=P^{\parallel}{\bf x}$.  Any point ${\bf x}$ in the higher-dimensional space $\mathbb{Q}^{d}$ may be written as a $\mathbb{Q}$-linear combination of the ${\bf f}_{k}$:
\begin{equation}
  {\bf x}=\sum_{k=1}^{d}\varphi_{k}{\bf f}_{k}\qquad(\varphi_{k}\in\mathbb{Q});
\end{equation}
and when we orthogonally project this point onto the parallel space, the resulting point ${\bf x}^{\parallel}=P^{\parallel}{\bf x}$ is a $\mathbb{Q}$-linear combination of the projected fundamental roots $P^{\parallel}{\bf f}_{k}$ of $\theta$ or, equivalently, a $\mathbb{K}$-linear combination of the fundamental roots ${\bf f}_{k}^{\parallel}$ of $\theta^{\parallel}$ (where the field $\mathbb{K}$ is a degree $N$ extension of $\mathbb{Q}$ -- {\it i.e.} a field obtained by adjoining to the rational numbers an appropriate root of a $N$th-order polynomial):
\begin{equation}
  {\bf x}^{\parallel}=\sum_{k=1}^{d}\varphi_{k}P^{\parallel}{\bf f}_{k}=\sum_{k=1}^{d^{\parallel}}\varphi_{k}^{\parallel}{\bf f}_{k}^{\parallel}
  \qquad(\varphi_{k}\in\mathbb{Q},\;\varphi_{k}^{\parallel}\in\mathbb{K}).
\end{equation}
This map is invertible: given a point ${\bf x}^{\parallel}$ in the parallel space ({\it i.e.}\ a $\mathbb{K}$-linear combination $\sum\varphi_{k}^{\parallel}{\bf f}_{k}^{\parallel}$ or, equivalently, a $\mathbb{Q}$-linear combination $\sum\varphi_{k}P^{\parallel}{\bf f}_{k}$) it lifts to a unique point ${\bf x}$ in the embedding space (namely, the point ${\bf x}=\sum\varphi_{k}{\bf f}_{k}$).  

For example, the non-crystallographic root system $I_{2}^{5}$, with two fundamental roots ${\bf f}_{k}^{\parallel}$ ($k=1,2$), is paired with the crystallographic root system $A_{4}$, with four fundamental roots ${\bf f}_{k}$ ($k=1,\ldots,4$); and, under the maximally-symmetric orthogonal projection of the $A_{4}$ roots onto 2D ({\it i.e.} onto the ``Coxeter plane" -- see Appendix \ref{Finding_all_Coxeter_pairs}), any 4D point ${\bf x}$ that is a rational linear combination of the fundamental roots ${\bf f}_{k}$ of $A_{4}$ projects to a 2D point ${\bf x}^{\parallel}$ that may either be written as a rational linear combination of the projected fundamental roots of $A_{4}$, $P^{\parallel}{\bf f}_{k}$, or else as a $\mathbb{K}$-linear combination of the two fundamental roots of $I_{2}^{5}$, ${\bf f}_{k}^{\parallel}$, where in this case $\mathbb{K}=\mathbb{Q}(\sqrt{5})$ is a quadratic extension of the rationals; and the map is invertible, so any such 2D point ${\bf x}^{\parallel}$ lifts to a corresponding 4D point ${\bf x}$.

\subsection{Quadratic Coxeter Pairs}
\label{quadratic_coxeter_pairs}

Let us distinguish between the $N=2$ (or ``quadratic") Coxeter pairs, and the $N>2$ (or ``higher") Coxeter pairs.  From Table \ref{CoxeterPairs} we see that all of the non-crystallographic root systems in dimension $d^{\parallel}>2$, and three of the simplest in dimension $d^{\parallel}=2$ are ``quadratic", while the rest (which are all in dimension $d^{\parallel}=2$) are ``higher".  In the remainder of this paper, we will only need the quadratic Coxeter pairs: we will show how they may be used to elegantly construct all of the irreducible Ammann patterns and their dual Penrose tilings.  In future work, it will be interesting to study the ``higher" Coxeter pairs, and the possibility of using them to construct ``higher" Ammann patterns and ``higher" Penrose tilings in 2D.

In the $N=2$ case, $\mathbb{K}$ is a real quadratic field $\mathbb{Q}(\sqrt{D})$, where $D$ is a square-free positive integer (see Table \ref{QuadraticCoxeterPairs}) and the $d$-dimensional embedding space is split into two parts (the $\parallel$ and $\perp$ spaces), both of which have the same dimension ($d^{\parallel}=d/2$) and are simply related by Galois conjugation $\sqrt{D}\to-\sqrt{D}$.  For this reason, instead of using ``$\parallel$" and ``$\perp$" super/subscripts, it will be more convenient to use ``$+$" and ``$-$" super/subscripts (so that ``$+$" and ``$-$" stand for ``$\parallel$" and ``$\perp$", respective, just as in Table 1 of Ref.~\cite{BoyleSteinhardt1D}).  

In particular, if $P^{+}$ and $P^{-}$ denote the orthogonal projectors onto the $\parallel$ and $\perp$ spaces, respectively, and we split each fundamental root ${\bf f}_{k}$ of $\theta$ into its $\parallel$-space part $P^{+}{\bf f}_{k}$ and its $\perp$-space part $P^{-}{\bf f}_{k}$:
\begin{equation}
  {\bf f}_{k}=P^{+}{\bf f}_{k}+P^{-}{\bf f}_{k}
\end{equation}
and then express these parts in the original ${\bf f}_{k}$ basis
\begin{equation}
  P^{\pm}{\bf f}_{k}=\sum_{k'=1}^{d}\zeta_{k,k'}^{\pm}{\bf f}_{k'}\qquad\zeta_{k,k'}^{\pm}\in\mathbb{Q}(\sqrt{D})
\end{equation}
then the ``$+$" coefficients $\zeta_{k,k'}^{+}$ will be related to the ``$-$" coefficients $\zeta_{k,k'}^{-}$ by $\sqrt{D}\to-\sqrt{D}$.  And thus, if we consider any point ${\bf x}=\sum\varphi_{k}{\bf f}_{k}$ that is a rational linear combination of the ${\bf f}_{k}$, and express the $\parallel$ and $\perp$ parts in the ${\bf f}_{k}$ basis, 
\begin{equation}
  P^{\pm}{\bf x}=\sum_{k=1}^{d}\eta_{k}^{\pm}{\bf f}_{k}\qquad\eta_{k}^{\pm}\in\mathbb{Q}(\sqrt{D})
\end{equation}
then the ``$+$" coefficients $\eta_{k}^{+}$ will also be related to the ``$-$" coefficients $\eta_{k}^{-}$ by $\sqrt{D}\to-\sqrt{D}$.  

\begin{table}
\begin{center}
\begin{tabular}{c|c|c}
$\theta_{\parallel}$ & $\theta$ & $\mathbb{K}$ \\
\hline
$I_{2}^{5}$ & $A_{4}$ & $\mathbb{Q}(\sqrt{5})$ \\
$I_{2}^{8}$ & $B_{4}/C_{4}$ & $\mathbb{Q}(\sqrt{2})$ \\
$I_{2}^{12}$ & $F_{4}$ & $\mathbb{Q}(\sqrt{3})$ \\
$H_{3}$ & $D_{6}$ & $\mathbb{Q}(\sqrt{5})$ \\
$H_{4}$ & $E_{8}$ & $\mathbb{Q}(\sqrt{5})$
\end{tabular}
\end{center}
\caption{Quadratic $(N=2$) Coxeter pairs and their corresponding field extensions.  Here $\theta_{\parallel}$ is the non-crystallographic root system, $\theta$ is the crystallographic partner, and $\mathbb{K}$ is the field extension.}
\label{QuadraticCoxeterPairs}
\end{table}
Note that, in our definition of Coxeter pairs, we have taken both members of the pair, $\theta^{\parallel}$ and $\theta$, to be irreducible.  This is because, in this paper (as explained in the next section) we are interested in constructing all {\it irreducible} Ammann patterns (which is why $\theta^{\parallel}$ is irreducible and characterized by quadratic irrationalities); and in all such cases, $\theta^{\parallel}$ {\it does} have a higher-rank partner $\theta$ that is also irreducible, so this definition leads to no loss of generality in terms of constructing the irreducible Ammann patterns.  However, it might nevertheless be interesting to also consider {\it generalized} Coxeter pairs in which $\theta^{\parallel}$, and possibly also $\theta$, are reducible; and to study the corresponding reducible quasicrystals they generate.  We leave this as a subject for future work.

\section{Constructing all irreducible Ammann patterns}
\label{general_construction}

\subsection{Introducing irreducible Ammann patterns}
\label{Introducing_Ammann_patterns}

Ammann noticed \cite{GrunbaumShephard, SenechalOnAmmann} that the two Penrose tiles could each be decorated with a special pattern of five line segments so that, in a defect-free Penrose tiling, the line segments join together to form five infinite sets of unbroken straight lines, parallel to the five edges of a regular pentagon (see the thin blue lines in Fig.~\ref{2D_10fold_AmmannLines}).  The pattern formed by these five infinite sets of lines is the prototype for what we will call an ``Ammann pattern".  It has three key properties.  (i) First, if we focus on any one of the five sets of parallel lines, they turn out to be spaced according to a 1D quasiperiodic sequence of long and short intervals.  More specifically, they form the simplest possible type of self-similar 1D quasilattice: a self-similar 1D quasilattice of degree two \cite{BoyleSteinhardt1D}.  (ii) Second, the pattern's (10-fold) orientational symmetry is described by an irreducible non-crystallographic reflection group and all of its constituent 1D quasilattices are equivalent up to this symmetry.  (iii) Third, the pattern is self-similar: that is, each of its constituent 1D quasilattices is equipped with a natural self-similarity (or ``inflation") transformation and, moreover, if we inflate each of the 1D constituents simultaneously, the pattern formed by their superposition is {\it also} self-similar.  We would like to find and understand all patterns with these three properties.

In more detail, consider a collection of $J$ unit vectors ${\bf e}_{j}^{+}$ ($j=1,\ldots,J$) in $d^{+}$ dimensions.  We will say that these vectors form an {\it irreducible non-crystallographic star} if:
\begin{enumerate}
\item the set $S=\{\pm{\bf e}_{1}^{+},\ldots,\pm{\bf e}_{J}^{+}\}$ (of all the vectors ${\bf e}_{j}^{+}$ and their opposites) is invariant under the action of an irreducible non-crystallographic reflection group $G(\theta^{+})$; and
\item the symmetry group $G(\theta^{+})$ acts transitively on the elements in $S$ (so that any two vectors in the set are equivalent up to symmetry).
\end{enumerate}
First focus on one vector ${\bf e}_{j}^{+}$ in this star, and imagine that, along this direction, we have an infinite sequence of (codimension-one) planes that are perpendicular to ${\bf e}_{j}^{+}$, with their locations along ${\bf e}_{j}^{+}$ forming a self-similar 1D quasilattice of degree two; and now imagine an analogous 1D quasilattice along each direction ${\bf e}_{j}^{+}$ in the star (where these 1D quasilattices are all locally isomorphic to one another).  Next consider the resulting ``multi-grid" (formed from the superposition of these $J$ differently-oriented 1D quasilattices): since each of the constituent 1D quasilattices is equipped with its own ``inflation" transformation under which it is self-similar (see \cite{BoyleSteinhardt1D} for a more detailed explanation), it is natural to consider the operation where we inflate all the constituent 1D quasilattices simultaneously, and to ask whether the multi-grid as a whole is also self-similar under this operation.  As we shall see, if the phases of the constituent 1D quasilattices are chosen {\it generically}, then the multi-grid will {\it not} be self-similar under inflation; but for a {\it special} choice of phases, the multi-grid {\it will} be -- and in this special case, we call the multi-grid an {\it irreducible Ammann pattern}.  

In this paper, we focus on patterns with irreducible reflection symmetry, because this captures and generalizes a natural feature of the Penrose tiling.  If an Ammann pattern is characterized by an irreducible reflection symmetry, then its dual Penrose-like tiling will be as well (and, of course, the original Ammann pattern and Penrose tiling have this property).  If, instead, a pattern in Euclidean space $\mathbb{R}^{n}$ is characterized by the {\it reducible} reflection group $G_{1}\times G_{2}$ (where $G_{1}$ and $G_{2}$ are two irreducible reflection groups of rank $r_{1}$ and $r_{2}$, respectively, with $r_{1}+r_{2}=n$), it means that $G_{1}$ acts purely on the subspace $\mathbb{R}^{r_{1}}$, while leaving the orthogonal subspace $\mathbb{R}^{r_{2}}$ invariant, while $G_{2}$ acts purely on the subspace $\mathbb{R}^{r_{2}}$, while leaving the orthogonal subspace $\mathbb{R}^{r_{1}}$ invariant.  Indeed, given two quasicrystalline point sets $X_{1}=\{x_{1,i}\}$ (a quasicrystalline subset of points in $\mathbb{R}^{r_{1}}$ with irreducible orientational symmetry $G_{1}$) and $X_{2}=\{x_{2,i}\}$ (a quasicrystalline subset of points in $\mathbb{R}^{r_{2}}$ with irreducible orientational symmetry $G_{2}$), we can trivially construct a new quasicrystalline set with orientational symmetry $G_{1}\times G_{2}$ by simply taking the product of these two sets -- {\it i.e.}\ by considering the set $X =\{(x_{1,i},x_{2,j}\}$ of points in $\mathbb{R}^{r_{1}+r_{2}}$ consisting of all pairs of points from the original two sets.  We consider Ammann patterns with irreducible reflection symmetry to rule out quasicrystals like these that are decomposable into a product of two quasicrystals and, more generally, less tightly knit together by their orientational symmetry than an irreducible quasicrystal like the Penrose tiling is.

It is important to emphasize that our definition of irreducible Ammann patterns (and our definition of Penrose-like tilings as the duals of these patterns) represents a choice about how to define a class of objects that nicely generalize the original Ammann pattern and Penrose tiling, but this choice excludes other types of patterns that may also be very interesting,and represent generalizations of the Ammann pattern and Penrose tiling in a different sense: {\it e.g.}\ quasiperidic patterns with {\it crystallographic} orientational symmetries, quasiperiodic tilings with {\it reducible} non-crystallographic symmetries, quasiperiodic tilings whose orientational symmetry is described by a non-crystallographic {\it rotation} group rather than a non-crystallographic {\it reflection} group ({\it e.g.} because the pattern is {\it chiral}), quasiperiodic patterns that are characterized by cubic irrationalities or higher-order irrationalities, etc.  These are potentially interesting extensions of the present work that we leave as topics for future investigation.

As mentioned above, Ammann patterns are built from 1D self-similar quasilattices of the simplest possible kind: those consisting of just two different intervals ($L$ and $S$), with just two possible separations between successive $L$'s, and just two possible separations between successive $S$'s.  (These are the self-similar 1D quasilattices of degree two studied in \cite{BoyleSteinhardt1D}.)  Under a self-similarity transformation, such quasilattices rescale by a quadratic irrationality; and, as we will see, this will restrict us to those non-crystallographic root systems that belong to a quadratic Coxeter pair (and are hence characterized by quadratic irrationalities).  As explained in Subsection \ref{quadratic_coxeter_pairs}, in dimension $d^{\parallel}>2$ this is no restriction at all (since all the non-crystallographic root systems in those dimensions -- namely $H_{3}$ and $H_{4}$ -- do indeed belong to quadratic root systems); but in dimension $d^{\parallel}=2$, where not all Coxeter pairs are quadratic, it amounts to restricting our attention to the root systems $I_{2}^{5}$, $I_{2}^{8}$ and $I_{2}^{12}$ (which will yield 2D Ammann patterns with 5/10-fold, 8-fold and 12-fold symmetry, respectively).  It is worth noting that that these symmetries stand out as being of particular importance: (i) as the symmetry axes that have been experimentally observed in physical quasicrystals in the lab; and (ii) also in the study of mathematical quasicrystals and quasicrystalline tilings, where many of the most interesting and widely studied examples have these symmetries (in addition to the Penrose tiling with 10-fold order in 2D, these include the Ammann-Beenker tiling with 8-fold order in 2D; the Ammann-Socolar tiling \cite{Socolar89}, Schlottmann's square-triangle tiling \cite{HermissonEtAl, BaakeGrimme}, and the shield tiling \cite{NiizekiMitani, GahlerPhD, BaakeGrimme} with 12-fold symmetry in 2D; the Socolar-Steinhardt tiling \cite{SocolarSteinhardt86}, Ammann's rhombohedral tiling \cite{GardnerBook} and Danzer's tetrahedral tiling \cite{Danzer89} with icosahedral ($H_{3}$) order in 3D; and the Elser-Sloane quasicrystal with hyper-icosahedral ($H_{4}$) order in 4D \cite{ElserSloane, SadocMosseri1, SadocMosseri2, MoodyPatera, BaakeGahler, MoodyWeiss}).  These symmetries are also distinguished for other reasons: they are the ones expected to arise from ``strong" local matching rules \cite{Levitov1987, Levitov1988b} or from the stable ground state of a local Hamiltonian \cite{Levitov1988a}.

In an Ammann pattern, each of the constituent 1D quasilattices is characterized by two ``phase" parameters (a translational phase and a ``phason" phase); and, as explained above, these phase parameters must be carefully chosen in order to ensure that the overall multigrid is self-similar when a simultaneous inflation transformation is applied to all of the 1D constituents.  How are these special phases to be found?  Prior to this work, the correct phases were only found in a few cases, using techniques which applied on a case-by-case basis: as described above, the original (2D 10-fold) Ammann pattern was obtained by stumbling on a clever decoration of the original Penrose tiles; and several more Ammann patterns (a 2D 8-fold pattern, a 2D 12-fold pattern, and a 3D icosahedral pattern) were obtained by noticing a particularly symmetric special case where it was easy enough to solve for the requisite phases directly \cite{SocolarSteinhardt86, Socolar89}.  But these techniques have the drawback that they only work in some cases (for example, in the particularly fascinating $H_{4}$ case in four dimensions, it does not seem possible to find a configuration that is sufficiently simple to allow one to solve for the phases directly); and, moreover, they do not yield any insight into the underlying meaning of the special phase arrangement, or where it comes from, thus making it difficult to answer various interesting follow-up questions.  

In this section, we present a new, unified geometric construction that yields all irreducible Ammann patterns directly.   The correct phases are obtained automatically, without having to solve for them, and are described by a simple analytical formula that reveals their underlying geometric meaning, and makes them particularly easy to work with.

\subsection{Constructing irreducible Ammann patterns}
\label{constructing_ammann_patterns}

Here is our recipe for generating any irreducible Ammann pattern:

\begin{enumerate}

\item Choose a (quadratic) Coxeter pair $\{\theta^{+},\theta\}$ (the non-crystallographic root system $\theta^{+}$ has lower rank $d^{+}$ and the crystallographic root system $\theta$ has higher rank $d$).

\item Pick a ($d^{+}$-dimensional) vector ${\bf a}_{0}^{+}$, which is a $\mathbb{Q}(\sqrt{D})$-linear combination of the simple roots of $\theta^{+}$, and act on it with all the elements of the reflection group $G(\theta^{+})$ to obtain a $G(\theta^{+})$-symmetric collection of ($d^{+}$-dimensional) vectors -- the ``star" $s({\bf a}_{0}^{+})$.  Let ${\bf a}_{j}^{+}$ denote the $j$th vector in this star.  (Actually, if the vectors come in pairs $\pm{\bf a}_{j}^{+}$, we can simplify our life by arbitrarily deleting from the star one member of each pair of vectors.  Note that this deletion step is optional -- the formalism in the remainder of the paper works whether we do it or not, as long as we remember to be consistent about always using either all of the vectors in the original star, or half of them.)

\item Now pick an appropriate value $m_{2}^{-}/m_{1}^{-}$ from Table 1 in \cite{BoyleSteinhardt1D} ({\it i.e.}\ a value corresponding to one of the $\theta^{+}$ scale factors -- see Table 1 in \cite{ReflectionQuasilattices}) and define the new $d^{+}$-dimensional vectors ${\bf b}_{j}^{+}=-(m_{1}^{-}/m_{2}^{-}){\bf a}_{j}^{+}$, so that we can write:
\begin{subequations}
  \begin{eqnarray}
    {\bf a}_{j}^{+}&=&a_{}^{+}{\bf e}_{j}^{+} \\
    {\bf b}_{j}^{+}&=&b_{}^{+}{\bf e}_{j}^{+}
  \end{eqnarray}
\end{subequations} 
where the unit vectors ${\bf e}_{j}^{+}$ form (all or half of) a $G(\theta^{+})$-symmetric star $s({\bf e}_{0}^{+})$, while the magnitudes $a^{+}\equiv({\bf a}_{j}^{+}\cdot{\bf a}_{j}^{+})^{1/2}$ and $b^{+}\equiv({\bf b}_{j}^{+}\cdot{\bf b}_{j}^{+})^{1/2}$ obey $b^{+}=-(m_{1}^{-}/m_{2}^{-})a^{+}$.

\item As described in Section \ref{Coxeter_Pairs}, the $d^{+}$-dimensional vectors ${\bf a}_{j}^{+}$ and ${\bf b}_{j}^{+}$ lift to $d$-dimensional vectors ${\bf a}_{j}$ and ${\bf b}_{j}$, respectively, which are $\mathbb{Q}$-linear combinations of the simple roots of $\theta$.  Note that, for fixed $j$, the $d^{+}$-dimensional vectors ${\bf a}_{j}^{+}$ and ${\bf b}_{j}^{+}$ are parallel, but their $d$-dimensional counterparts ${\bf a}_{j}$ and ${\bf b}_{j}$ are not.

\item Along each $d$-dimensional vector ${\bf a}_{j}$, construct an infinite sequence of codimension-one hyperplanes, with the $A$th hyperplane ($A\in\mathbb{Z}$) defined by:
\begin{subequations}
  \begin{equation}
    {\bf a}_{j}\cdot{\bf x}=A+\alpha,
  \end{equation}
  where ${\bf x}$ is the position coordinate in $d$-dimensional space.  In other words, these ``${\bf a}_{j}$ hyperplanes" are perpendicular to ${\bf a}_{j}$, and evenly spaced  
  along ${\bf a}_{j}$.     In order for the $a_{j}$ hyperplanes to have $\theta$ symmetry, we should take $\{\alpha\}=\alpha-\lfloor\alpha\rfloor$ ({\it i.e.}\ the fractional part of $\alpha$) to be either $0$ (so that $A+\alpha=\ldots,-2,-1,0,+1,+2,\ldots$) or $1/2$ (so that $A+\alpha=\ldots,-3/2,-1/2,+1/2,+3/2,\ldots$).  Similarly, along each $d$-dimensional vector ${\bf b}_{j}$, construct a sequence of  ``${\bf b}_{j}$ hyperplanes":
  \begin{equation}
    {\bf b}_{j}\cdot{\bf x}=B+\beta,
  \end{equation}
\end{subequations}
where, again, $\{\beta\}=\beta-\lfloor\beta\rfloor$ is either $0$ or $1/2$.  The collection of all the ${\bf a}_{j}$ and ${\bf b}_{j}$ hyperplanes (for all $A,B,j\in\mathbb{Z}$) together form a (crystallographic) $d$-dimensional pattern: a $d$-dimensional honeycomb (with symmetry described by $\theta$).  

\item Now consider a $d^{+}$-dimensional slice through this honeycomb -- the slice is parallel to the original maximally-symmetric $d^{+}$-dimensional space on which $\theta^{+}$ lives and acts, but its origin is displaced by an arbitrary $d$-dimensional vector ${\bf q}_{0}$ (relative to the origin of the coordinate ${\bf x}$); so the $d^{+}$-dimensional coordinate ${\bf x}^{+}$ on the hyperplane is related to the $d$-dimensional coordinate ${\bf x}$ by
\begin{equation}
  {\bf x}={\bf q}_{0}+{\bf x}^{+}.
\end{equation}
Let us call this $d^{+}$-dimensional hyperplane the ``Coxeter slice".

\item Consider the intersection of this Coxeter slice with the $d$-dimensional honeycomb.   Along each direction ${\bf e}_{j}^{+}$, we obtain a 1D {\it bi-grid} -- {\it i.e.}\ a superposition of two infinite sequences of $(d^{+}\!-1)$ dimensional hyperplanes that are perpendicular to ${\bf e}_{j}^{+}$, and evenly spaced along ${\bf e}_{j}^{+}$.  (In particular, the first sequence, produced by the ${\bf a}_{j}$ planes, has regular spacing $1/a^{+}$, while the second sequence, produced by the ${\bf b}_{j}$ planes, has regular spacing $1/b^{+}$.)  The $(d^{+}\!-1)$ dimensional planes perpendicular to the direction ${\bf e}_{j}^{+}$ are at a location ${\bf e}_{j}^{+}\cdot{\bf x}^{+}$ given by
\begin{subequations}
  \label{bi_grid}
  \begin{eqnarray}
    x_{j,A}^{(a)}&=&(A-{\bf a}_{j}\cdot{\bf q}_{0}+\alpha)/a^{+}, \\
    x_{j,B}^{(b)}&=&(B-{\bf b}_{j}\cdot{\bf q}_{0}+\beta)/b^{+}.
  \end{eqnarray}
\end{subequations}

\item In Ref.~\cite{BoyleSteinhardt1D}, we explain that there is a canonical pairing between a 1D bi-grid like this one and an associated 1D quasilattice: in geometrical terms, if we consider a 1D line ${\bf q}(t)$ which slices through a 2D lattice $\Lambda$ equipped with an integer basis $\{{\bf m}_{1},{\bf m}_{2}\}$, then the integer grid lines associated with the $\{{\bf m}_{1},{\bf m}_{2}\}$ basis slice up the 2D plane into parallelograms, and the 1D bi-grid corresponds to the intersection of the line ${\bf q}(t)$ with the edges of these parallelograms (see Subsection 2.2 in \cite{BoyleSteinhardt1D}), while the 1D quasilattice corresponds to the sequence of points obtained by orthogonally projecting onto ${\bf q}(t)$ the center of every parallelogram that is intersected by ${\bf q}(t)$ (see Subsection 2.3 in \cite{BoyleSteinhardt1D}).  In particular, although there might seem to be an overall translational ambiguity in relating the 1D bi-grid to the dual 1D quasilattice, as explained in Section 2 of \cite{BoyleSteinhardt1D}, this ambiguity may be canonically fixed by the requirement that the 1D bi-grid is reflection symmetric if and only if the 1D quasilattice is, too.  It follows that the 1D bi-grid along the ${\bf e}_{j}^{+}$ direction that is described by Eq.~(\ref{bi_grid}) corresponds to a 1D quasilattice $x_{j,n}$ that may be described in the following two equivalent forms for $x_{j,n}$:
  \begin{eqnarray}
    \label{xjn_1}
    \!&\!=\!&\!m_{1}^{+}(n\!-\!\chi_{1,j}^{+})\!+\!(m_{2}^{+}\!-\!m_{1}^{+})
    (\lfloor\kappa_{1}(n\!-\!\chi_{1,j}^{-})\rfloor\!+\!\frac{1}{2})\;\; \\
    \label{xjn_2}
    \!&\!=\!&\!m_{2}^{+}(n\!-\!\chi_{2,j}^{+})\!+\!(m_{1}^{+}\!-\!m_{2}^{+})
    (\lfloor\kappa_{2}(n\!-\!\chi_{2,j}^{-})\rfloor\!+\!\frac{1}{2})\;\;
  \end{eqnarray}
where the parameters are given as follows.  First, the ratios $m_{2}^{+}/m_{1}^{+}$ and $m_{2}^{-}/m_{1}^{-}$ are fixed by our choice of a row from Table 1 in Ref.~\cite{BoyleSteinhardt1D}; the parameters $\kappa_{1}$ and $\kappa_{2}$ are
\begin{equation}
  \label{kappa}
  \kappa_{1}=\frac{1}{1-(m_{2}^{-}/m_{1}^{-})},\qquad\kappa_{2}=\frac{1}{1-(m_{1}^{-}/m_{2}^{-})};
\end{equation}
the parameters  $m_{1}^{+}$ and $m_{2}^{+}$ are
\begin{subequations}
  \label{m1_m2_para}
  \begin{eqnarray}
    \label{m1_para}
    m_{1}^{+}&=&\left[\left(1-\frac{m_{1}^{-}}{m_{2}^{-}}\frac{m_{2}^{+}}{m_{1}^{+}}\right)a^{+}\right]^{-1}, \\
    \label{m2_para}
    m_{2}^{+}&=&\left[\left(1-\frac{m_{2}^{-}}{m_{1}^{-}}\frac{m_{1}^{+}}{m_{2}^{+}}\right)b^{+}\right]^{-1};
  \end{eqnarray}
\end{subequations}
and the parameters $\chi_{1,j}^{\pm}$ and $\chi_{2,j}^{\pm}$ are given by:
\begin{equation}
  \label{chi}
    m_{1}^{\pm}\chi_{1,j}^{\pm}=m_{2}^{\pm}\chi_{2,j}^{\pm}=
    (m_{1}^{\pm}{\bf a}_{j}^{\pm}+m_{2}^{\pm}{\bf b}_{j}^{\pm})
    \cdot{\bf q}_{0}^{\pm}-(m_{1}^{\pm}\alpha+m_{2}^{\pm}\beta),
\end{equation}
where, in this last equation, we have split the $d$-dimensional vectors ${\bf a}_{j}$, ${\bf b}_{j}$ and ${\bf q}_{0}$ into the parts that are parallel and perpendicular to the Coxeter slice
\begin{subequations}
  \begin{eqnarray}
    {\bf a}_{j}&=&{\bf a}_{j}^{+}+{\bf a}_{j}^{-} \\
    {\bf b}_{j}&=&{\bf b}_{j}^{+}+{\bf b}_{j}^{-} \\
    {\bf q}_{0}&=&{\bf q}_{0}^{+}+{\bf q}_{0}^{-}
  \end{eqnarray}
\end{subequations}
and used the fact that, as a consequence of our original definition ${\bf b}_{j}^{+}=-(m_{1}^{-}/m_{2}^{-}){\bf a}_{j}^{+}$, the parallel and perpendicular components satisfy the identities:
\begin{equation}
    m_{1}^{\pm}{\bf a}_{j}^{\mp}+m_{2}^{\pm}{\bf b}_{j}^{\mp}=0.
\end{equation}

\item At first glance, we seem to have constructed four different patterns, depending on whether the fractional parts of $\alpha$ and $\beta$ are given by $(\{\alpha\},\{\beta\})=(0,0)$, $(1/2,0)$, $(0,1/2)$ or $(1/2,1/2)$.  But, often, by shifting the origin of the $d$-dimensional coordinate $x$, we may find that some of these four options turn out to be equivalent to one another -- {\it i.e.}\ two options $(\{\alpha\},\{\beta\})$ and $(\{\alpha'\},\{\beta'\})$ may secretly define the same $d$-dimensional honeycomb, expressed with respect to two different origins.

\item Now, we know from Ref.~\cite{BoyleSteinhardt1D} that, under an inflation transformation, all the 1D phase parameters $\chi_{1,j}^{\pm}$ and $\chi_{2,j}^{\pm}$ given by (\ref{chi}) should transform to new parameters $\chi_{1,j}^{\pm}{}'$ and $\chi_{2,j}^{\pm}{}'$ given by
\begin{equation}
  \chi_{j}^{\pm}{}'=\chi_{j}^{\pm}/\lambda_{\pm},
\end{equation}
where the parameters $\lambda_{\pm}$ are again obtained from our chosen row in Table 1 of Ref.~\cite{BoyleSteinhardt1D}.  If the new inflated parameters $\chi_{j}^{\pm}{}'$ could alternatively be obtained by shifting the origin of the Coxeter slice from its initial position ${\bf q}_{0}$ to a new position ${\bf q}_{0}'$ ({\it i.e.}\ if the new parameters $\chi_{j}^{\pm}{}'$ can be obtained from the simple substitution ${\bf q}_{0}\to{\bf q}_{0}'$ in Eq.~(\ref{chi})), then the pattern is self-similar -- it is an Ammann pattern.  On the other hand, if the new inflated parameters $\chi_{j}^{\pm}{}'$ could alternatively be obtained by shifting the origin of the Coxeter slice from its initial position ${\bf q}_{0}$ to a new position ${\bf q}_{0}'$ and simultaneously shifting the choice $(\{\alpha\},\{\beta\})$ to a new {\it inequivalent} choice $(\{\alpha'\},\{\beta'\})$, 
then, instead of an Ammann pattern, we have a member of an ``Ammann cycle," since we must repeat the inflation two or more times before obtaining a pattern that is 
similar to the original.

\end{enumerate}

We again illustrate with our two basic examples:
\begin{itemize}
\item Example 1 (Penrose): To obtain the original Ammann pattern (Fig.~\ref{2D_10fold_PureAmmann}), whose dual is the Penrose tiling (Fig.~\ref{2D_10fold_PurePenrose}), we take the Coxeter pair $\{\theta^{+},\theta\}=\{I_{2}^{5},A_{4}\}$, the minimal 5-fold-symmetric star ${\bf a}_{j}^{+}=({\rm cos}\,\frac{2\pi j}{5},{\rm sin}\,\frac{2\pi j}{5}\}$ ($j=1,\ldots,5$), and the values $m_{2}^{\pm}/m_{1}^{\pm}=\frac{1}{2}(1\pm\sqrt{5})$ and $\lambda_{\pm}=\frac{1}{2}(1\pm\sqrt{5})$ from row 1 of Table 1 in Ref.~\cite{BoyleSteinhardt1D}.  It follows that $a^{+}=1$, $b^{+}=\frac{1}{2}(1+\sqrt{5})$, $\kappa_{1}=\frac{1}{2}(-1+\sqrt{5})$, $\kappa_{2}=\frac{1}{2}(3-\sqrt{5})$, $m_{1}^{+}=(5-\sqrt{5})/10$, and $m_{2}^{+}=1/\sqrt{5}$.  Taking $(\{\alpha\},\{\beta\})=(0,0)$ for the fractional parts of $\alpha$ and $\beta$, and carrying out the above procedure then yields the Ammann pattern Fig.~\ref{2D_10fold_PureAmmann} (whose dual is the Penrose tiling Fig.~\ref{2D_10fold_PurePenrose}).
\item Example 2 (Ammann-Beenker): To obtain the Ammann pattern (Fig.~\ref{2D_8foldA1_PureAmmann}), whose dual is the Ammann-Beenker tiling (Fig.~\ref{2D_8foldA1_PurePenrose}), we take the Coxeter pair $\{\theta^{+},\theta\}=\{I_{2}^{8},B_{4}\}$, the minimal 8-fold-symmetric star ${\bf a}_{j}^{+}=({\rm cos}\,\frac{2\pi j}{8},{\rm sin}\,\frac{2\pi j}{8}\}$ ($j=1,\ldots,8$), and the values $m_{2}^{\pm}/m_{1}^{\pm}=\pm\sqrt{2}$ and $\lambda_{\pm}=1\pm\sqrt{2}$ from row 2a of Table 1 in Ref.~\cite{BoyleSteinhardt1D}.  It follows that $a^{+}=1$, $b^{+}=1/\sqrt{2}$, $\kappa_{1}=-1+\sqrt{2}$, $\kappa_{2}=2-\sqrt{2}$, $m_{1}^{+}=1/2$, and $m_{2}^{+}=1/\sqrt{2}$.  Taking $(\{\alpha\},\{\beta\})=(0,0)$ for the fractional parts of $\alpha$ and $\beta$, and carrying out the above procedure yields the Ammann pattern Fig.~\ref{2D_8foldA1_PureAmmann} (whose dual is the Ammann-Beenker tiling, Fig.~\ref{2D_8foldA1_PurePenrose}).
\end{itemize}

\section{Constructing the corresponding Penrose Tilings}
\label{Penrose_Tilings}

In this section, we explain how to dualize any Ammann pattern to obtain the associated Penrose tiling (along with its Ammann decoration and inflation rule).

The idea that a Penrose tiling could be generated by dualizing a {\it periodic} (penta-)grid was first systematically developed and studied by de Bruijn \cite{deBruijn81b, SocolarLevineSteinhardt85, GahlerRhyner86} (and it is worth noting that the basic idea was also described informally by Ammann in earlier correspondence to Gardner \cite{SenechalOnAmmann}).  Socolar and Steinhardt subsequently discovered \cite{SocolarSteinhardt86} that a Penrose tiling could, instead, be obtained by dualizing an {\it Ammann} (penta-)grid in an analogous fashion.  Moreover, they realized that this approach had a fundamental advantage: it allowed one to {\it derive} two key properties of the Penrose tiling (the Ammann decoration and the inflation rule) which had previously been obtained by inspired guesswork.  

Here we build upon this second approach.  In order for the Socolar-Steinhardt approach to work, the Penrose tiling must be correctly {\it scaled} and {\it translated} so that it is properly situated relative to the original Ammann pattern.  In earlier work, the correct scaling and translation were achieved by inspection; but in Subsection \ref{dualization} we derive a simple ``dualization formula" (\ref{dualization_formula}) that yields the correct scaling and translation {\it automatically}.  Then, in Subsections \ref{AmmannDecoration} and \ref{InflationDecoration}, we explain how this formula may be used to {\it systematically} generate the Ammann decoration and the inflation rule for the Penrose tiling, and to ensure that they are in one-to-one correspondance with one another.  Our approach is completely systematic, and does not involve any inspection or guess work (which is particularly important in more complicated cases -- like the 4D case -- where inspection or guesswork become impractical).  

\subsection{The dualization formula}
\label{dualization}

Consider an Ammann pattern in which the Ammann planes are arrayed along the $J$ different directions ${\bf e}_{j}$ $(j=1,\ldots,J)$.  These planes slice up $d$-dimensional Euclidean space into open $d$-dimensional regions (``cells").  To each cell, we assign a set of $J$ integer coordinates $\{n_{1},\ldots,n_{J}\}$: the $j$th coordinate $n_{j}$ indicates that, along the ${\bf e}_{j}$ direction, the cell lies between the hyperplanes labelled $n_{j}$ and $n_{j}+1$ ({\it i.e.}\ the position ${\bf x}_{\parallel}$ of any point in the cell satisfies $x_{j,n_{j}^{}}<{\bf e}_{j}\cdot{\bf x}_{\parallel}<x_{j,n_{j}^{}+1}$).

The dualization procedure maps each cell in the Ammann pattern to a vertex in the corresponding Penrose tiling.  In particular, any point (with position ${\bf x}_{\parallel}$) in the cell with integer coordinates $\{n_{1},\ldots,n_{J}\}$ gets mapped to a point (with position ${\bf x}_{\parallel}'$) in the Penrose tiling via an equation of the form
\begin{equation}
  \label{dualization_formula_prelim}
  {\bf x}_{\parallel}'({\bf x}_{\parallel})={\bf z}+C\sum_{j=1}^{J}n_{j}{\bf e}_{j}.
\end{equation}
We now want to determine what the scaling parameter $C$ and overall translation ${\bf z}$ should be in order to ensure that the Penrose tiling produced by this formula is always properly situated relative to the original Ammann pattern from which it came.

To determine the formula for the scaling parameter $C$, first note that in the 1D quasilattice $x_{j,n}$, a fraction $\kappa_{1}$ of the steps have length $m_{2}^{+}$, and a fraction $\kappa_{2}$ have length $m_{1}^{+}$, so that the average step size is
\begin{equation}
  \label{def_mp}
  \langle m^{+}\rangle=\kappa_{1} m_{2}^{+}+\kappa_{2} m_{1}^{+}.
\end{equation}
The point with position ${\bf x}_{\parallel}$ lies within an Ammann-pattern cell whose $j$th integer coordinate is roughly 
$n_{j}\approx({\bf x}_{\parallel}\cdot\hat{{\bf e}}_{j})/\langle m^{+}\rangle$ (with an error that doesn't grow with $|{\bf x}_{\parallel}|$).  So, from (\ref{dualization_formula_prelim}), dualization maps the point ${\bf x}_{\parallel}$ to a point ${\bf x}_{\parallel}'$ roughly given by
\begin{equation}
  \label{x_prime_estimate}
  {\bf x}_{\parallel}'\approx{\bf z}+C\sum_{j=1}^{J}\frac{({\bf x}_{\parallel}\cdot{\bf e}_{j})}{\langle m^{+}\rangle}{\bf e}_{j}
  ={\bf z}+\frac{C\gamma}{\langle m^{+}\rangle}{\bf x}_{\parallel}
\end{equation}
where, in the last equality, we have used the fact that if we take the outer product of ${\bf e}_{j}$ with itself, and then sum over $j$, we obtain a multiple of the identity matrix $\delta^{\alpha\beta}$:
\begin{equation}
  \label{def_gamma}
  \sum_{j=1}^{J}{\bf e}_{j}^{\alpha}{\bf e}_{j}^{\beta}=\gamma\delta^{\alpha\beta},
\end{equation}
where the coefficient $\gamma$ depends on the star formed by the $\pm{\bf e}_{j}$.  If the Penrose tiling is correctly situated with respect to the original Ammann pattern, the position ${\bf x}_{\parallel}'$ estimated in (\ref{x_prime_estimate}) should be close to the original position ${\bf x}_{\parallel}$ (with an error that does not grow with $|{\bf x}_{\parallel}|$).  This requires the coefficient $C\gamma/\langle m^{+}\rangle$ on the right-hand side of (\ref{x_prime_estimate}) to be unity:
\begin{equation}
  C=\frac{\langle m^{+}\rangle}{\gamma}.
\end{equation}

To determine the overall translation ${\bf z}$, first note that if we change ${\bf q}_{0}$ in the ``parallel" direction (${\bf q}_{0}\to\bar{{\bf q}}_{0}={\bf q}_{0}+\Delta {\bf q}_{0}^{\parallel}$) while holding everything else fixed, this leaves the Ammann pattern unchanged apart from an overall translation ${\bf x}_{\parallel}\to\bar{{\bf x}}_{\parallel}={\bf x}_{\parallel}-\Delta{\bf q}_{0}^{\parallel}$; and since the Penrose tiling must shift in lock-step with the Ammann pattern, the overall translation ${\bf z}$ must depend on ${\bf q}_{0}$ as follows: ${\bf z}=-{\bf q}_{0}^{\parallel}+\delta{\bf z}$, where $\delta{\bf z}$ is ${\bf q}_{0}$-independent.  

Finally, the precise form of $\delta{\bf z}$ may then be determined from the requirement that, when the Ammann pattern is inversion symmetric (under ${\bf x}_{\parallel}\to-{\bf x}_{\parallel}$), the corresponding Penrose-like tiling should be, too (under ${\bf x}_{\parallel}'\to-{\bf x}_{\parallel}'$).  Now, the Ammann pattern as a whole is inversion symmetric if and only if each 1D quasilattice $x_{j,n}$ is separately inversion-symmetric: {\it i.e.}\ if $-x_{j,n}$ (the inversion of the $j$th 1D quasilattice) is the same as $x_{j,N-n}$ (the $j$th 1D quasilattice with the points labelled in the opposite order).  Using the ``umklaap" formulae from Section 2.3 of \cite{BoyleSteinhardt1D}, one can check that the 1D quasilattice $x_{j,n}$ is inversion-symmetric if and only if
\begin{equation}
  \label{chi_inversion_symmetric}
  m_{1}^{\pm}\chi_{1,j}^{\pm}=m_{2}^{\pm}\chi_{2,j}^{\pm}=\frac{1}{2}(p_{1,j}^{}m_{1}^{\pm}+p_{2,j}^{}m_{2}^{\pm}),
\end{equation}
where $p_{1,j}^{}$ and $p_{2,j}^{}$ are integers.  Specifically, in this case, one has
\begin{equation}
  \label{x_inversion_symmetric}
  -x_{j,n}=x_{j,p_{1,j}^{}+p_{2,j}^{}-n},
\end{equation}
and if we compare Eqs.~(\ref{chi}) and (\ref{chi_inversion_symmetric}), we find that the integers $p_{1,j}^{}$ and $p_{2,j}^{}$ are given by
\begin{subequations}
  \begin{eqnarray}
    p_{1,j}^{}&=&2({\bf a}_{j}\cdot{\bf q}_{0}-\alpha), \\
    p_{2,j}^{}&=&2({\bf b}_{j}\cdot{\bf q}_{0}-\beta).
  \end{eqnarray}
\end{subequations}
Now, if the point ${\bf x}_{\parallel}$ in the Ammann pattern satisfies $x_{j,n_{j}}<{\bf x}_{\parallel}\cdot{\bf e}_{j}<x_{j,n_{j}+1}$ (so that it lies in the Ammann cell with integer coordinates $\{n_{j}\}$) then $-{\bf x}_{\parallel}$ (the inversion of the original point) satisfies $-x_{j,n_{j}+1}<-{\bf x}_{\parallel}\cdot{\bf e}_{j}<-x_{j,n}$ or, equivalently [using Eq.~(\ref{x_inversion_symmetric})], $x_{j,p_{1,j}+p_{2,j}-n_{j}-1}<-{\bf x}_{\parallel}\cdot{\bf e}_{j}<x_{p_{1,j}+p_{2,j}-n_{j}}$ (so that it lies in the Ammann cell with integer coordinates $\{p_{1,j}+p_{2,j}-n_{j}-1\}$).  So, while the dualization formula (\ref{dualization_formula_prelim}) maps the original point ${\bf x}_{\parallel}$ to a Penrose vertex at position
\begin{subequations}
  \begin{equation}
    \label{x_prime_from_x}
    {\bf x}_{\parallel}'({\bf x}_{\parallel})=-{\bf q}_{0}^{\parallel}+\delta{\bf z}+\frac{\langle m^{+}\rangle}{\gamma}\sum_{j=1}^{J}
    n_{j}{\bf e}_{j}
  \end{equation}
  it maps the inverted point $-{\bf x}_{\parallel}$ to a Penrose vertex at position
  \begin{equation}
    \label{x_prime_from_minus_x}
    {\bf x}_{\parallel}'(-{\bf x}_{\parallel})=-{\bf q}_{0}^{\parallel}+\delta{\bf z}+\frac{\langle m^{+}\rangle}{\gamma}\sum_{j=1}^{J}
    (p_{1,j}+p_{2,j}-n_{j}-1){\bf e}_{j}.
  \end{equation}
\end{subequations}
Now the above requirement (that an inversion-symmetric Ammann pattern should map two points $\pm{\bf x}_{\parallel}$ related by inversion to two Penrose vertices related by inversion) means that we should have ${\bf x}_{\parallel}'({\bf x}_{\parallel})=-{\bf x}_{\parallel}'(-{\bf x}_{\parallel})$.  Using Eqs.~(\ref{x_prime_from_x}, \ref{x_prime_from_minus_x}) and simplifying, this becomes
\begin{subequations}
  \begin{eqnarray}
    \delta{\bf z}\!&\!=\!&\!{\bf q}_{0}^{\parallel}\!+\!\frac{\langle m^{+}\rangle}{2\gamma}\!\sum_{j}(1-p_{1,j}-p_{2,j}){\bf e}_{j} \\
    \!&\!=\!&\!{\bf q}_{0}^{\parallel}\!+\!\frac{\langle m^{+}\rangle}{\gamma}\!\sum_{j}(\frac{1}{2}\!+\!\alpha\!+\!\beta\!-\!{\bf a}_{j}\!\cdot\!{\bf q}_{0}\!-\!{\bf b}_{j}\!\cdot\!{\bf q}_{0}){\bf e}_{j}\;\; \\
    \!&\!=\!&\!\frac{\langle m^{+}\rangle}{\gamma}\sum_{j}(\frac{1}{2}+\alpha+\beta){\bf e}_{j}
  \end{eqnarray}
\end{subequations}

Finally notice that it is natural to re-absorb $\delta{\bf z}$ into a re-labelling of the Ammann cells: if, along the ${\bf e}_{j}$ direction, an Ammann cell lies between the Ammann planes $x_{j,n_{j}}$ and $x_{j,n_{j}+1}$, instead of assigning it the integer coordinate $n_{j}$, let us assign it the half-integer coordinate $\nu_{j}=n_{j}+1/2+\alpha+\beta$, so that the Ammann cell is labelled by $J$ such integer or half-integer coordinates $\{\nu_{1},\ldots,\nu_{J}\}$.  

Then what we have shown is that the dualization formula (\ref{dualization_formula_prelim}) finally assumes the following simple form:
\begin{equation}
  \label{dualization_formula}
  {\bf x}_{\parallel}'({\bf x}_{\parallel})=-{\bf q}_{0}^{\parallel}+\frac{\langle m^{+}\rangle}{\gamma}\sum_{j=1}^{J}\nu_{j}{\bf e}_{j}
\end{equation}

Once again, let us illustrate with our two basic examples:
\begin{itemize}
\item Example 1 (Penrose): Using Eqs.~(\ref{def_mp}, \ref{def_gamma}), along with the values summarized in Example 1 of Subsection \ref{constructing_ammann_patterns}, we obtain $\langle m^{+}\rangle=\frac{1}{2}(3-\sqrt{5})$ and $\gamma=5/2$.  Apply dualization formula (\ref{dualization_formula}) with these constants to the Ammann pattern in Fig.~\ref{2D_10fold_PureAmmann} yields the dual Penrose tiling (Fig.~\ref{2D_10fold_PurePenrose}), its Ammann decoration (Fig.~\ref{2D_10fold_AmmannLines}),
and its inflation (Fig.~\ref{2D_10fold_Inflation}), as further explained in the following two subsections.
\item Example 2 (Ammann-Beenker): Using Eqs.~(\ref{def_mp}, \ref{def_gamma}), along with the values summarized in Example 2 of Subsection \ref{constructing_ammann_patterns}, we obtain $\langle m^{+}\rangle=2-\sqrt{2}$ and $\gamma=4$.  Apply dualization formula (\ref{dualization_formula}) with these constants to the Ammann pattern in Fig.~\ref{2D_8foldA1_PureAmmann} yields the dual Ammann-Beenker tiling (Fig.~\ref{2D_8foldA1_PurePenrose}), its Ammann decoration (Fig.~\ref{2D_8foldA1_AmmannLines}), and its inflation (Fig.~\ref{2D_8foldA1_Inflation}), as explained in the following two subsections.
\end{itemize}

\subsection{Obtaining the Ammann decorations}
\label{AmmannDecoration}

In Subsection \ref{dualization}, we explained how to start with the Ammann pattern $x_{j,n}$ given by Eq.~(\ref{xjn_1}) [with parameters given by Eqs.~(\ref{kappa}, \ref{m1_m2_para}, \ref{chi})] and use the dualization formula (\ref{dualization_formula}) to obtain the dual Penrose tiling with vertices ${\bf x}_{\parallel}'({\bf x}_{\parallel})$.

Now let us consider a new Ammann pattern $\bar{x}_{j,n}$ which is obtained from the original Ammann pattern $x_{j,n}$ by a single inflation (so that $\bar{x}_{j,n}$ is one level {\it denser} or more {\it refined} than ${\bf x}_{j,n}$).  In other words, using the results of Section 4 in Ref.~\cite{BoyleSteinhardt1D}, we find that the refined Ammann pattern is described by the following equation for $\bar{x}_{j,n}$:
\begin{subequations}
  \label{xjn_bar}
  \begin{eqnarray}
    \!&\!=\!&\!\frac{1}{\lambda^{+}\!}[m_{1}^{+}(n\!-\!\bar{\chi}_{1,j}^{+})\!+\!(m_{2}^{+}\!-\!m_{1}^{+})
    (\lfloor\kappa_{1}(n\!-\!\bar{\chi}_{1,j}^{-})\rfloor\!+\!\frac{1}{2})]\quad \\
    \!&\!=\!&\!\frac{1}{\lambda^{+}\!}[m_{2}^{+}(n\!-\!\bar{\chi}_{2,j}^{+})\!+\!(m_{1}^{+}\!-\!m_{2}^{+})
    (\lfloor\kappa_{2}(n\!-\!\bar{\chi}_{2,j}^{-})\rfloor\!+\!\frac{1}{2})]\quad
  \end{eqnarray}
\end{subequations}
where the parameters $m_{1,2}^{+}$ and $\kappa_{1,2}$ are the same as before, while the new ``phases" $\bar{\chi}_{1,2}^{\pm}$ are related to the original phases $\chi_{1,2}^{\pm}$ by
\begin{equation}
  \bar{\chi}_{1}^{\pm}=\lambda_{\pm}\chi_{1}^{\pm}\qquad\bar{\chi}_{2}^{\pm}=\lambda_{\pm}\chi_{2}^{\pm}.
\end{equation}
and $\lambda_{\pm}$ are obtained from the appropriate row of Table 1 in Ref.~\cite{BoyleSteinhardt1D} (the same row as the ratios $m_{2}^{\pm}/m_{1}^{\pm}$ were obtained from in specifying the original Ammann pattern in Sec.~\ref{general_construction}).  

Now the Ammann decoration is automatically obtained by simply superposing the refined Ammann pattern $\bar{x}_{j,n}$ directly on top of the Penrose tiling ${\bf x}_{\parallel}'({\bf x}_{\parallel})$.

\subsection{Obtaining the inflation rules}
\label{InflationDecoration}

In Subsection \ref{AmmannDecoration}, we started with the original Ammann pattern $x_{j,n}$ and explained how to obtain the more refined Ammann pattern $\bar{x}_{j,n}$.  In Subsection \ref{dualization}, we explained how to use Eq.~(\ref{dualization_formula}) to dualize the original Ammann pattern $x_{j,n}$ to obtain the corresponding Penrose tiling ${\bf x}_{\parallel}'({\bf x}_{\parallel})$.  Similarly, we can dualize the refined Ammann pattern $\bar{x}_{j,n}$ to obtain a refined Penrose tiling $\bar{{\bf x}}_{\parallel}'({\bf x}_{\parallel})$: if, along the ${\bf e}_{j}$ direction, a point ${\bf x}_{\parallel}$ in a cell of the refined Ammann pattern lies between the Ammann planes $\bar{x}_{j,\bar{n}_{j}}$ and $\bar{x}_{j,\bar{n}_{j}+1}$, then we say its $j$th integer or half-integer coordinate is $\bar{\nu}_{j}=\bar{n}_{j}+1/2+\bar{\alpha}+\bar{\beta}$, and we map this point to a vertex in the refined Penrose tiling with position
\begin{equation}
  \label{dualization_formula_refined}
  \bar{{\bf x}}_{\parallel}'({\bf x}_{\parallel})=-{\bf q}_{0}^{\parallel}+\frac{1}{\lambda^{+}}\frac{\langle m^{+}\rangle}{\gamma}\sum_{j=1}^{J}
  \bar{\nu}_{j}{\bf e}_{j}.
\end{equation}
Now we automatically obtain the inflation rule for the Penrose tiling by simply superposing the new refined tiling (\ref{dualization_formula_refined}) on the original Penrose tiling (\ref{dualization_formula}).

In the next section, we will apply these techniques to generate all of the minimal Penrose tilings as well as their Ammann decorations and inflation rules.  But here we make a remark.  The procedures outlined in this section reproduce the standard 10-fold Penrose tiling, along with its standard Ammann decoration and inflation rule.  In Ref.~\cite{Socolar89}, a similar technique was used to produce an an 8-fold tiling and a 12-fold tiling; but note that in those cases, the suggested Ammann decoration was obtained by taking the dual of the original Ammann pattern $x_{j,n}$ to obtain a Penrose tiling, and then superposing on this tiling the {\it same} Ammann pattern $x_{j,n}$ (rather than its refinement $\bar{x}_{j,n}$).  But note that the decoration obtained by superposing the original Ammann pattern $x_{j,n}$ will, in general, be too sparse: tiles that are {\it distinct} ({\it i.e.}\ they have the same shape, but distinct inflation rules) can wind up receiving the {\it same} decoration, while tiles that are actually {\it identical} ({\it i.e.}\ they have the same shape and the same inflation rule) can wind up receiving {\it distinct} decorations.  By contrast, if we superpose the refined Ammann pattern $\bar{x}_{j,n}$ we obtain an Ammann decoration that is in one-to-one correspondence with the tiles and their inflation rules: {\it i.e.} two tiles receive the same Ammann decoration if and only if they have the same inflation rule.  

\section{The minimal Ammann patterns and Penrose-like tilings}
\label{Minimal_Penrose_Tilings}

We now apply the approach developed in the previous sections to construct all the {\it minimal} Penrose-like tilings -- {\it i.e.}\ all tilings obtained by dualizing a {\it minimal} Ammann pattern (an Ammann pattern with the minimal star compatible with its orientational symmetry).  

Since the formalism for constructing such tilings has already been explained in detail in Sections \ref{general_construction} and \ref{Penrose_Tilings}, we will be brief in this section, just providing the information needed to specify each tiling, along with a few other relevant facts.

As we have already explained, each of the tilings corresponds to one of the five quadratic Coxeter pairs listed in Table \ref{QuadraticCoxeterPairs}, and we will present them in this same order.  

At the end of the section, we present figures illustrating all eleven of the 2D patterns/tilings, as well as their Ammann decorations and inflation rules; and we also display 
the four ``Ammann cycles" (one 10-fold-symmetric "Ammann trio" and three 12-fold-symmetric "Ammann duos").  The figures are arranged in the following order.  For each Coxeter pair, we begin by depicting it via its root diagram -- {\it i.e.} by showing the maximally-symmetric 2D projection ({\it i.e.}\ the Coxeter-plane projection) of the $\theta$ roots onto 2D (yielding two concentric copies of the $\theta^{\parallel}$ roots).  Following each root diagram, we present the associated self-similar patterns/tilings, with the relevant Cases ordered as in Table 1 of Ref.~\cite{BoyleSteinhardt1D}.  For each pattern/tiling, we:
\begin{itemize}
\item (i) display the (undecorated) Ammann pattern and the (undecorated) Penrose-like tiling (on the same page); and   
\item (ii) display a decorated Penrose-like tiling (first with its Ammann decoration and then with its inflation decoration, on the same page).
\end{itemize}
Then, in Figs. (50-67), we present all four Ammann cycles (the 10-fold-symmetric Ammann trio, and three 12-fold-symmetric Ammann duos); and for each of the three members of the 10-fold-symmetric Ammann trio, we also show the dual Penrose-like tiling, its Ammann decoration, and its inflation. 

{\it Singular and Non-Singular Ammann patterns.}  Note that, in a $d^{\parallel}$-dimensional Ammann pattern, since each Ammann plane has codimension one, we generically expect $d^{\parallel}$ such planes to intersect at a point.  But some Ammann patterns are ``singular" in the sense that they contain points where more than $d^{\parallel}$ Ammann planes intersect.  In particular, a 2D Ammann pattern is non-singular if at most two Ammann lines intersect at any point, and singular if there are points where three or more Ammann lines intersect.  When we dualize an Ammann pattern, each intersection point in the pattern is in one-two-one correspondance with a tile in the dual tiling.  In particular, in 2D, if $n$ Ammann lines intersect at a point, then the tile dual to that intersection (in the dual Penrose-like tiling) will have $2n$ sides.  Hence, a non-singular 2D Ammann pattern will have a dual Penrose-like tiling in which all the tiles are four-sided rhombs, while a singular 2D Ammann pattern in which three, four, or more Ammann lines intersect at a point will have a dual Penrose-like tiling in which the tiles will be corresponding polygons with six, eight or more sides.  For example, the 8-fold-symmetric ``A2" and ``B1" Ammann patterns described below are singular: they contain points where four Ammann lines intersect, and hence the dual Penrose-like tilings contain octagonal tiles.

\subsection{The $I_{2}^{5}$ (2D 10-fold) tiling}

Here the relevant Coxeter pair is $\{\theta^{\parallel},\theta\}=\{I_{2}^{5},A_{4}\}$.  The $A_{4}$ root system has 20 roots: all vectors obtained from $\{+1,-1,0,0,0\}$ by allowing all permutations of the coordinates.  The maximally-symmetric orthogonal projection onto $d^{\parallel}=2$ dimensions may be found by the Coxeter-plane construction (see Appendix \ref{Finding_all_Coxeter_pairs} in this paper, or Section 4.2 in Ref.~\cite{ConwaySloane}).  The columns of the following matrix are a standard choice for the fundamental roots ${\bf f}_{k}$ (see {\it e.g.}\ Section 6.1 in \cite{ConwaySloane}):
\begin{equation}
  \left({\bf f}_{1}\;\;{\bf f}_{2}\;\;{\bf f}_{3}\;\;{\bf f}_{4}\right)
  =\left(\!\begin{array}{rrrr} 
    \;-1 & 0 & 0 & 0 \\
    \;+1 & \;-1 & 0 & 0 \\
    0 & \;+1 & \;-1 & 0 \\
    0 & 0 & \;+1 & \;-1 \\
    0 & 0 & 0 & \;+1 
  \end{array}\;\;\right).
\end{equation}
(Note that the angles between these fundamental roots agree with the $A_{4}$ diagram in Fig.~\ref{CoxeterDynkinDiagrams}.)
From these we can compute the corresponding Coxeter element:
\begin{equation}
  C=F_{1}F_{2}F_{3}F_{4}=\left(\begin{array}{rrrrr}
  0 & 0 & 0 & 0 & \;\;1 \\
  \;\;1 & 0 & 0 & 0 & 0 \\
  0 & \;\;1 & 0 & 0 & 0 \\
  0 & 0 & \;\;1 & 0 & 0 \\
  0 & 0 & 0 & \;\;1 & 0 \end{array}\;\;\right).
\end{equation}
We can project the 20 roots of $A_{4}$ onto the Coxeter plane spanned by the eigenvectors $u_{\pm}$ of $C$ (corresponding to the eigenvalues ${\rm e}^{\pm 2\pi i/5}$); the result is shown in Figure \ref{A4rootfig}.

To proceed we just need to choose: (i) the underlying star $\{{\bf e}_{j}^{+}\}$ (which, in this case, points to the 5 vertices of a regular pentagon); and (ii) a relevant row from Table 1 in Ref.~\cite{BoyleSteinhardt1D} (the only relevant row in this case is Row 1).

Carrying out the procedure described in Sections \ref{general_construction} and \ref{Penrose_Tilings}, when $(\{\alpha\},\{\beta\})=(0,0)$ we obtain the self-similar pattern/tiling shown in Figs.~\ref{2D_10fold_PureAmmann}, \ref{2D_10fold_PurePenrose}, \ref{2D_10fold_AmmannLines} and \ref{2D_10fold_Inflation}.  Comparing these figures with Refs.~\cite{GrunbaumShephard, SocolarSteinhardt86}, we see that this is precisely the original 10-fold Penrose tiling, with its standard Ammann decoration and inflation rule.  

\subsection{The $I_{2}^{8}$ (2D 8-fold) tilings (A1, A2, B1, B2)}

Here the relevant Coxeter pair is $\{\theta^{\parallel},\theta\}=\{I_{2}^{8},B_{4}\}$ (or, equivalently, $\{I_{2}^{8},C_{4}\}$).  The $B_{4}$ root system has 32 roots: all vectors obtained from $(\pm1,0,0,0)$ or $(\pm1,\pm1,0,0)$ by allowing all combinations of signs, and all permutations of the coordinates.  Again, the maximally-symmetric 2D orthogonal projection may be found by the Coxeter-plane construction.  The columns of the following matrix are a standard choice for the fundamental roots ${\bf f}_{k}$:
\begin{equation}
  \left({\bf f}_{1}\;\;{\bf f}_{2}\;\;{\bf f}_{3}\;\;{\bf f}_{4}\right)=
  \left(\!\begin{array}{rrrr} 
  \;+1 & \;-1 & 0 & 0 \\
  0 & \;+1 & \;-1 & 0 \\
  0 & 0 & \;+1 & \;-1 \\
  0 & 0 & 0 & \;+1 \end{array}\;\;\right).
\end{equation}
(Note that the angles between these fundamental roots agree with the $B_{4}$ diagram in Fig.~\ref{CoxeterDynkinDiagrams}.)
From these we can compute the corresponding Coxeter element:
\begin{equation}
  C=F_{1}F_{2}F_{3}F_{4}=\left(\!\begin{array}{rrrr}
  0 & 0 & 0 & \;-1 \\
  \;+1 & 0 & 0 & 0 \\
  0 & \;+1 & 0 & 0 \\
  0 & 0 & \;+1 & 0 \end{array}\;\;\right).
\end{equation}
We can project the 32 roots of $B_{4}$ onto the Coxeter plane spanned by the eigenvectors $u_{\pm}$ of $C$ (corresponding to the eigenvalues ${\rm e}^{\pm 2\pi i/8}$); the result is shown in Fig.~\ref{B4rootfig}.

To proceed we just need to choose: (i) the underlying star $\{{\bf e}_{j}^{+}\}$ (the minimal star in this case is an 8-pointed star aligned with the ``long" roots of $I_{2}^{8}$; or, equivalently, an 8-pointed star aligned with the ``short" roots of $I_{2}^{8}$); and (ii) a relevant row from Table 1 in Ref.~\cite{BoyleSteinhardt1D} (the relevant rows are 2a and 2b).

{\it The 8-fold ``A1" and ``A2" tilings.} Consider Table 1, Row 2a in \cite{BoyleSteinhardt1D}.   Then, if we take $(\{\alpha\},\{\beta\})=(0,0)$ or, equivalently, $(\{\alpha\},\{\beta\})=(0,1/2)$, we obtain the self-similar 8-fold A1 tiling shown in Figs.~\ref{2D_8foldA1_PureAmmann}, \ref{2D_8foldA1_PurePenrose}, \ref{2D_8foldA1_AmmannLines} and \ref{2D_8foldA1_Inflation}.  Alternatively, if we take $(\{\alpha\},\{\beta\})=(1/2,0)$ or, equivalently, $(\{\alpha\},\{\beta\})=(1/2,1/2)$, we obtain the self-similar 8-fold A2 tiling shown in Figs.~\ref{2D_8foldA2_PureAmmann}, \ref{2D_8foldA2_PurePenrose}, \ref{2D_8foldA2_AmmannLines} and \ref{2D_8foldA2_Inflation}.   

{\it The 8-fold ``B1" and ``B2" tilings.} Consider Table 1, Row 2b in \cite{BoyleSteinhardt1D}.  Then, if we take $(\{\alpha\},\{\beta\})=(0,0)$ or, equivalently, $(\{\alpha\},\{\beta\})=(1/2,1/2)$, we obtain the self-similar 8-fold B1 tiling shown in Figs.~\ref{2D_8foldB1_PureAmmann}, \ref{2D_8foldB1_PurePenrose}, \ref{2D_8foldB1_AmmannLines} and \ref{2D_8foldB1_Inflation}.  Alternatively, if we take $(\{\alpha\},\{\beta\})=(1/2,0)$ or, equivalently, $(\{\alpha\},\{\beta\})=(0,1/2)$, we obtain the self-similar 8-fold B2 tiling shown in Figs.~\ref{2D_8foldB2_PureAmmann}, \ref{2D_8foldB2_PurePenrose}, \ref{2D_8foldB2_AmmannLines} and \ref{2D_8foldB2_Inflation}.  

Comparing Figs.~\ref{2D_8foldA1_PureAmmann}, \ref{2D_8foldA1_PurePenrose}, \ref{2D_8foldA1_AmmannLines} and \ref{2D_8foldA1_Inflation} with Refs.~\cite{GrunbaumShephard, Socolar89}, we see the 8-fold A1 tiling is the well-known Ammann-Beenker tiling, with its standard inflation rule.  And the Ammann decoration of this tiling produced by our procedure is closely related to the one suggested in \cite{Socolar89}, but differs as follows: our decoration produces an Ammann pattern which is precisely the inflation of the Ammann pattern produced by the decoration suggestion in \cite{Socolar89} ({\it i.e.}\ our decoration is ``denser" by one level of inflation). The 8-fold A2, B1 and B2 tilings are new.

\subsection{The $I_{2}^{12}$ (2D 12-fold) tilings (A1, A2, B1, B2, C1, C2)}

Here the relevant Coxeter pair is $\{\theta^{\parallel},\theta\}=\{I_{2}^{12},F_{4}\}$.  The $F_{4}$ root system has 48 roots: the union of the vertices of a 24-cell and the vertices of the dual 24-cell \cite{RegularPolytopes}.  The vertices of the first 24-cell are obtained from $\frac{1}{2}(\pm1, \pm1, \pm1, \pm1)$ and $(\pm1, 0, 0,0)$ by allowing all combinations of $\pm$ signs and all permutations of the coordinates; and, similarly, the vertices of the dual 24-cell are obtained from $(\pm1,\pm1, 0, 0)$ by allowing all combinations of $\pm$ signs and all permutations of the coordinates.  Again, the maximally-symmetric 2D orthogonal projection may be found by the Coxeter-plane construction.  The columns of the following matrix are a choice for the fundamental roots ${\bf f}_{k}$ of $F_{4}$:
\begin{equation}
  \left({\bf f}_{1}\;\;{\bf f}_{2}\;\;{\bf f}_{3}\;\;{\bf f}_{4}\right)=
  \left(\!\begin{array}{rrrr}
  \;+1 & 0 & 0 & \;-1/2 \\
  \;-1 & \;+1 & 0 & \;-1/2 \\
  0 & \;-1 & \;+1 & \;-1/2 \\
  0 & 0 & 0 & \;-1/2 \end{array}\;\;\right).
\end{equation}
(Note that the angles between these fundamental roots agree with the $F_{4}$ diagram in Fig.~\ref{CoxeterDynkinDiagrams}.)
From these we can compute the corresponding Coxeter element:
\begin{equation}
  C=R_{1}R_{2}R_{3}R_{4}=\frac{1}{2}\left(\!\begin{array}{rrrr}
  \;+1 & \;+1 & \;-1 & \;+1 \\
  \;+1 & \;-1 & \;-1 & \;-1 \\
  \;-1 & \;+1 & \;-1 & \;-1 \\
  \;-1 & \;-1 & \;-1 & \;+1 \end{array}\;\;\right).
\end{equation}
We can project the 48 roots of $F_{4}$ onto the Coxeter plane spanned by the eigenvectors $u_{\pm}$ of $C$ (corresponding to the eigenvalues ${\rm e}^{\pm 2\pi i/12}$); the result is shown in Fig.~\ref{F4rootfig}.

To proceed we just need to choose: (i) the underlying star $\{{\bf e}_{j}^{+}\}$ (the minimal star in this case is a 12-pointed star aligned with the ``long" roots of $I_{2}^{12}$; or, equivalently, a 12-pointed star aligned with the ``short" roots of $I_{2}^{12}$); and (ii) a relevant row from Table 1 in Ref.~\cite{BoyleSteinhardt1D} (the relevant rows are 3a, 3b and 3c).

{\it The 12-fold ``A1" and ``A2" tilings.} Consider Table 1, Row 3a in \cite{BoyleSteinhardt1D}.   Then, if we take $(\{\alpha\},\{\beta\})=(0,0)$, we obtain the self-similar 12-fold A1 tiling shown in Figs.~\ref{2D_12foldA1_PureAmmann}, \ref{2D_12foldA1_PurePenrose}, \ref{2D_12foldA1_AmmannLines} and \ref{2D_12foldA1_Inflation}.   Alternatively, if we take $(\{\alpha\},\{\beta\})=(1/2,0)$, we obtain the self-similar 12-fold A2 tiling shown in Figs.~\ref{2D_12foldA2_PureAmmann}, \ref{2D_12foldA2_PurePenrose}, \ref{2D_12foldA2_AmmannLines} and \ref{2D_12foldA2_Inflation}.  

{\it The 12-fold ``B1" and ``B2" tilings.} Consider Table 1, Row 3b in \cite{BoyleSteinhardt1D}.   Then, if we take $(\{\alpha\},\{\beta\})=(0,0)$, we obtain the self-similar 12-fold B1 tiling shown in Figs.~\ref{2D_12foldB1_PureAmmann}, \ref{2D_12foldB1_PurePenrose}, \ref{2D_12foldB1_AmmannLines} and \ref{2D_12foldB1_Inflation}.   Alternatively, if we take $(\{\alpha\},\{\beta\})=(1/2,1/2)$, we obtain the self-similar 12-fold B2 tiling shown in Figs.~\ref{2D_12foldB2_PureAmmann}, \ref{2D_12foldB2_PurePenrose}, \ref{2D_12foldB2_AmmannLines} and \ref{2D_12foldB2_Inflation}.  

{\it The 12-fold ``C1" and ``C2" tilings.} Consider Table 1, Row 3c in \cite{BoyleSteinhardt1D}.   Then, if we take $(\{\alpha\},\{\beta\})=(0,0)$, we obtain the self-similar 12-fold C1 tiling shown in Figs.~\ref{2D_12foldC1_PureAmmann}, \ref{2D_12foldC1_PurePenrose}, \ref{2D_12foldC1_AmmannLines} and \ref{2D_12foldC1_Inflation}.   Alternatively, if we take $(\{\alpha\},\{\beta\})=(0,1/2)$, we obtain the self-similar 12-fold C2 tiling shown in Figs.~\ref{2D_12foldC2_PureAmmann}, \ref{2D_12foldC2_PurePenrose}, \ref{2D_12foldC2_AmmannLines} and \ref{2D_12foldC2_Inflation}.  

Comparing with Ref.~\cite{Socolar89}, we see that our 12-fold A1 tiling corresponds to the 12-fold Ammann-Socolar tiling \cite{Socolar89} (including the inflation rule and Ammann decoration).  Note that, as for the 8-fold ``A" tiling, our approach produces an Ammann decoration that is closely related to the one suggested in \cite{Socolar89}, but one level of inflation ``denser".  The 12-fold A2 tiling was first discovered by Socolar \cite{SocolarPrivate}.  The 12-fold B1, B2, C1, and C2 tilings are new, as are Ammann duos A, B and C.

\subsection{The 2D Ammann Cycles: 10-fold Ammann trio, 12-fold Ammann duos (A,B,C)}

The remaining choices for $(\{\alpha\},\{\beta\})$ all yield ``Ammann cycles" which, instead of inflating into themselves, inflate into one another in a cycle.  We collect these interesting objects in this subsection.

{\it The 10-fold Ammann trio.} In the $I_{2}^{5}$ case, consider Table 1, Row 1 in \cite{BoyleSteinhardt1D}: if we take $(\{\alpha\},\{\beta\})$ to be $(1/2,0)$ or $(0,1/2)$ or $(1/2,1/2)$, we obtain the three different Ammann patterns show in Figs.~\ref{2D_10foldTrio1_PureAmmann}, \ref{2D_10foldTrio2_PureAmmann}, and \ref{2D_10foldTrio3_PureAmmann}, which inflate into one another in an ``Ammann 3-cycle" or ``Ammann trio."  This Ammann trio was first discovered by Socolar \cite{SocolarPrivate}.  The three Penrose-like-tilings dual to these three Ammann patterns are shown in Figs.~\ref{2D_10foldTrio1_PurePenrose}, \ref{2D_10foldTrio2_PurePenrose}, and \ref{2D_10foldTrio3_PurePenrose}, respectively; the Ammann decorations of these three tilings are shown in 
Figs.~\ref{2D_10foldTrio1_AmmannLines}, \ref{2D_10foldTrio2_AmmannLines} and \ref{2D_10foldTrio3_AmmannLines}, respectively; and the inflations of these three tilings are shown in Figs.~\ref{2D_10foldTrio1_Inflation}, \ref{2D_10foldTrio2_Inflation}, and \ref{2D_10foldTrio3_Inflation}, respectively.  It would be interesting to understand the relationship of these tilings to those obtained by dualizing de Bruijn pentagrids with $\sum_{j} \gamma_{j}\neq 0$ ({\it e.g.}  $\sum_{j} \gamma_{j}= 1/7, 3/7, 2/7$).  (We thank the referee for pointing this out.)

{\it The 12-fold Ammann duo ``A."}  In the $I_{2}^{12}$ case, consider Table 1, Row 3A in \cite{BoyleSteinhardt1D}: if we take $(\{\alpha\},\{\beta\})=(0,1/2)$ or $(\{\alpha\},\{\beta\})=(1/2,1/2)$, we obtain the two different Ammann patterns show in Figs.~\ref{2D_12foldA_AmmannDuo_1} and \ref{2D_12foldA_AmmannDuo_2} which inflate into one another in an ``Ammann 2-cycle": Ammann duo A. 

{\it The 12-fold Ammann duo ``B."}  In the $I_{2}^{12}$ case, consider Table 1, Row 3B in \cite{BoyleSteinhardt1D}: if we take $(\{\alpha\},\{\beta\})=(0,1/2)$ or $(\{\alpha\},\{\beta\})=(1/2,0)$, we obtain the two different Ammann patterns show in Figs.~\ref{2D_12foldB_AmmannDuo_1} and \ref{2D_12foldB_AmmannDuo_2} which inflate into one another, forming an Ammann 2-cycle: Ammann duo B.  

{\it The 12-fold Ammann duo ``C."} In the $I_{2}^{12}$ case, consider Table 1, Row 3C in \cite{BoyleSteinhardt1D}: if we take $(\{\alpha\},\{\beta\})=(1/2,0)$ or $(\{\alpha\},\{\beta\})=(1/2,1/2)$, we obtain the two different Ammann patterns show in Figs.~\ref{2D_12foldC_AmmannDuo_1} and \ref{2D_12foldC_AmmannDuo_2} which inflate into one another, forming an Ammann 2-cycle: Ammann duo C.

\subsection{The $H_{3}$ (3D icosahedral) tilings (A1, A2, B1, B2, C1, C2, D1, D2 and E)}
\label{3Dfrom6D}

Here the relevant Coxeter pair is $\{\theta^{\parallel},\theta\}=\{H_{3},D_{6}\}$.  The $D_{6}$ root system has 60 roots: all vectors obtained from $(\pm1,\pm1,0,0,0,0)$ by allowing all combinations of signs and all permuations of the coordinates.   The columns of the following matrix are a standard choice for the fundamental roots \cite{ConwaySloane}:
\begin{equation}
  \left({\bf f}_{1}\,{\bf f}_{2}\,{\bf f}_{3}\,{\bf f}_{4}\,{\bf f}_{5}\,{\bf f}_{6}\right)
  =\left(\!\!\begin{array}{cccccc}
  -1 & +1 & 0 & 0 & 0 & 0 \\
  -1 & -1 & +1 & 0 & 0 & 0 \\
  0 & 0 & -1 & +1 & 0 & 0 \\
  0 & 0 & 0 & -1 & +1 & 0 \\
  0 & 0 & 0 & 0 & -1 & +1 \\
  0 & 0 & 0 & 0 & 0 & -1 \end{array}\!\!\right).
\end{equation}
In this case, the maximally-symmetric 3D orthogonal projection of the $D_{6}$ roots may be achieved by taking the six columns of the $6\times6$ matrix
\begin{equation}
  \left({\bf v}_{1}^{+}\,{\bf v}_{2}^{+}\,{\bf v}_{3}^{+}\,{\bf v}_{1}^{-}\,{\bf v}_{2}^{-}\,{\bf v}_{3}^{-}\right)
  \!=\!\left(\!\!\begin{array}{cccccc}
  \tau & +1 & 0 & \sigma & +1 & 0 \\
  0 & \tau & +1 & 0 & \sigma & +1 \\
  +1 & 0 & \tau & +1 & 0 & \sigma \\
  \tau & -1 & 0 & \sigma & -1 & 0 \\
  0 & \tau & -1 & 0 & \sigma & -1 \\
  -1 & 0 & \tau & -1 & 0 & \sigma \end{array}\!\!\right).
\end{equation}
as an orthogonal basis in six dimensions and choosing $\{{\bf v}_{1}^{+}, {\bf v}_{2}^{+}, {\bf v}_{3}^{+}\}$ as a basis for the $\parallel$ space, while $\{{\bf v}_{1}^{-}, {\bf v}_{2}^{-}, {\bf v}_{3}^{-}\}$ are a basis for the $\perp$ space.  It is easy to check that, with this choice, the 12 faces of the 6-cube in 6D [{\it i.e.}\ the 12 vectors obtained from $(\pm1,0,0,0,0,0)$ by allowing both signs and permutations of the coordinates] are project onto the $\parallel$ space to yield the 12 vertices of the icosahedron in 3D, while the 60 $D_{6}$ roots project to two copies of the 30 $H_{3}$ roots (an inner copy and an outer copy that is longer by $\tau$).

To proceed we just need to choose: (i) the underlying star $\{{\bf e}_{j}^{+}\}$ (the minimal star in this case is a 12-pointed star pointing towards the vertices of the icosahedron); and (ii) a relevant row from Table 1 in Ref.~\cite{BoyleSteinhardt1D} (the relevant rows are 4a, 4b, 4c, 4d and 1).

{\it The icosahedral ``A1" and ``A2" tilings.}  Consider Table 1, Row 4a in \cite{BoyleSteinhardt1D}.   Then, if we take $(\{\alpha\},\{\beta\})=(0,0)$ or, equivalently,  $(\{\alpha\},\{\beta\})=(0,1/2)$, we obtain the self-similar icosahedral A1 tiling.   Alternatively, if we take $(\{\alpha\},\{\beta\})=(1/2,0)$ or, equivalently,  $(\{\alpha\},\{\beta\})=(1/2,1/2)$, we obtain the self-similar icosahedral A2 tiling. 

{\it The icosahedral ``B1" and ``B2" tilings.}  Consider Table 1, Row 4b in \cite{BoyleSteinhardt1D}.   Then, if we take $(\{\alpha\},\{\beta\})=(0,0)$ or, equivalently,  $(\{\alpha\},\{\beta\})=(1/2,1/2)$, we obtain the self-similar icosahedral B1 tiling.   Alternatively, if we take $(\{\alpha\},\{\beta\})=(1/2,0)$ or, equivalently,  $(\{\alpha\},\{\beta\})=(0,1/2)$, we obtain the self-similar icosahedral B2 tiling. 

{\it The icosahedral ``C1" and ``C2" tilings.}  Consider Table 1, Row 4c in \cite{BoyleSteinhardt1D}.   Then, if we take $(\{\alpha\},\{\beta\})=(0,0)$ or, equivalently,  $(\{\alpha\},\{\beta\})=(0,1/2)$, we obtain the self-similar icosahedral C1 tiling.   Alternatively, if we take $(\{\alpha\},\{\beta\})=(1/2,0)$ or, equivalently,  $(\{\alpha\},\{\beta\})=(1/2,1/2)$, we obtain the self-similar icosahedral C2 tiling. 

{\it The icosahedral ``D1" and ``D2" tilings.}  Consider Table 1, Row 4d in \cite{BoyleSteinhardt1D}.   Then, if we take $(\{\alpha\},\{\beta\})=(0,0)$ or, equivalently,  $(\{\alpha\},\{\beta\})=(1/2,1/2)$, we obtain the self-similar icosahedral D1 tiling.   Alternatively, if we take $(\{\alpha\},\{\beta\})=(1/2,0)$ or, equivalently,  $(\{\alpha\},\{\beta\})=(0,1/2)$, we obtain the self-similar icosahedral D2 tiling. 

{\it The icosahedral ``E" tiling.}  Consider Table 1, Row 1 in \cite{BoyleSteinhardt1D}.   Then the cases where $(\{\alpha\},\{\beta\})$ are $(0,0)$, $(1/2,0)$, $(0,1/2)$ or, $1/2,1/2)$ are all equivalent, and all yield the self-similar icosahedral E tiling.

We note that in the first eight cases (A1, A2, B1, B2, C1, C2, D1, and D2), we obtain Ammann patterns and Penrose tilings that exhibit $\tau^{3}$ scaling and are all new.  In the fifth case (E), we obtain an Ammann pattern and Penrose tiling with $\tau$ scaling: this is precisely the icosahedral tiling found in by Socolar and Steinhardt in \cite{SocolarSteinhardt86}.  

Although all nine Ammann patterns are directly obtained from our construction, we will leave for future work the explicit presentation of the corresponding Penrose tiles, inflation rules and Ammann decorations, but we emphasize that they are completely specified by the above information and, although they are harder to display than the 2D cases, working them out is ultimately just a matter of turning the same crank that we used to produce the 2D tilings above.    

We also emphasize that, in each of these nine icosahedral cases, the Ammann pattern that is directly produced by our construction is already an icosahedral quasicrystalline tiling in its own right, with a finite number of tiles, and self-similarity; and, in contrast to the dual Penrose tiling, it has the advantage that it is much easier to work with analytically -- an advantage that becomes more useful as we get to higher dimensions, where it becomes harder and harder to study the tiles by inspection.

\subsection{The $H_{4}$ (4D hyper-icosahedral) tiling}
\label{4Dfrom8D}

Here the relevant Coxeter pair is $\{\theta^{\parallel},\theta\}=\{H_{4},E_{8}\}$.  The $E_{8}$ root system has 240 roots: all 128 vectors of the form $(1/2)(\pm1,\pm1,\pm1,\pm1,\pm1,\pm1,\pm1,\pm1)$ (with an even number of minus signs), along with all 112 vectors of the form $(\pm1,\pm1,0,0,0,0,0,0)$ (including all sign combinations and permutations of the coordinates).  The columns of the following matrix are a standard choice for the fundamental roots ${\bf f}_{k}$ \cite{ConwaySloane}:
\begin{equation}
  \left(\begin{array}{cccccccc}
  +2 & -1 & 0 & 0 & 0 & 0 & 0 & 1/2 \\
  0 & +1 & -1 & 0 & 0 & 0 & 0 & 1/2 \\
  0 & 0 & +1 & -1 & 0 & 0 & 0 & 1/2 \\
  0 & 0 & 0 & +1 & -1 & 0 & 0 & 1/2 \\
  0 & 0 & 0 & 0 & +1 & -1 & 0 & 1/2 \\
  0 & 0 & 0 & 0 & 0 & +1 & -1 & 1/2 \\
  0 & 0 & 0 & 0 & 0 & 0 & +1 & 1/2 \\
  0 & 0 & 0 & 0 & 0 & 0 & 0 & 1/2 
  \end{array}\right).
\end{equation}
In this case, if we take $I$ to be the $4\times4$ unit matrix and $H$ to be the Hadamard matrix:
\begin{equation}
  H=\frac{1}{2}\left[\begin{array}{rrrr}
  -1 & -1 & -1 & -1 \\
  1 & -1 & -1 & 1 \\
  1 & 1 & -1 & -1 \\
  1 & -1 & 1 & -1 \end{array}\right]
\end{equation}
then the maximally-symmetric 4D orthogonal projection of the $E_{8}$ roots may be achieved by taking the eight columns of the $8\times8$ matrix
\begin{equation}
  \left({\bf v}_{1}^{+}{\bf v}_{2}^{+}{\bf v}_{3}^{+}{\bf v}_{4}^{+}{\bf v}_{1}^{-}{\bf v}_{2}^{-}{\bf v}_{3}^{-}{\bf v}_{4}^{-}\right)
  =\left[\!\!\begin{array}{cc} 
  (I\!+\!\sigma H)\; & \;(I\!+\!\tau H) \\
  (I\!-\!\sigma H)\; & \;(I\!-\!\tau H) \end{array}\!\!\right]
\end{equation}
as an orthogonal basis in eight dimensions, and choosing $\{{\bf v}_{1}^{+}, {\bf v}_{2}^{+}, {\bf v}_{3}^{+}, {\bf v}_{4}^{+}\}$ as a basis for the $\parallel$ space, while $\{{\bf v}_{1}^{-}, {\bf v}_{2}^{-}, {\bf v}_{3}^{-}, {\bf v}_{4}^{-}\}$ are a basis for the $\perp$ space.  With this choice, the 240 $E_{8}$ roots project onto the parallel space to yield two copies of the 120 $H_{4}$ roots (an inner copy and an outer copy that is longer by $\tau$).

To proceed we just need to choose: (i) the underlying star $\{{\bf e}_{j}^{+}\}$ (the minimal star in this case is a 120-pointed star pointing towards the vertices of the 600-cell \cite{RegularPolytopes}); and (ii) a relevant row from Table 1 in Ref.~\cite{BoyleSteinhardt1D} (the only relevant row is Row 1).  

The resulting Ammann pattern and the dual Penrose tiling are both new. As in the icosahedral case, we emphasize that the Ammann pattern is directly obtained from our construction, whereas we will leave for future work the explicit presentation of the corresponding Penrose tiles, inflation rules and Ammann decorations (but, as before, we emphasize that they are completely specified by the above information and, although they are now {\it much} harder to display than the 2D cases, working them out is ultimately just a matter of turning the same crank that we used to produce the 2D tilings above).  

\begin{center}
  \includegraphics[width=2.4in]{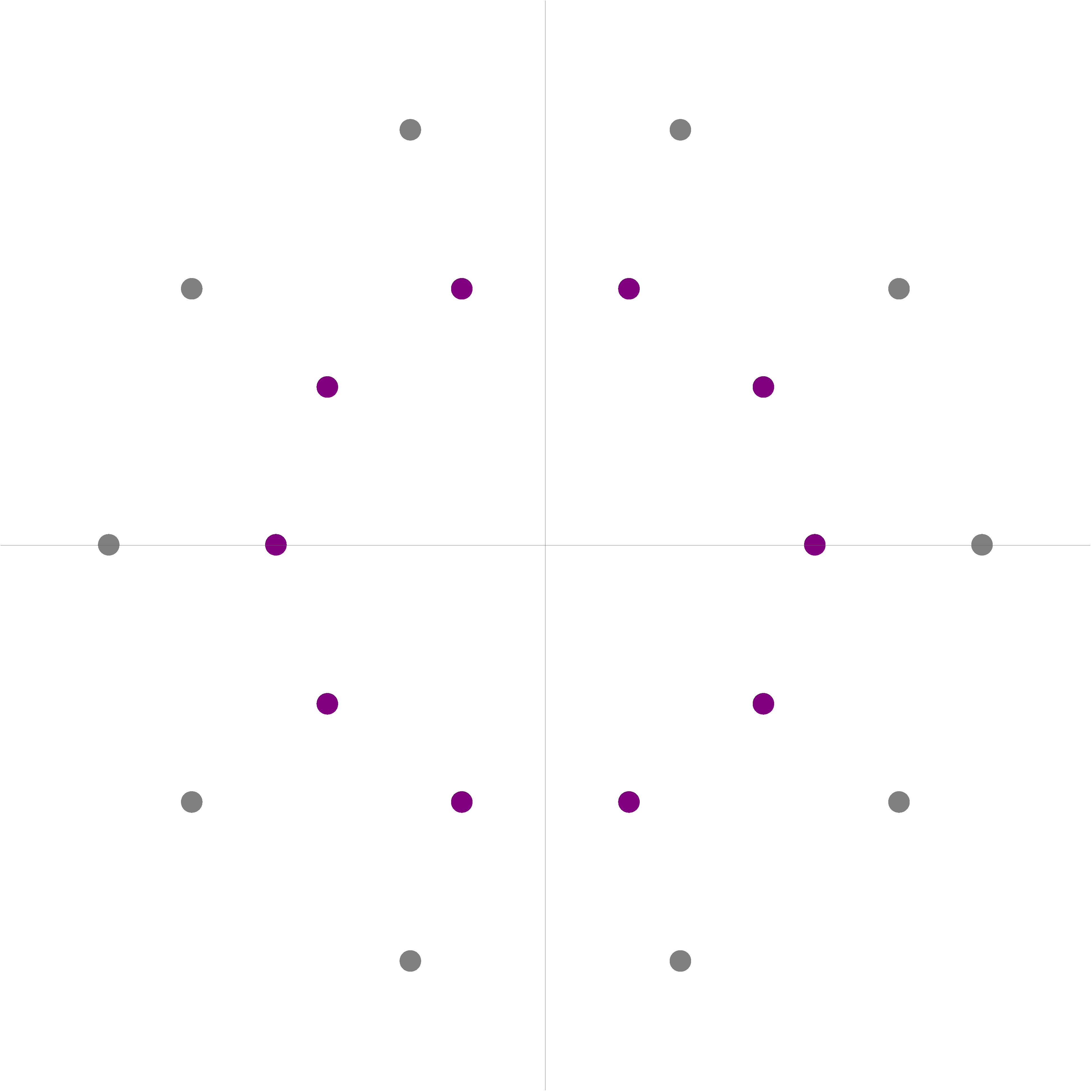}
  \captionof{figure}{The 20 $A_{4}$ roots, projected on the Coxeter plane.}
  \label{A4rootfig}
\end{center}

\begin{center}
  \includegraphics[width=2.4in]{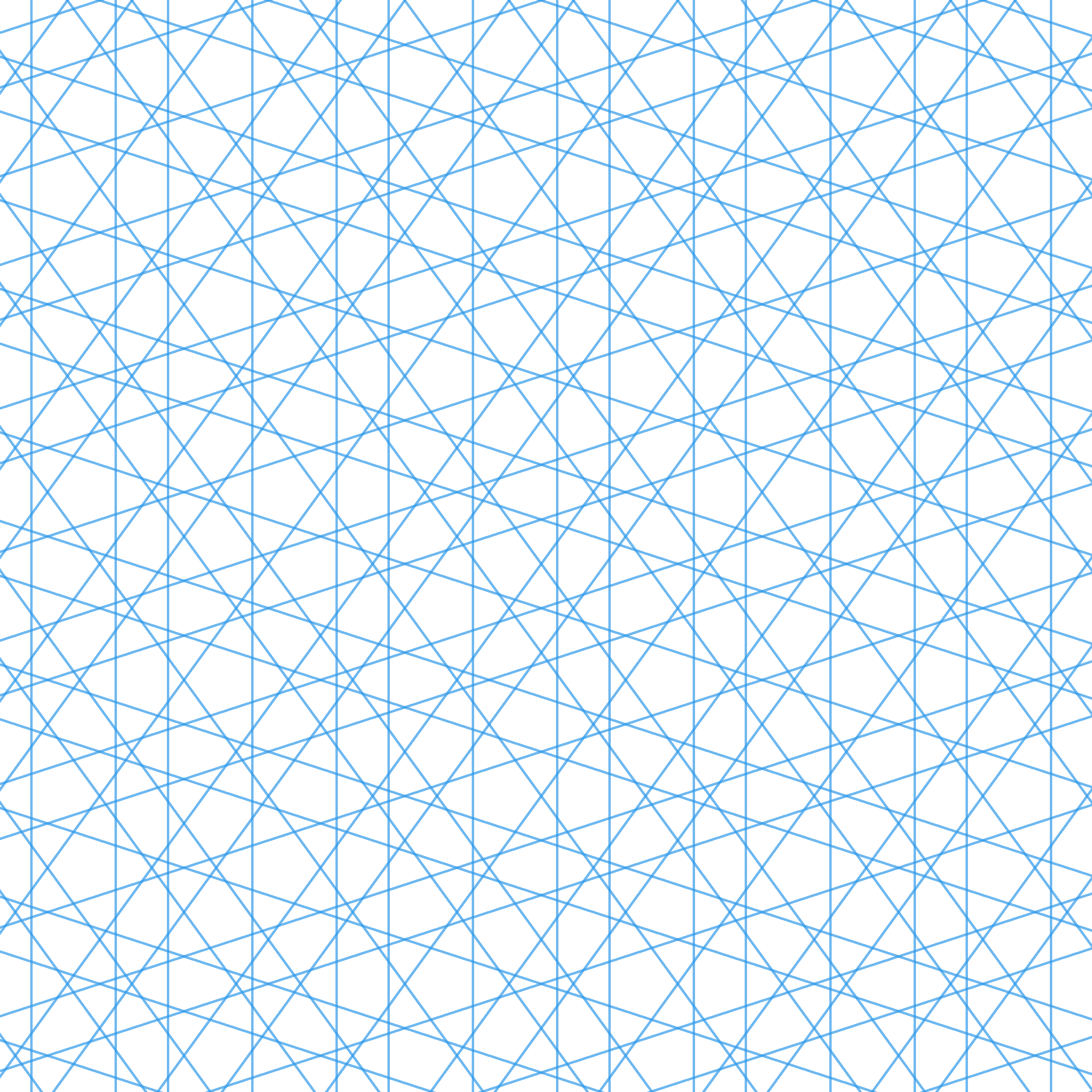}
  \captionof{figure}{The 2D 10-fold Ammann pattern.}
  \label{2D_10fold_PureAmmann}
  \vspace{10mm}
  \includegraphics[width=2.4in]{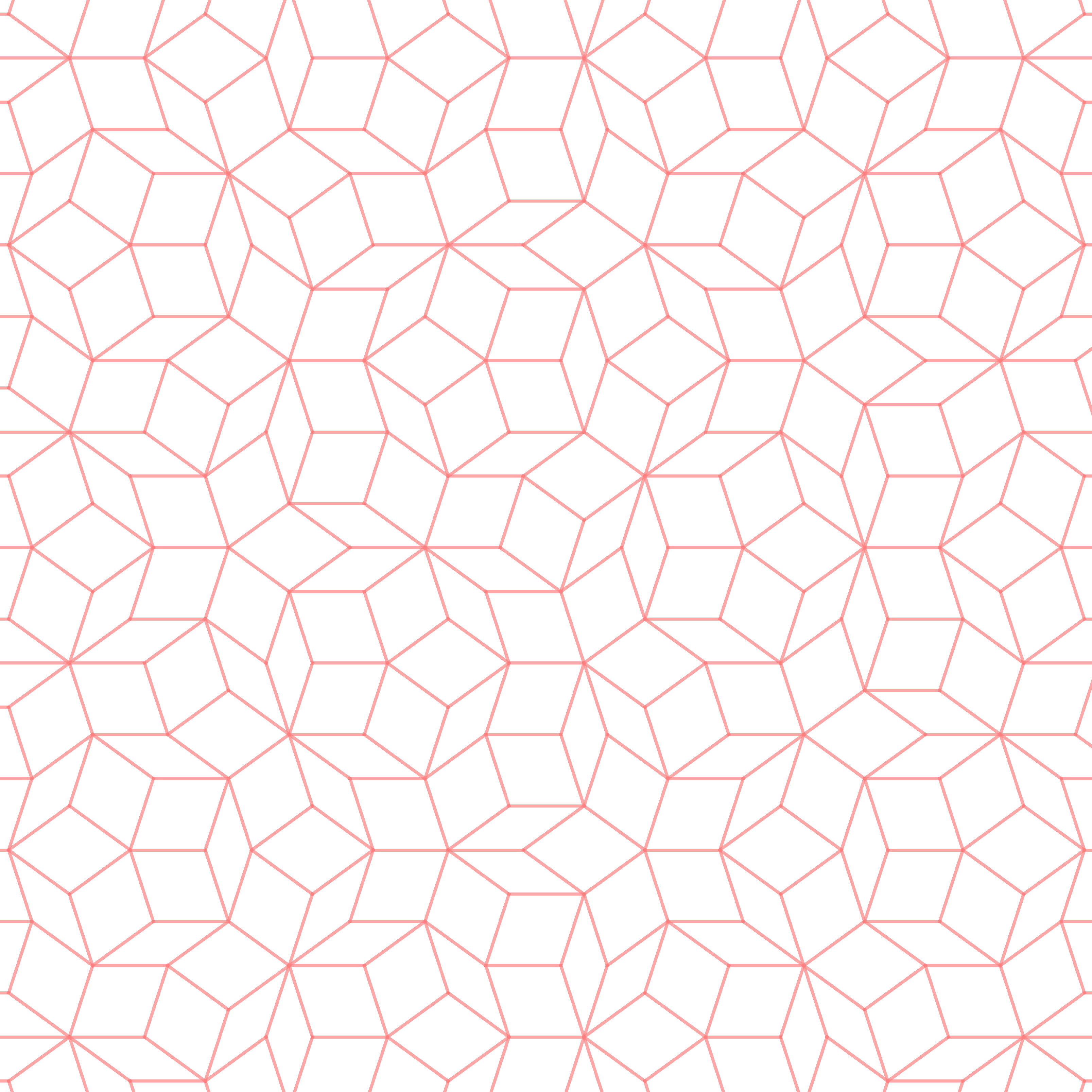}
  \captionof{figure}{The dual 2D 10-fold (Penrose) tiling.}
   \label{2D_10fold_PurePenrose}
\end{center}

\begin{center}
  \includegraphics[width=2.4in]{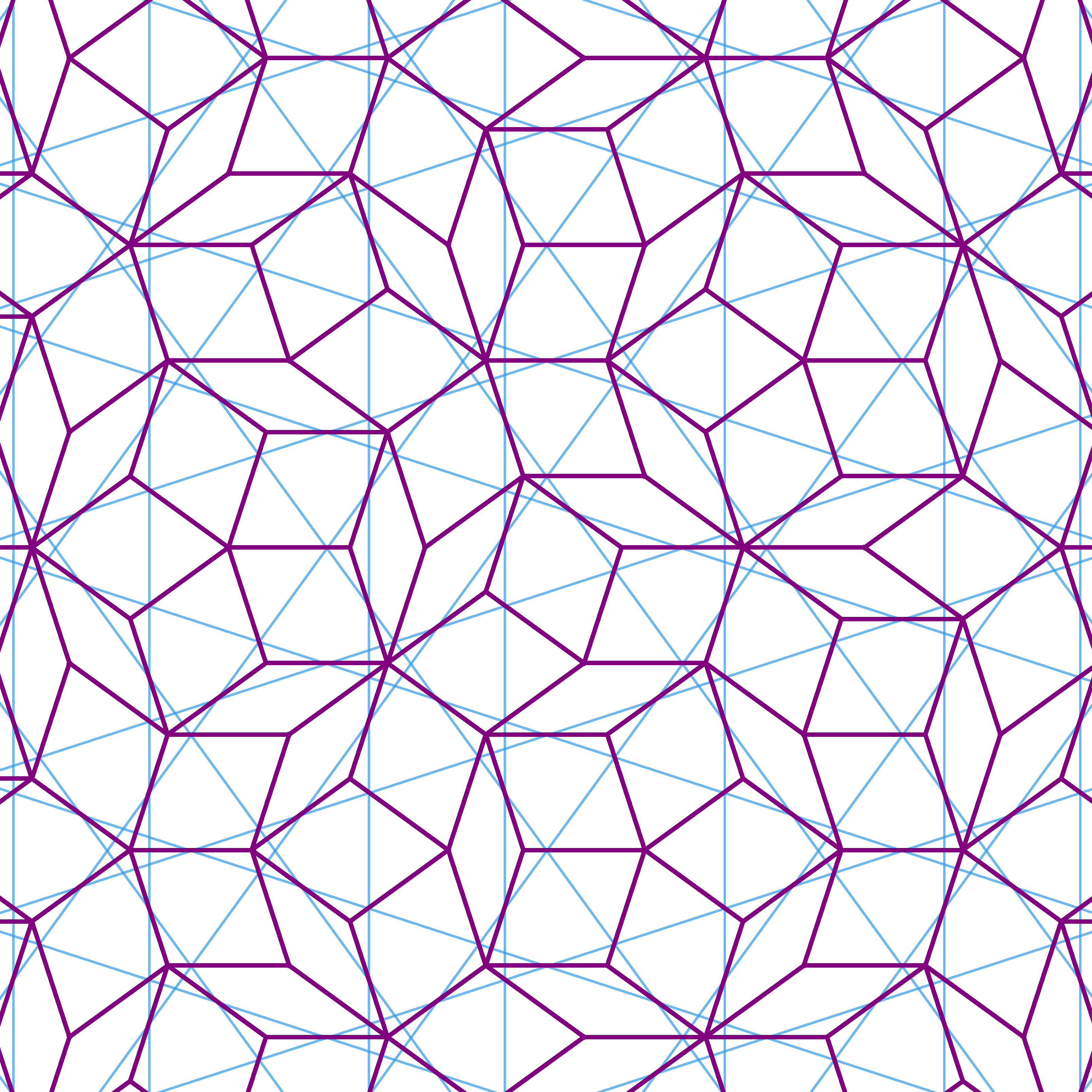}
  \captionof{figure}{The 2D 10-fold tiling (thick, purple), with Ammann lines (thin, blue).}
  \label{2D_10fold_AmmannLines}
  \vspace{10mm}
  \includegraphics[width=2.4in]{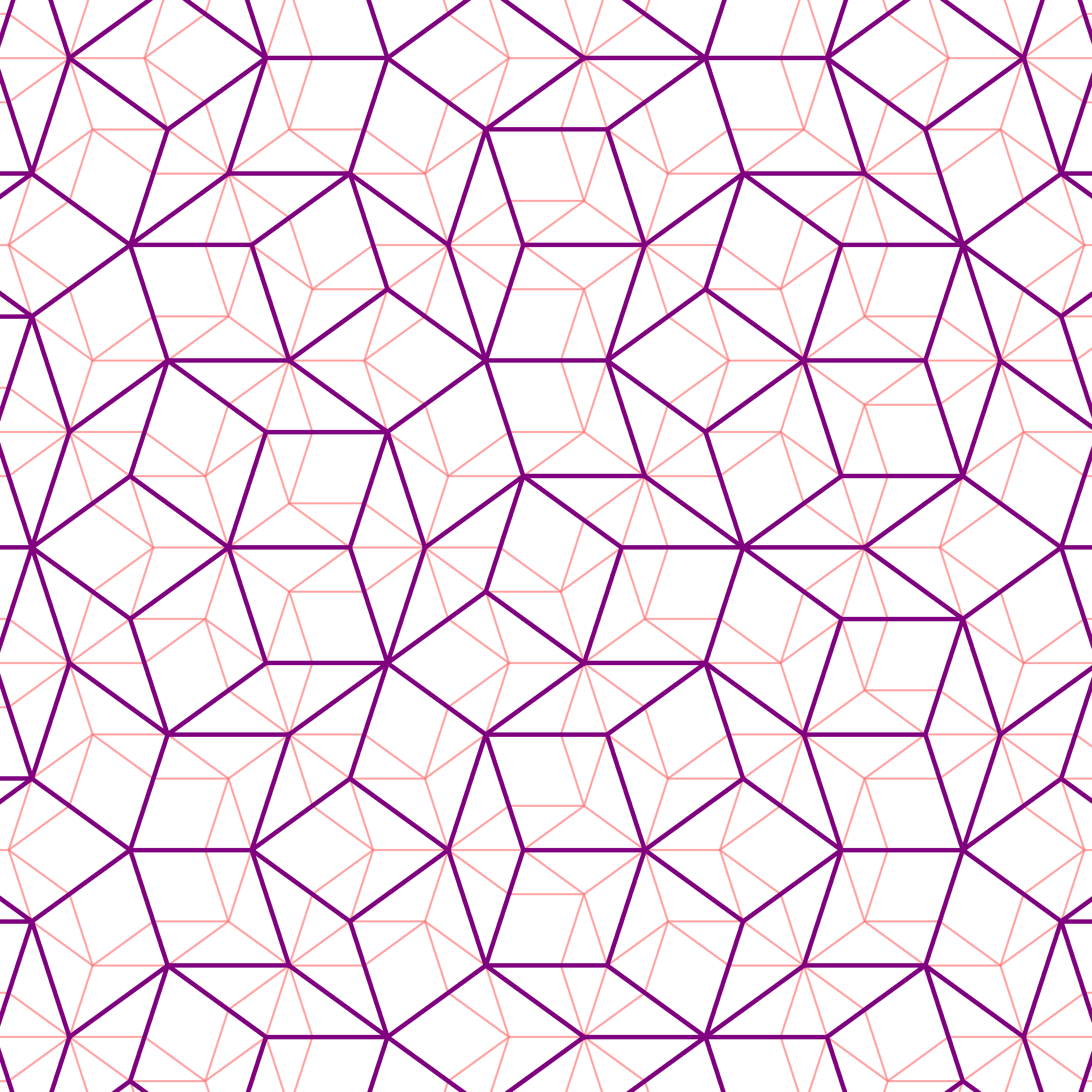}
  \captionof{figure}{The 2D 10-fold tiling (thick, purple), and its inflation (thin, pink).}
   \label{2D_10fold_Inflation}
\end{center}

\begin{center}
  \includegraphics[width=2.4in]{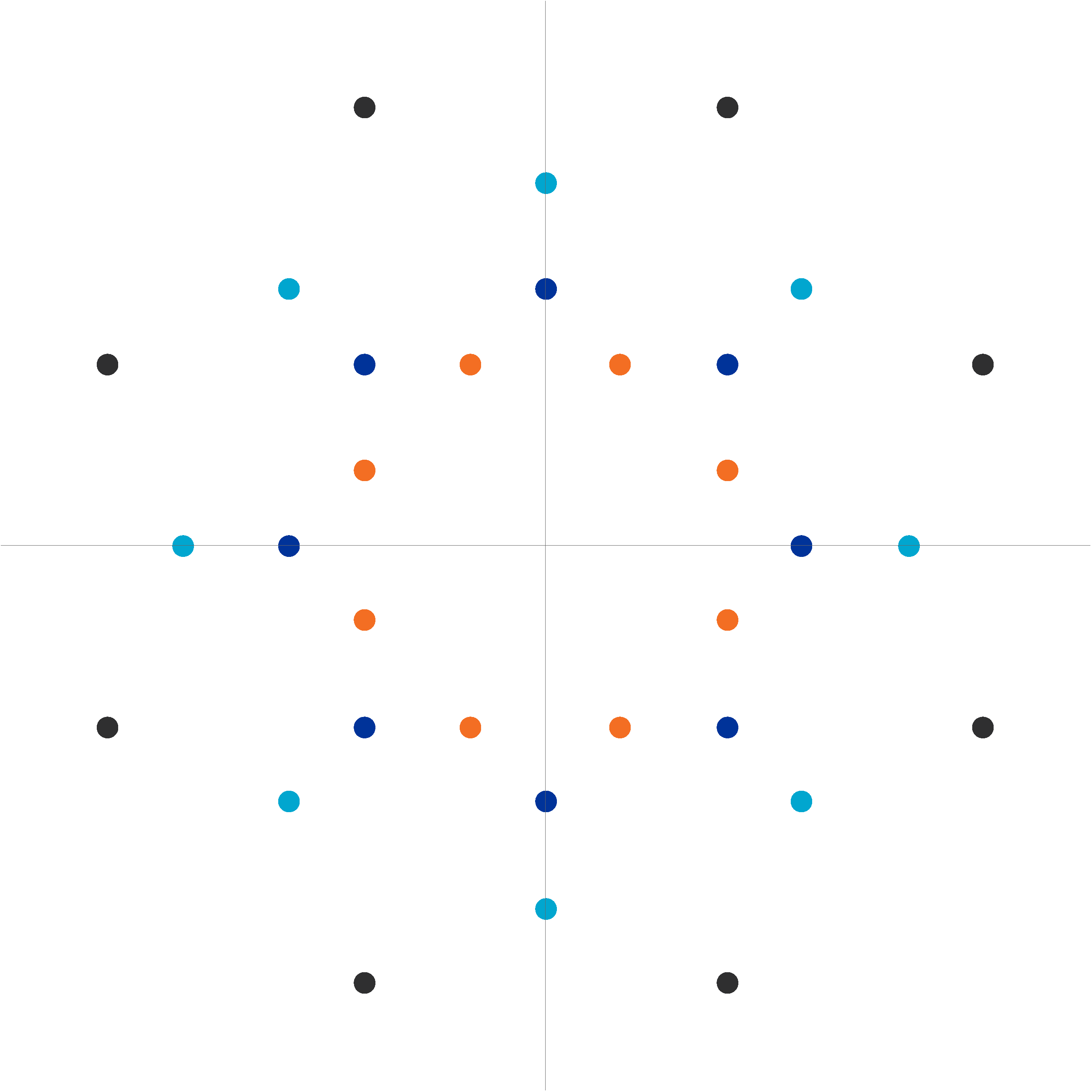}
  \captionof{figure}{The 32 $B_{4}$ roots, projected on the Coxeter plane.}
  \label{B4rootfig}
\end{center}

\begin{center}
  \includegraphics[width=2.4in]{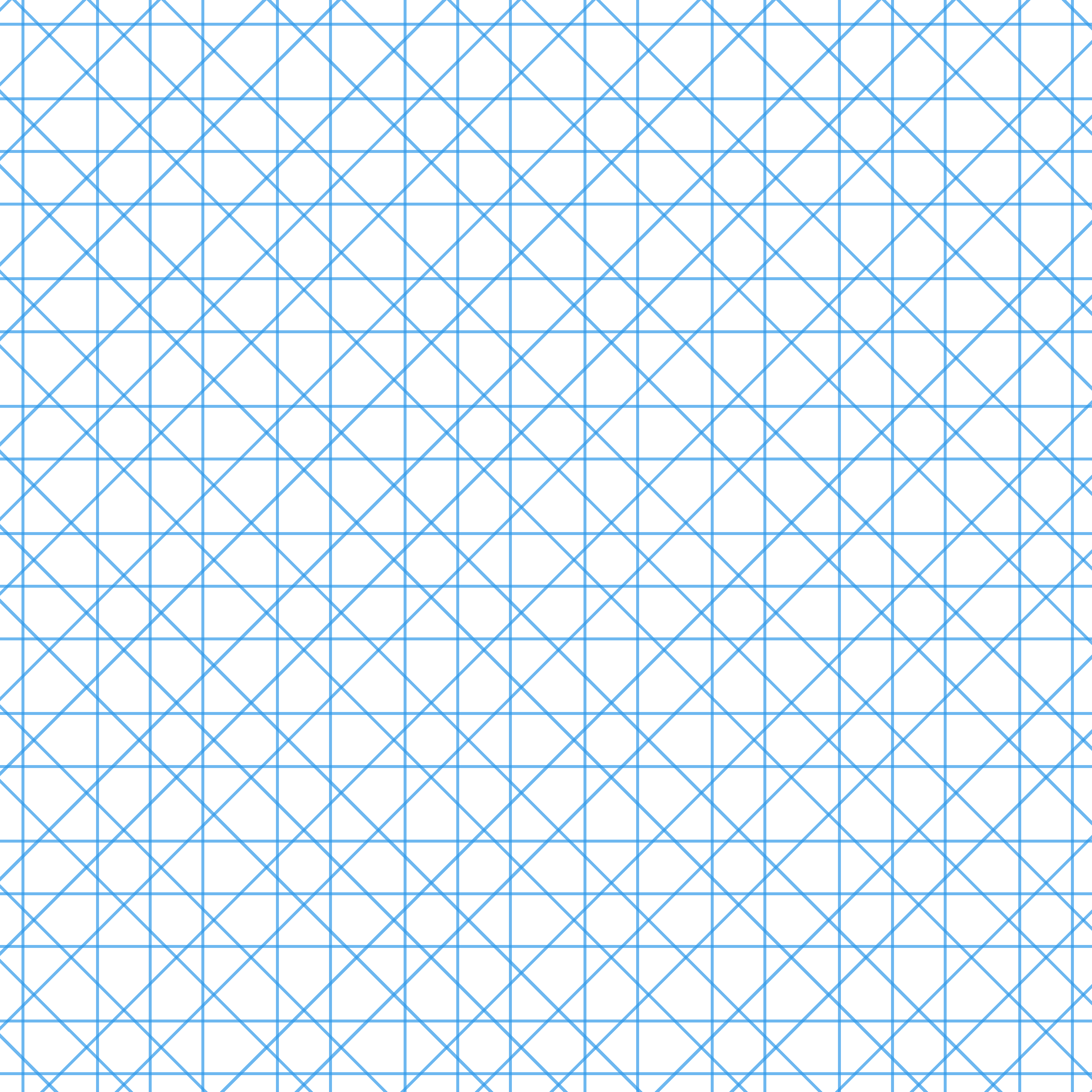}
  \captionof{figure}{The 2D 8-fold A1 Ammann pattern.}
  \label{2D_8foldA1_PureAmmann}
  \vspace{10mm}
  \includegraphics[width=2.4in]{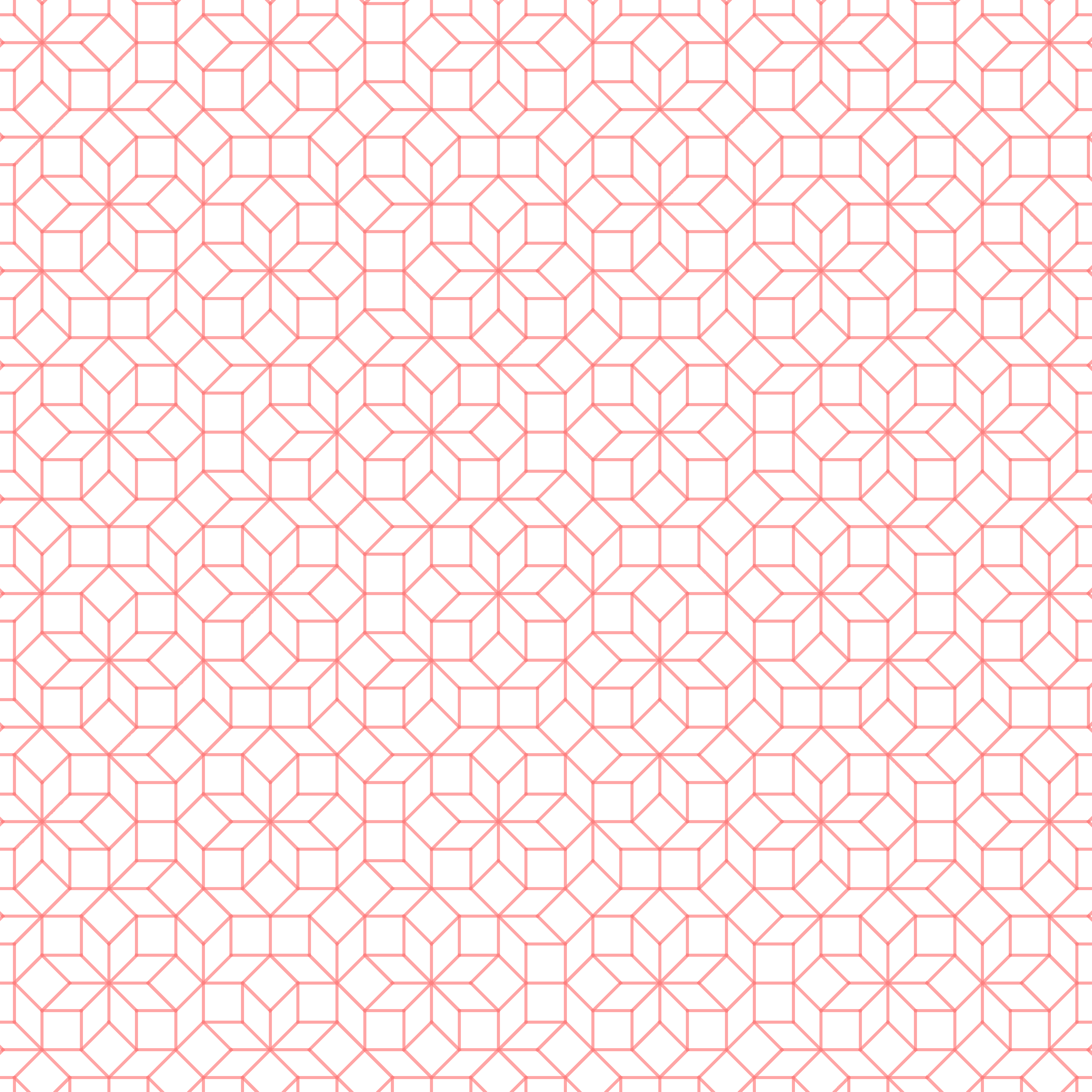}
  \captionof{figure}{The dual 2D 8-fold A1 (Ammann-Beenker) tiling.}
  \label{2D_8foldA1_PurePenrose}
\end{center}

\begin{center}
  \includegraphics[width=2.4in]{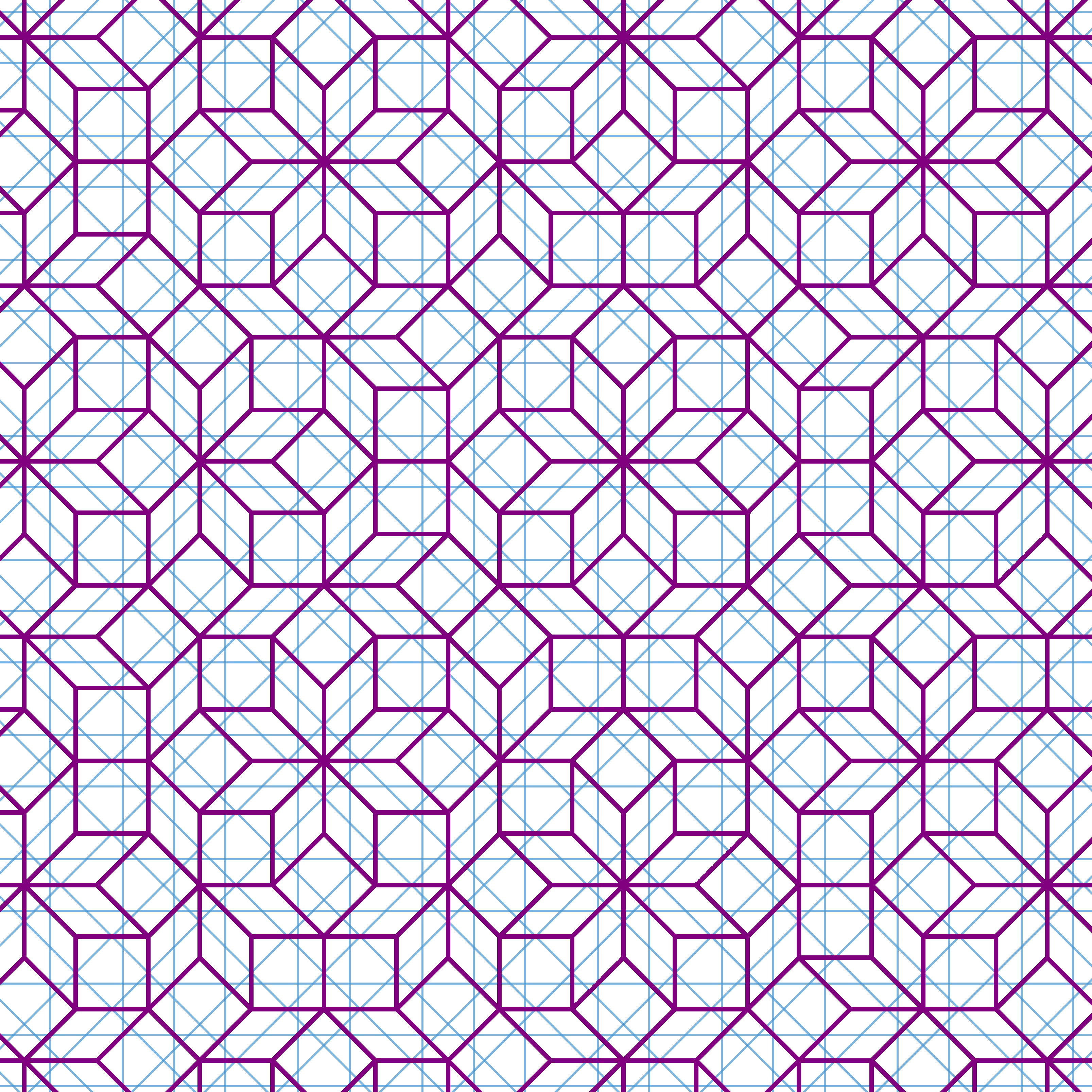}
  \captionof{figure}{The 2D 8-fold A1 (Ammann-Beenker) tiling (thick, purple), with Ammann lines (thin, blue).}
  \label{2D_8foldA1_AmmannLines}
  \vspace{10mm}
  \includegraphics[width=2.4in]{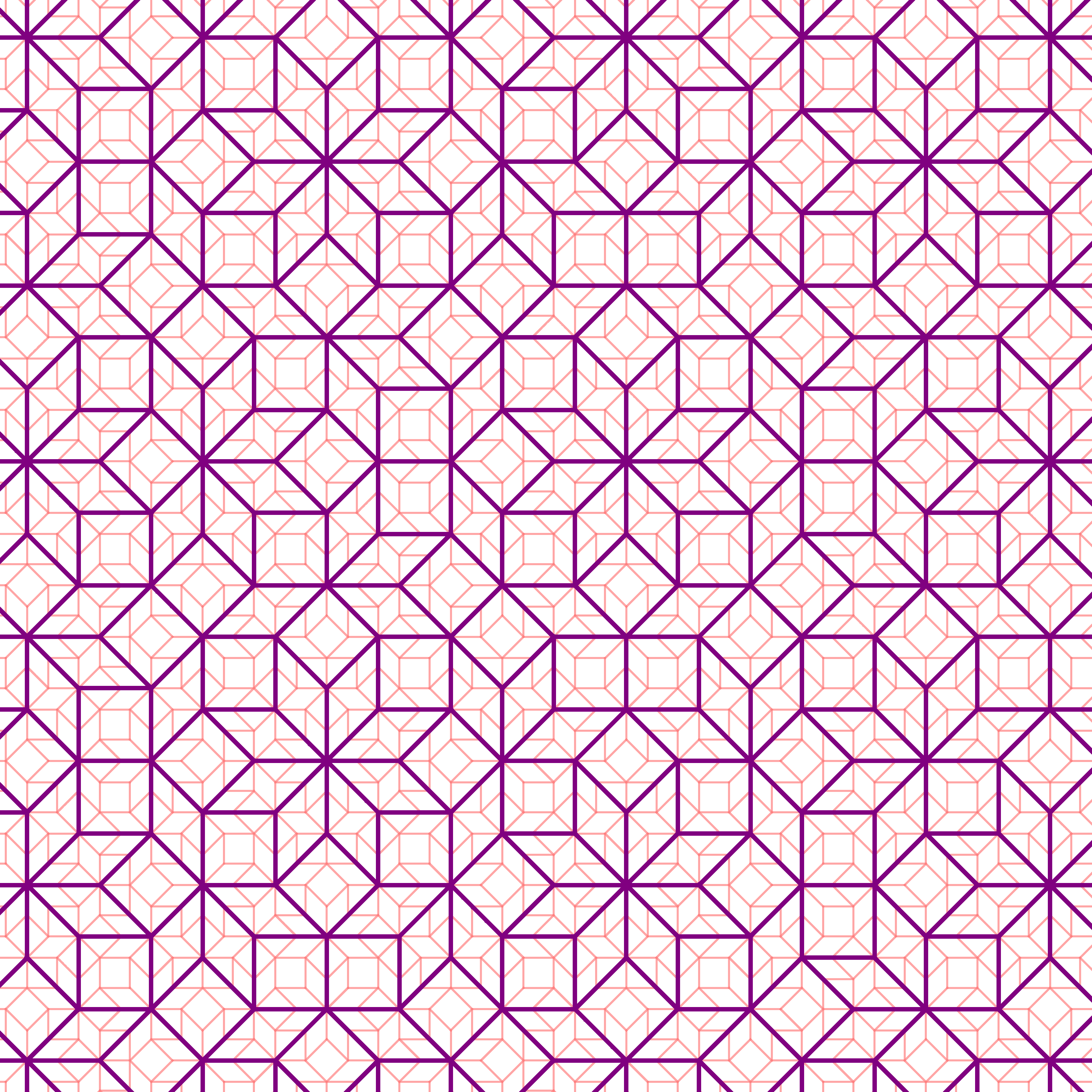}
  \captionof{figure}{The 2D 8-fold A1 (Ammann-Beenker) tiling (thick, purple), and its inflation (thin, pink).}
  \label{2D_8foldA1_Inflation}
\end{center}

\begin{center}
  \includegraphics[width=2.4in]{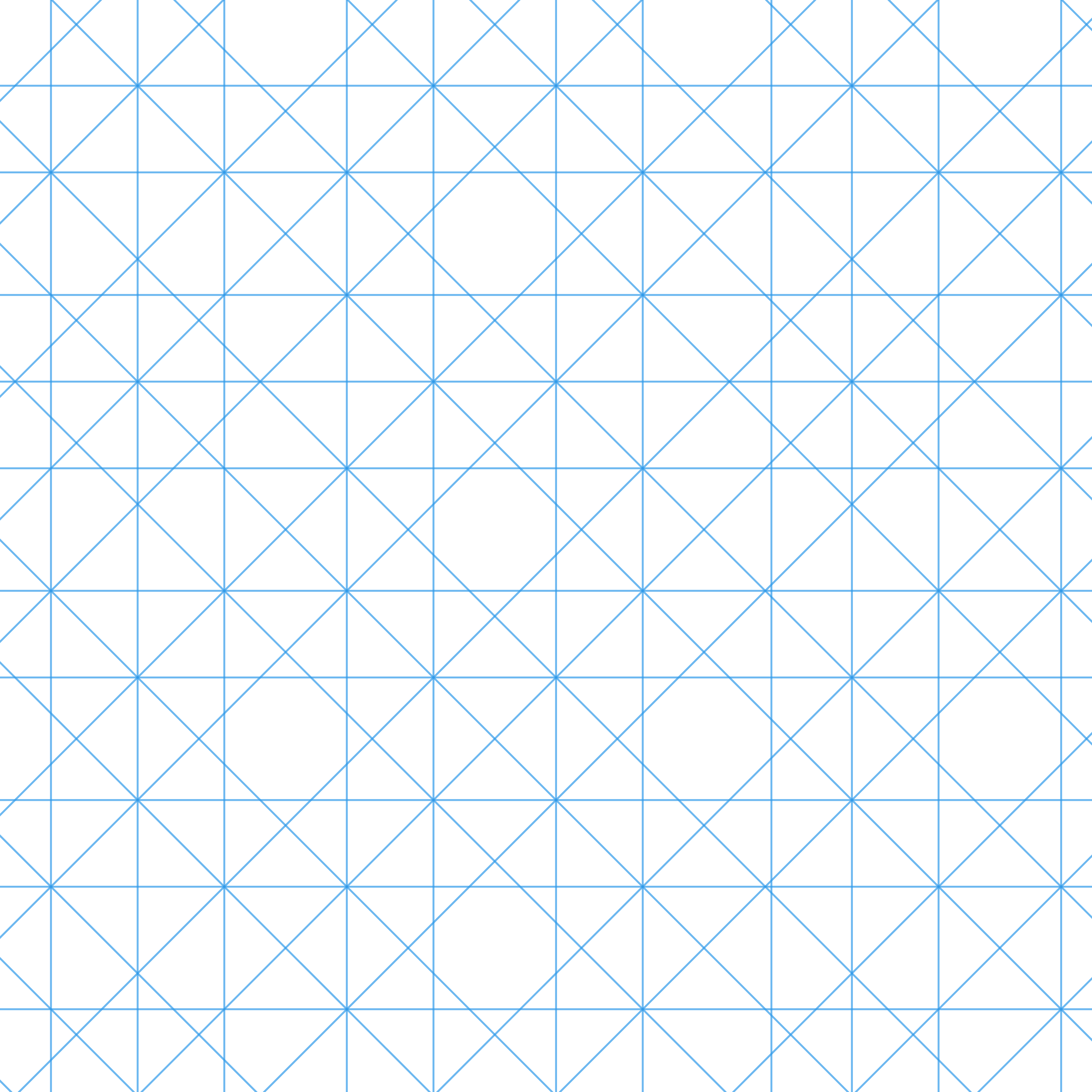}
  \captionof{figure}{The 2D 8-fold A2 Ammann pattern.}
  \label{2D_8foldA2_PureAmmann}
  \vspace{10mm}
  \includegraphics[width=2.4in]{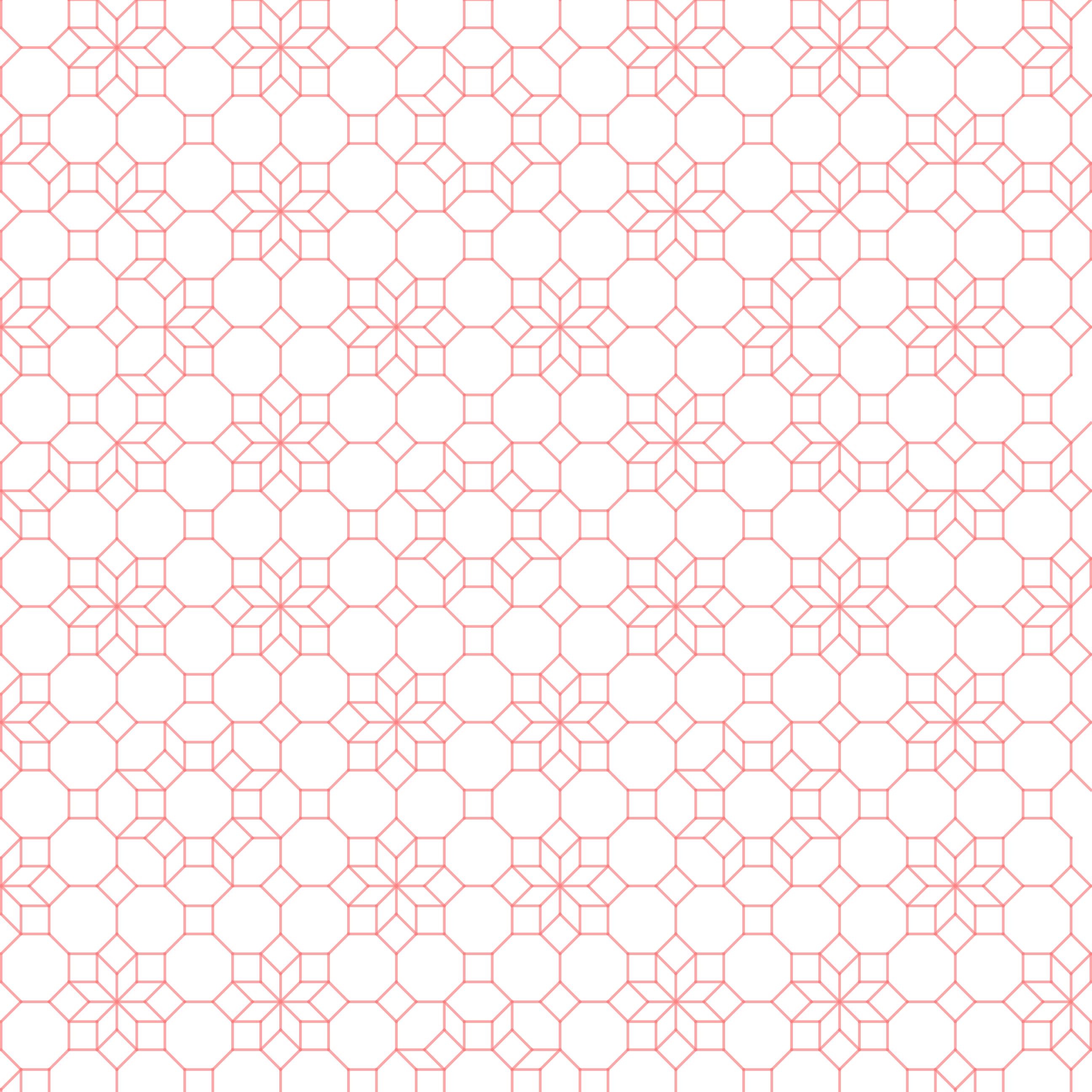}
  \captionof{figure}{The dual 2D 8-fold A2 tiling.}
  \label{2D_8foldA2_PurePenrose}
\end{center}

\begin{center}
  \includegraphics[width=2.4in]{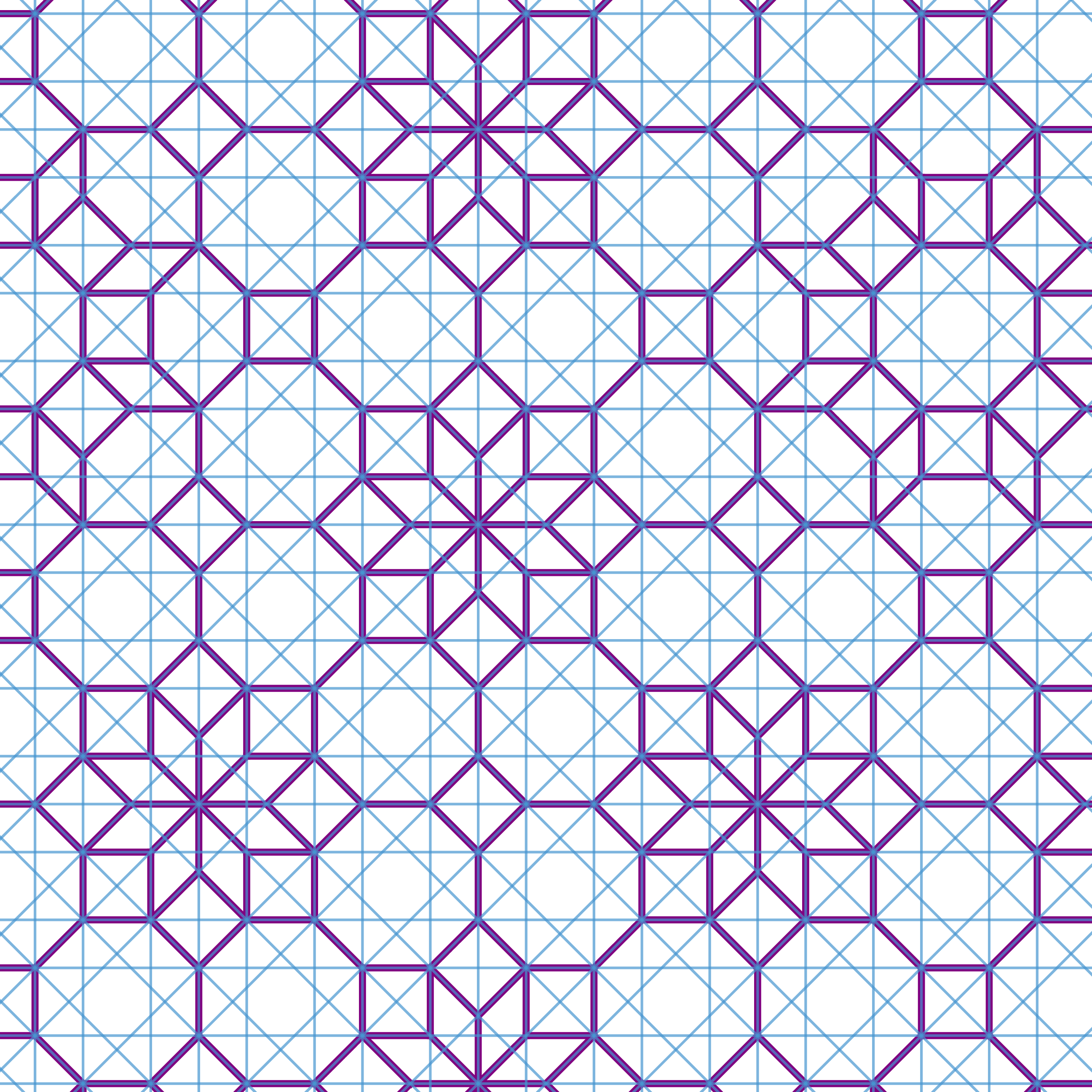}
  \captionof{figure}{The 2D 8-fold A2 tiling (thick, purple), with Ammann lines (thin, blue).}
  \label{2D_8foldA2_AmmannLines}
  \vspace{10mm}
  \includegraphics[width=2.4in]{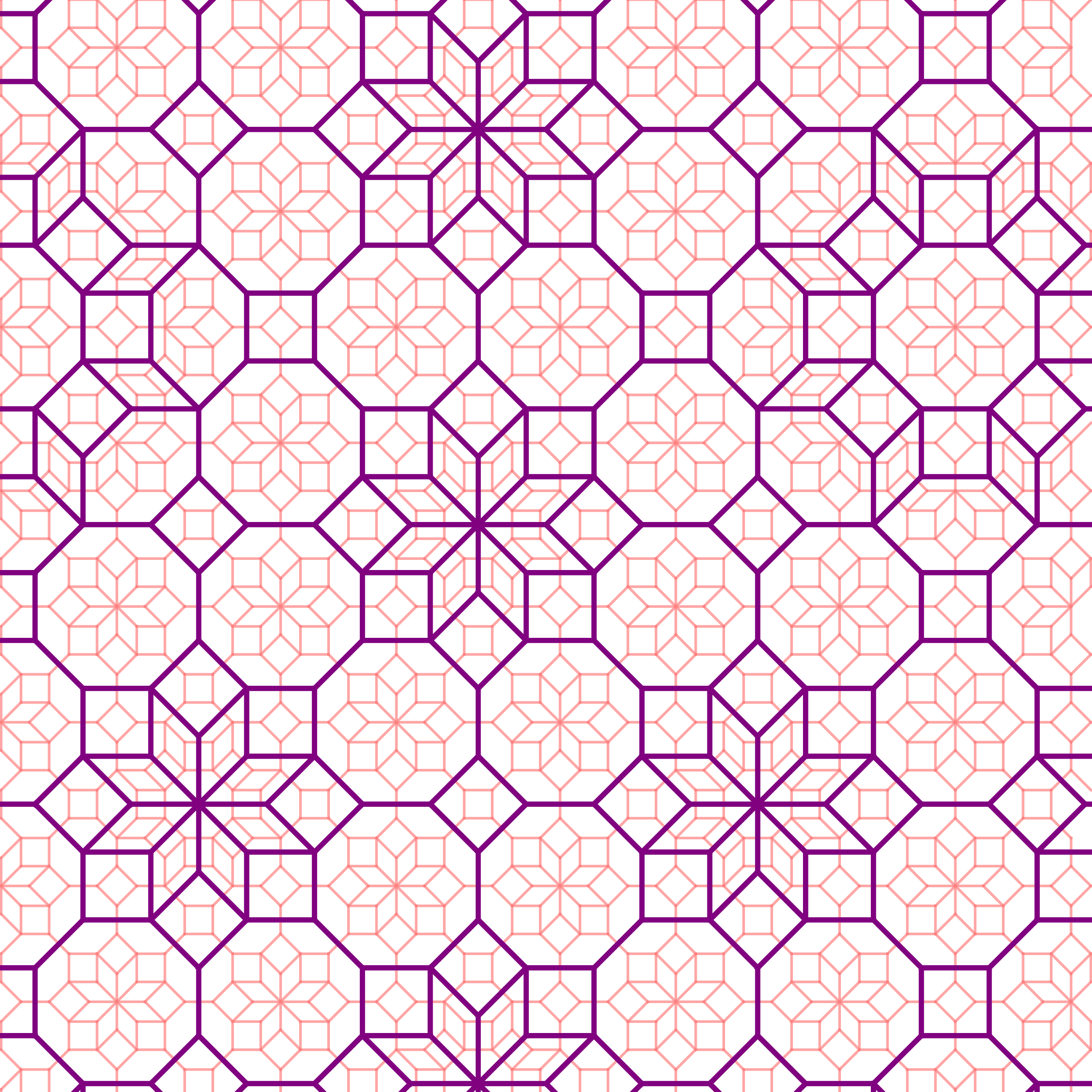}
  \captionof{figure}{The 2D 8-fold A2 tiling (thick, purple), and its inflation (thin, pink).}
  \label{2D_8foldA2_Inflation}
\end{center}

\begin{center}
  \includegraphics[width=2.4in]{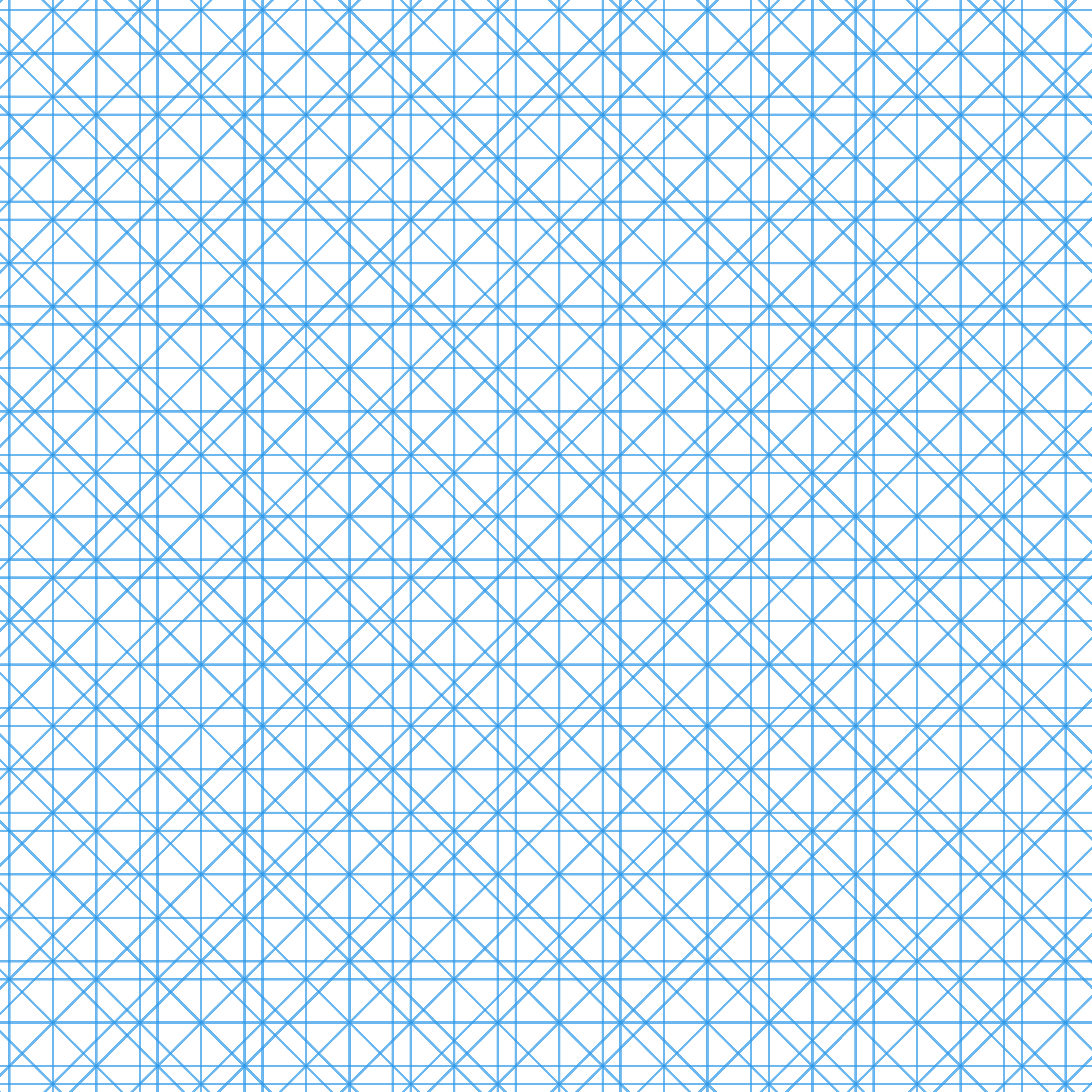}
  \captionof{figure}{The 2D 8-fold B1 Ammann pattern.}
  \label{2D_8foldB1_PureAmmann}
  \vspace{10mm}
  \includegraphics[width=2.4in]{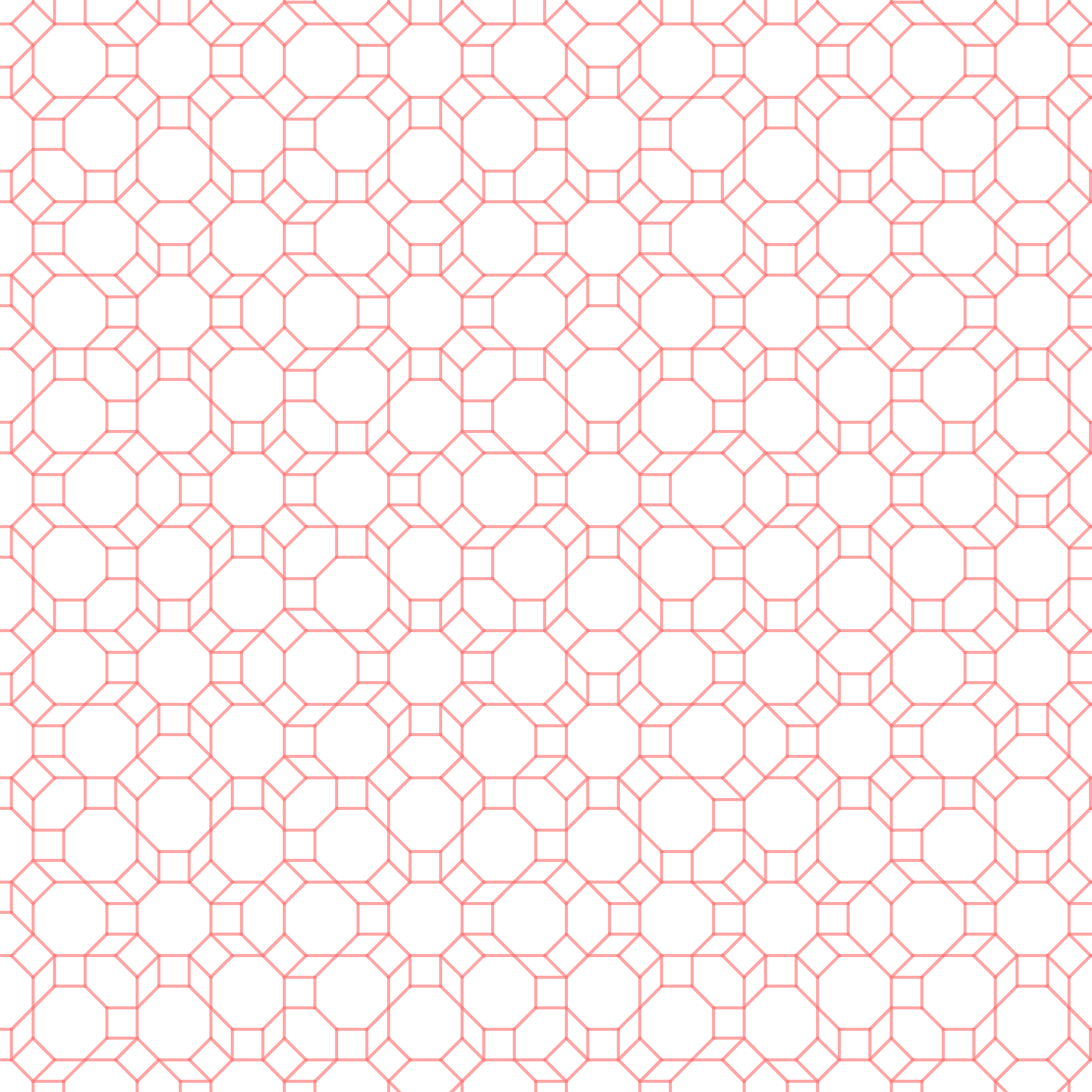}
  \captionof{figure}{The dual 2D 8-fold B1 tiling.}
  \label{2D_8foldB1_PurePenrose}
\end{center}

\begin{center}
  \includegraphics[width=2.4in]{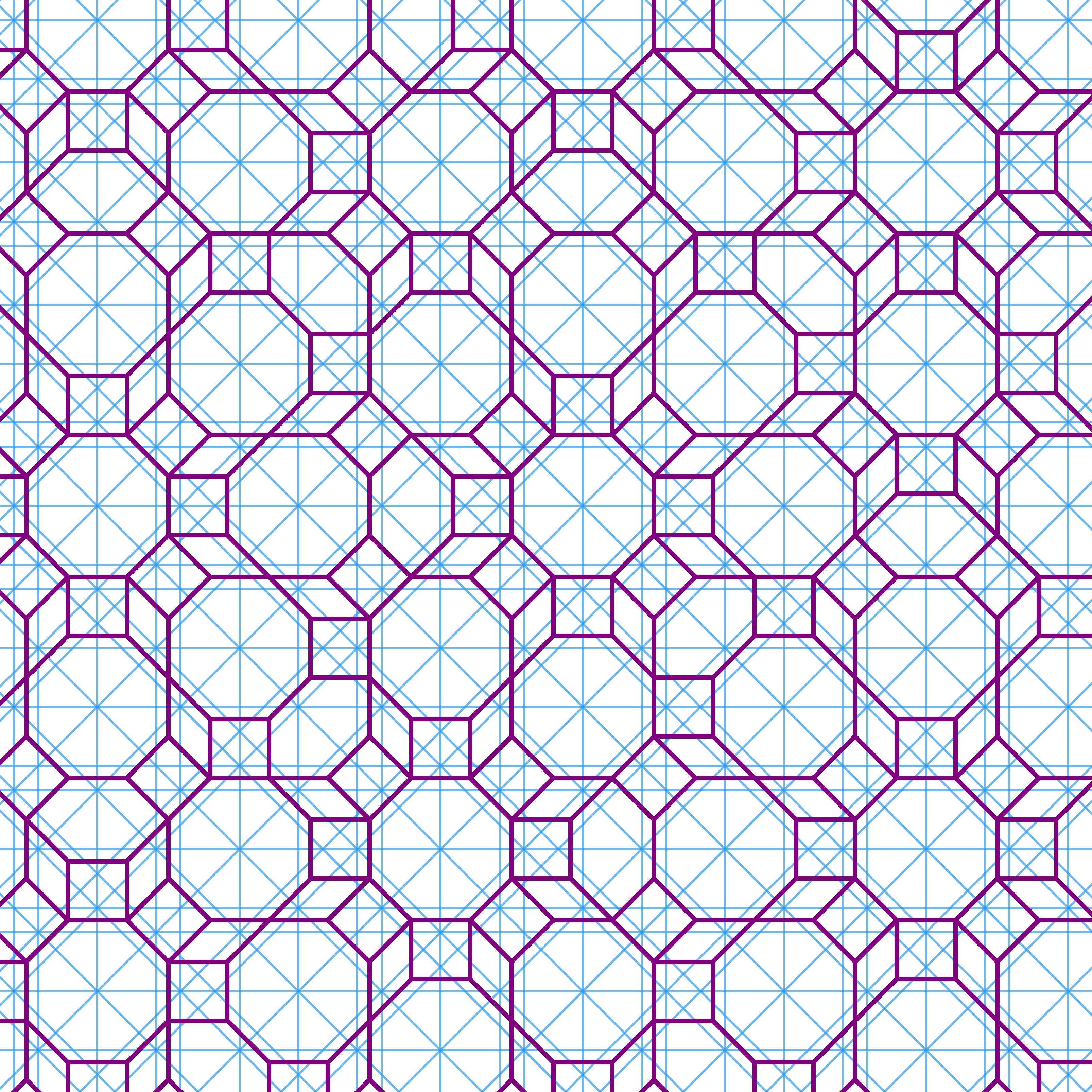}
  \captionof{figure}{The 2D 8-fold B1 tiling (thick, purple), with Ammann lines (thin, blue).}
  \label{2D_8foldB1_AmmannLines}
  \vspace{10mm}
  \includegraphics[width=2.4in]{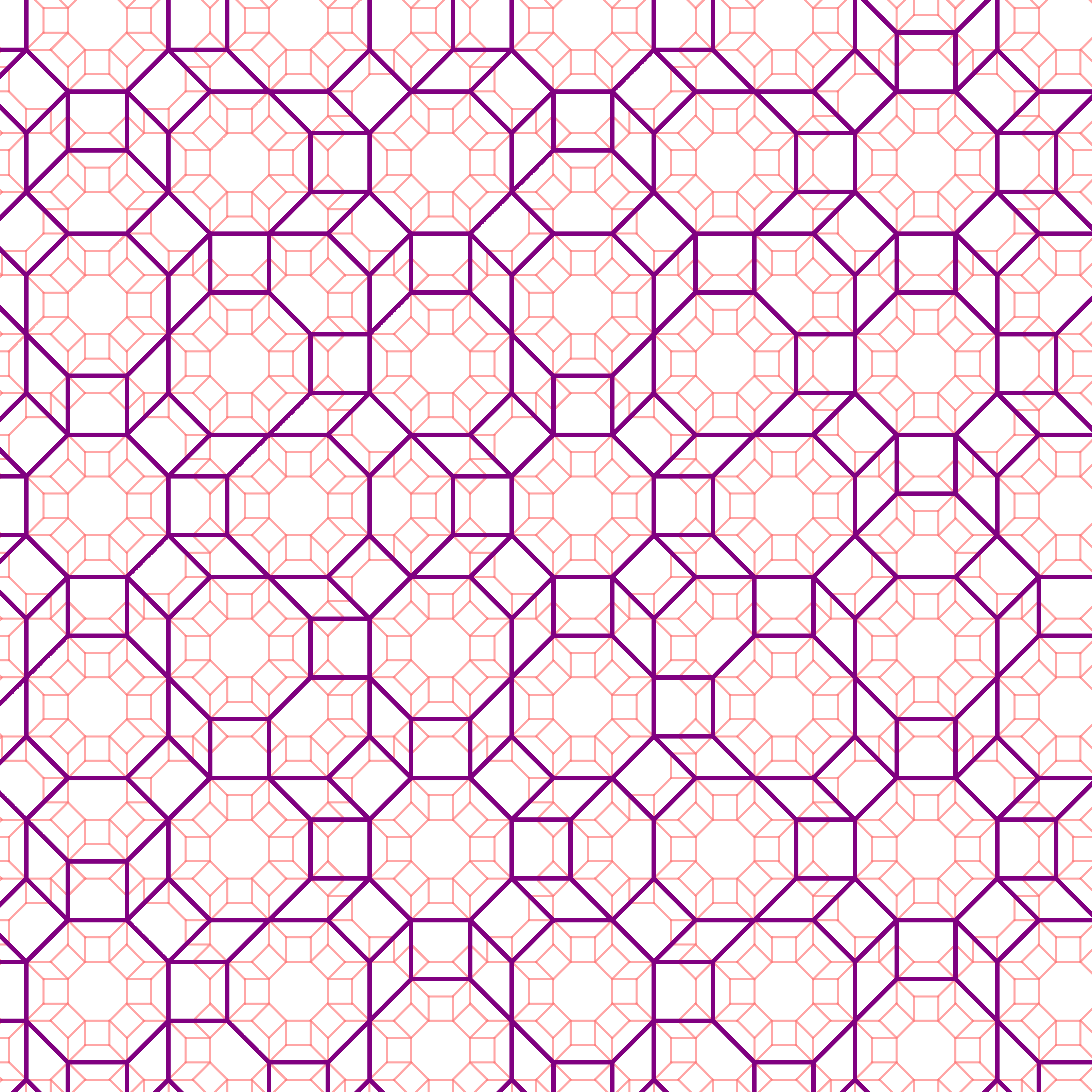}
  \captionof{figure}{The 2D 8-fold B1 tiling (thick, purple), and its inflation (thin, pink).}
  \label{2D_8foldB1_Inflation}
\end{center}

\begin{center}
  \includegraphics[width=2.4in]{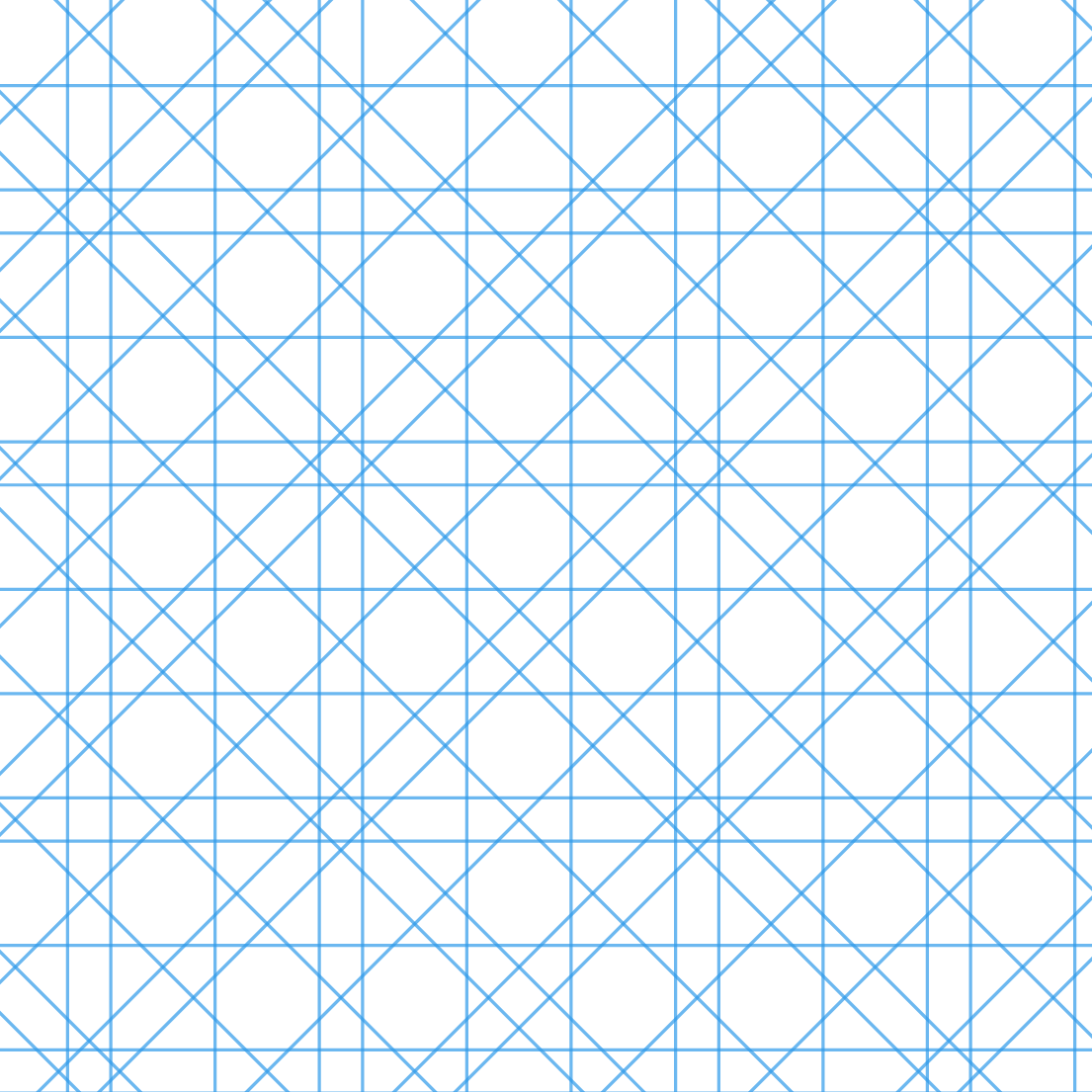}
  \captionof{figure}{The 2D 8-fold B2 Ammann pattern.}
  \label{2D_8foldB2_PureAmmann}
  \vspace{10mm}
  \includegraphics[width=2.4in]{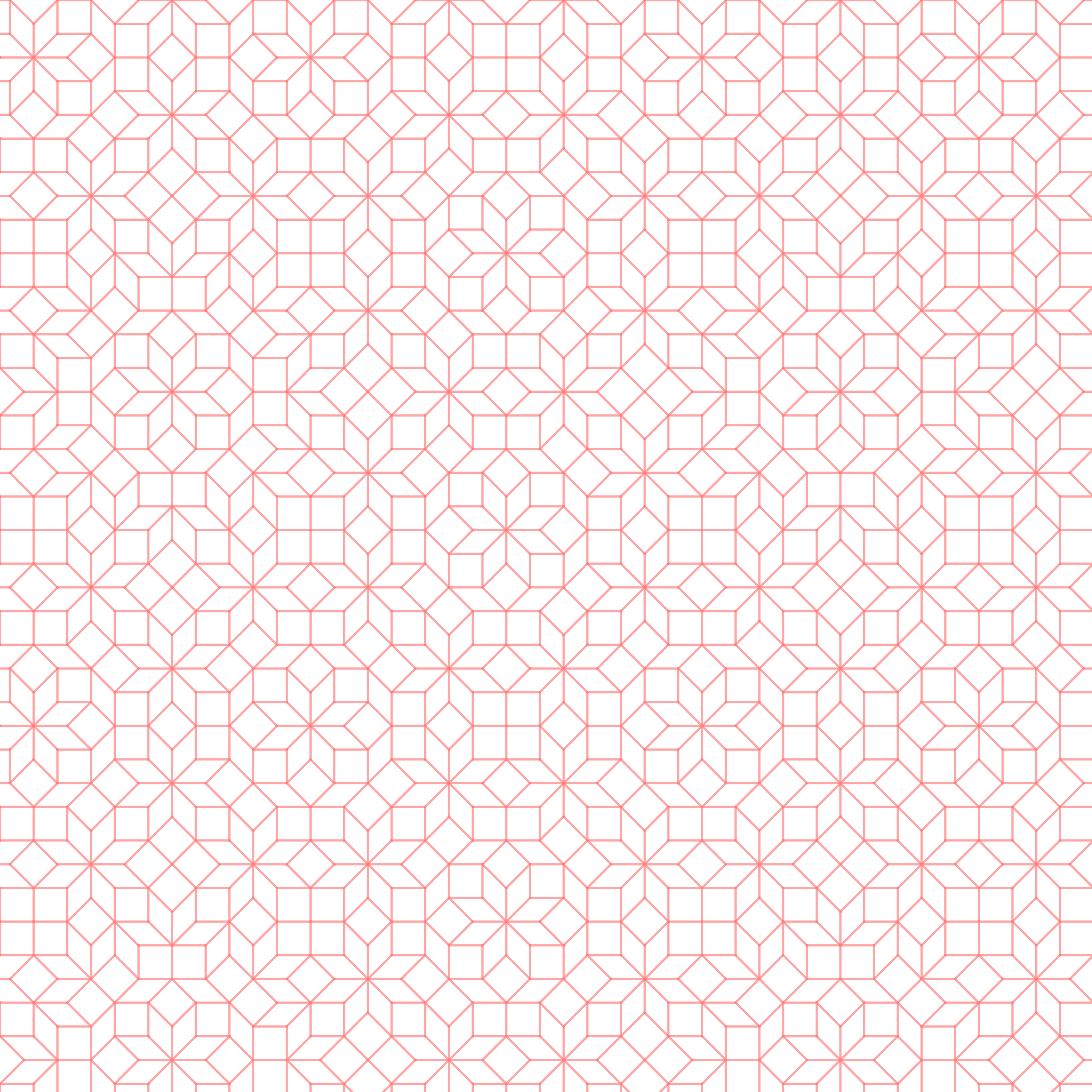}
  \captionof{figure}{The dual 2D 8-fold B2 tiling.}
  \label{2D_8foldB2_PurePenrose}
\end{center}

\begin{center}
  \includegraphics[width=2.4in]{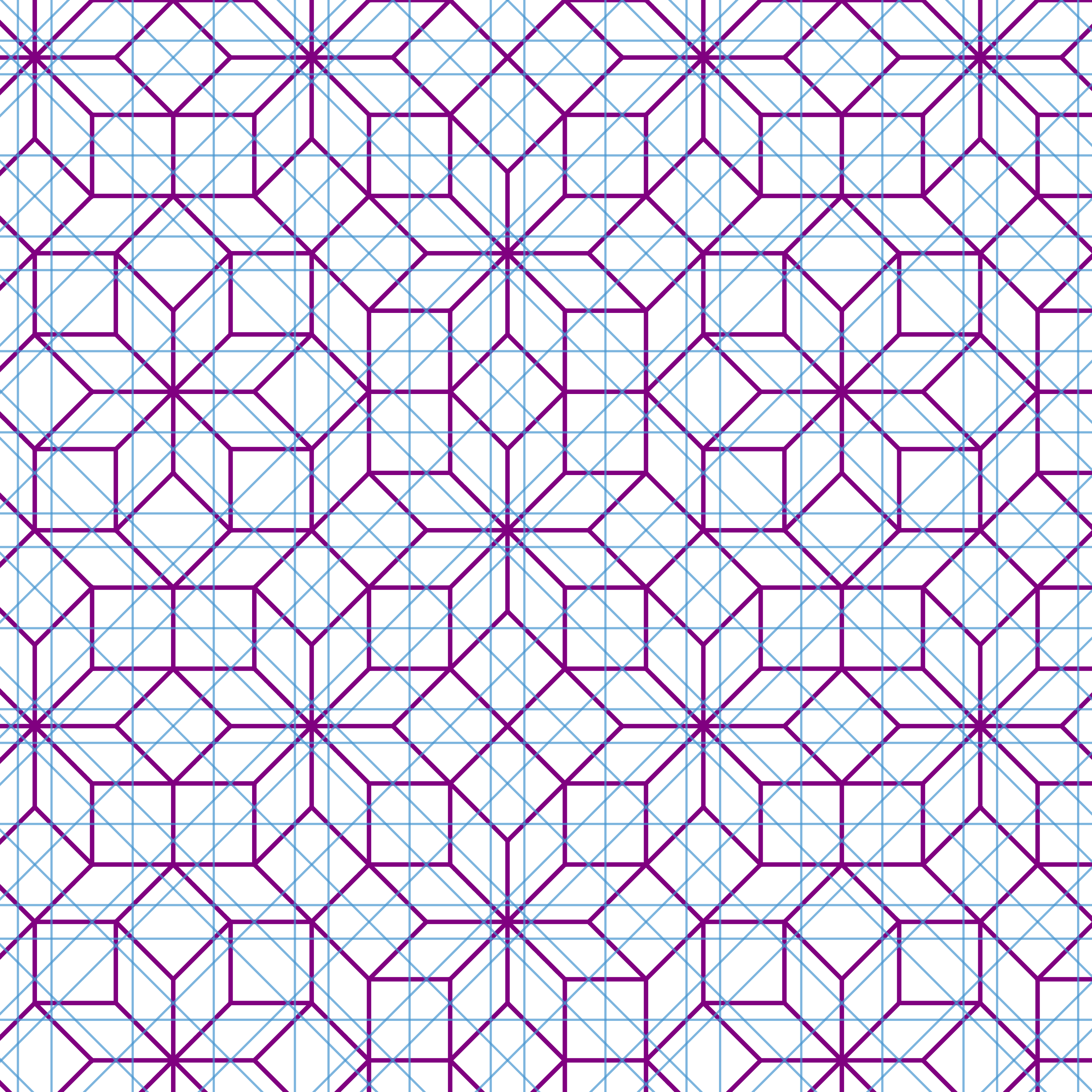}
  \captionof{figure}{The 2D 8-fold B2 tiling (thick, purple), with Ammann lines (thin, blue).}
  \label{2D_8foldB2_AmmannLines}
  \vspace{10mm}
  \includegraphics[width=2.4in]{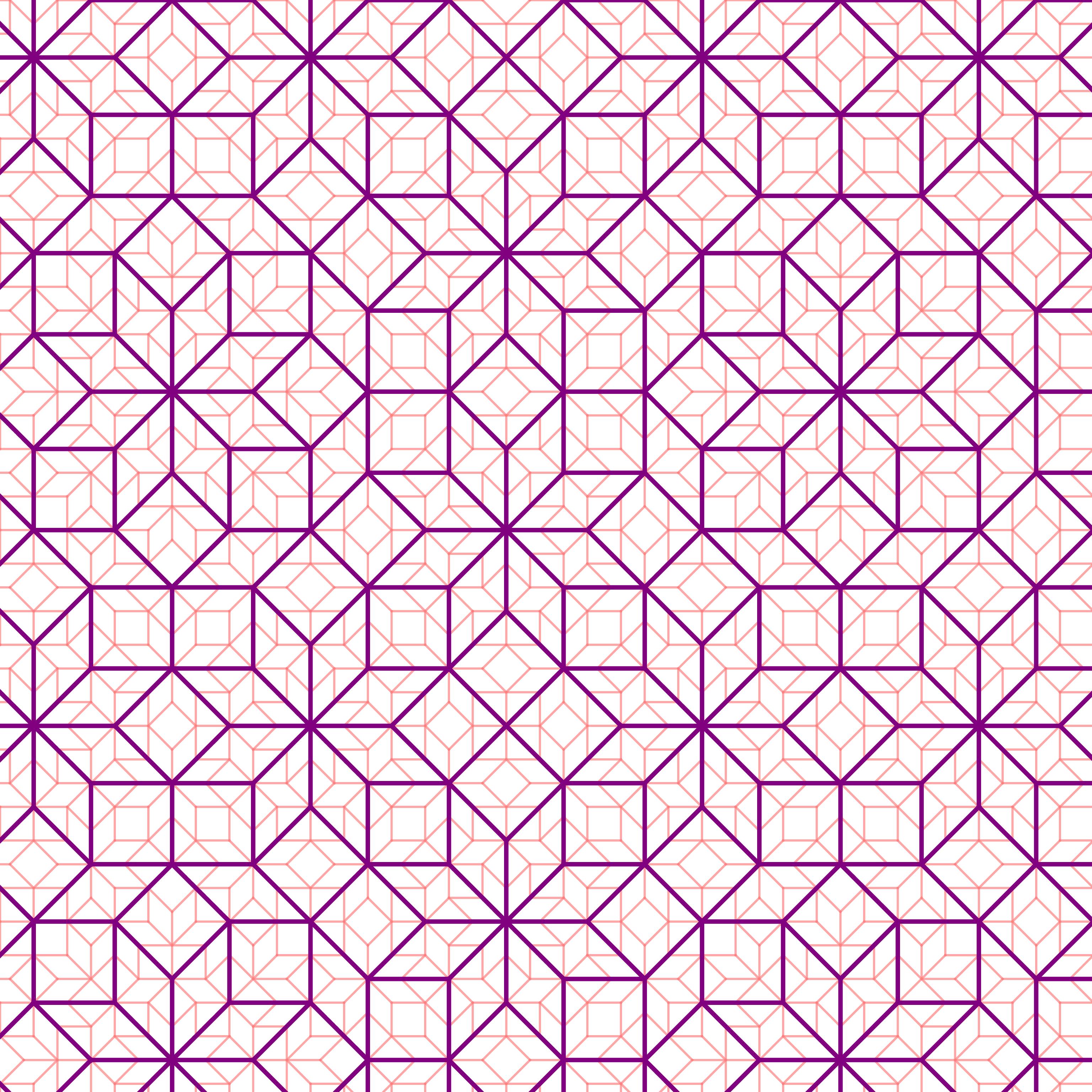}
  \captionof{figure}{The 2D 8-fold B2 tiling (thick, purple), and its inflation (thin, pink).}
  \label{2D_8foldB2_Inflation}
\end{center}

\begin{center}
  \includegraphics[width=2.4in]{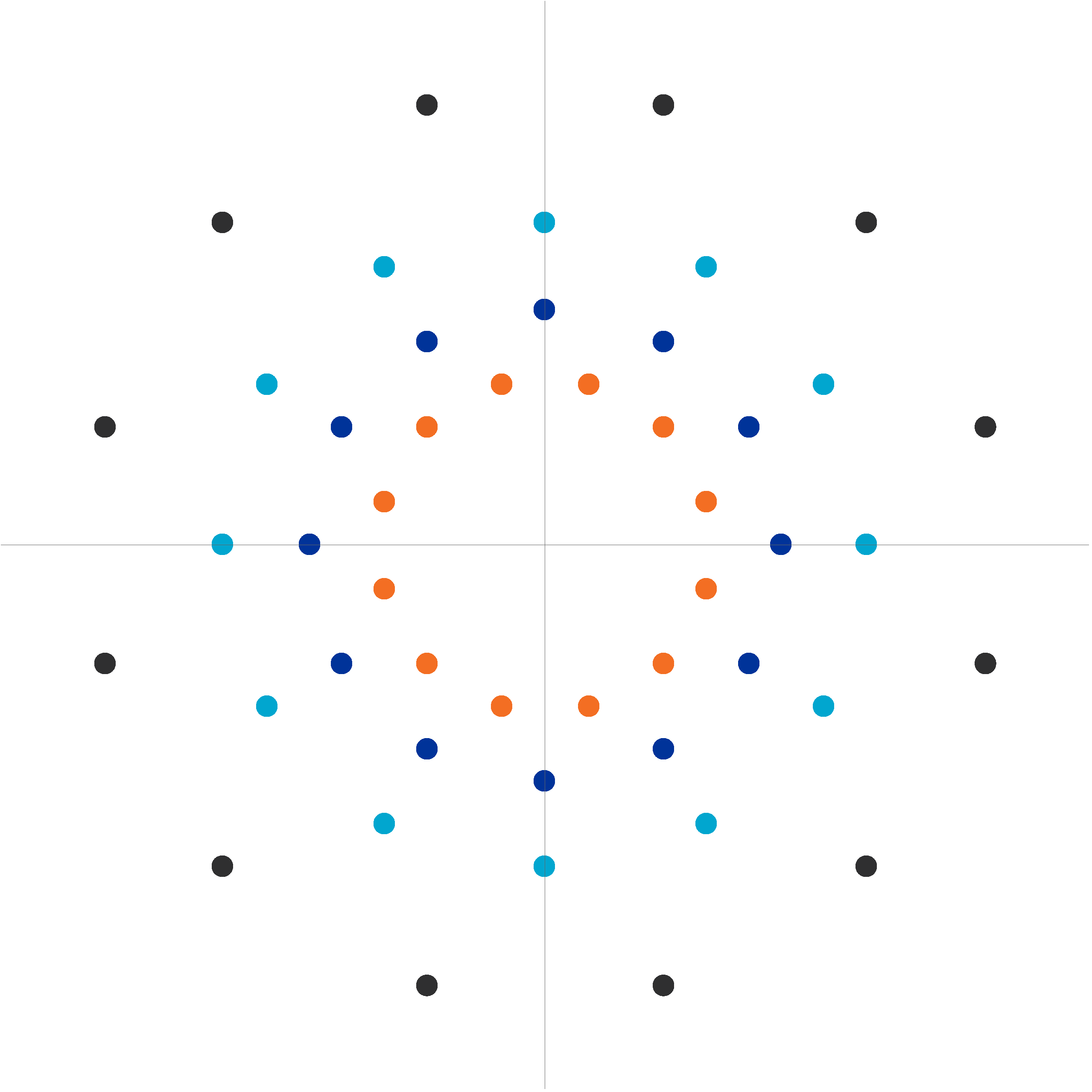}
  \captionof{figure}{The 48 $F_{4}$ roots, projected on the Coxeter plane.}
  \label{F4rootfig}
\end{center}

\begin{center}
  \includegraphics[width=2.4in]{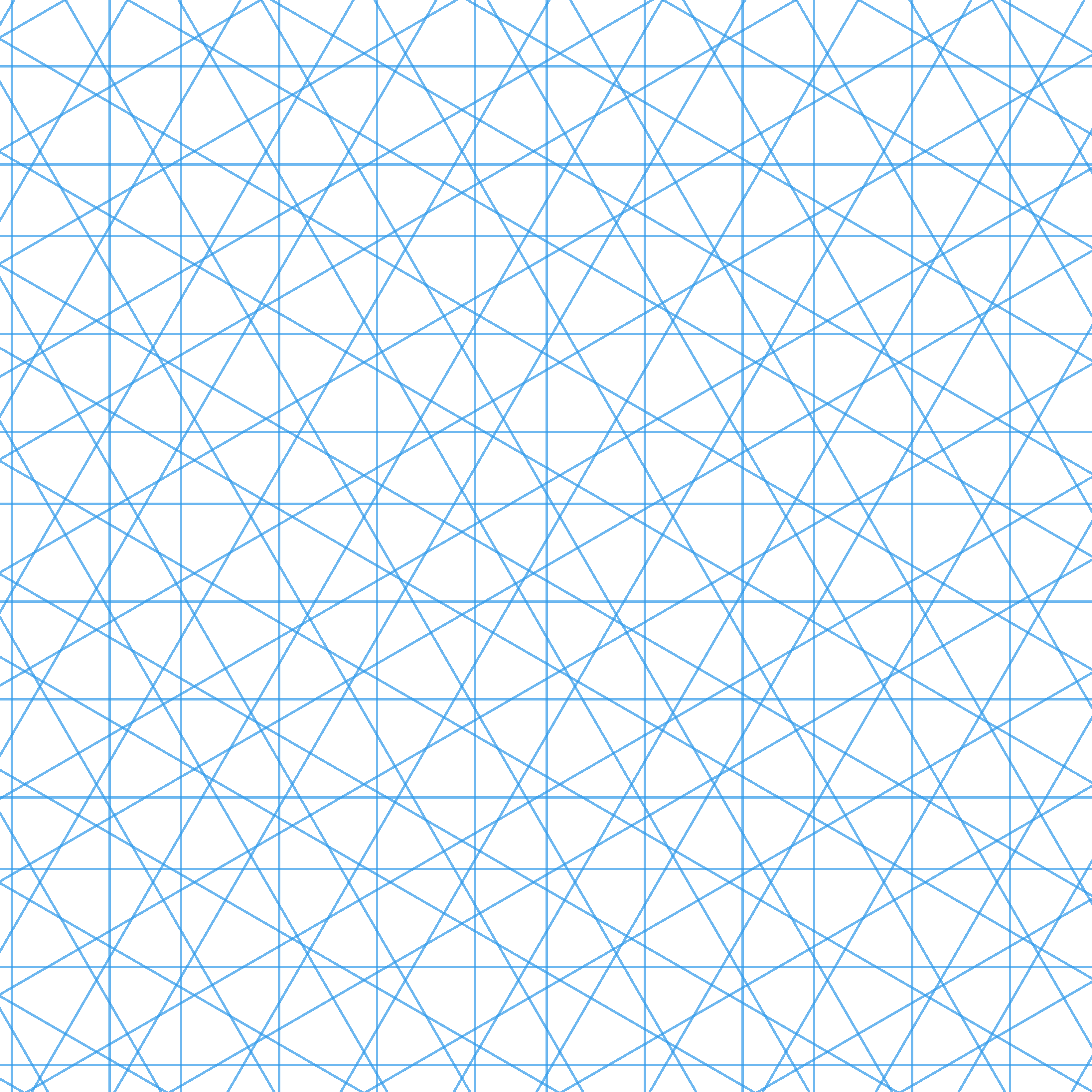}
  \captionof{figure}{The 2D 12-fold A1 Ammann pattern.}
  \label{2D_12foldA1_PureAmmann}
  \vspace{10mm}
  \includegraphics[width=2.4in]{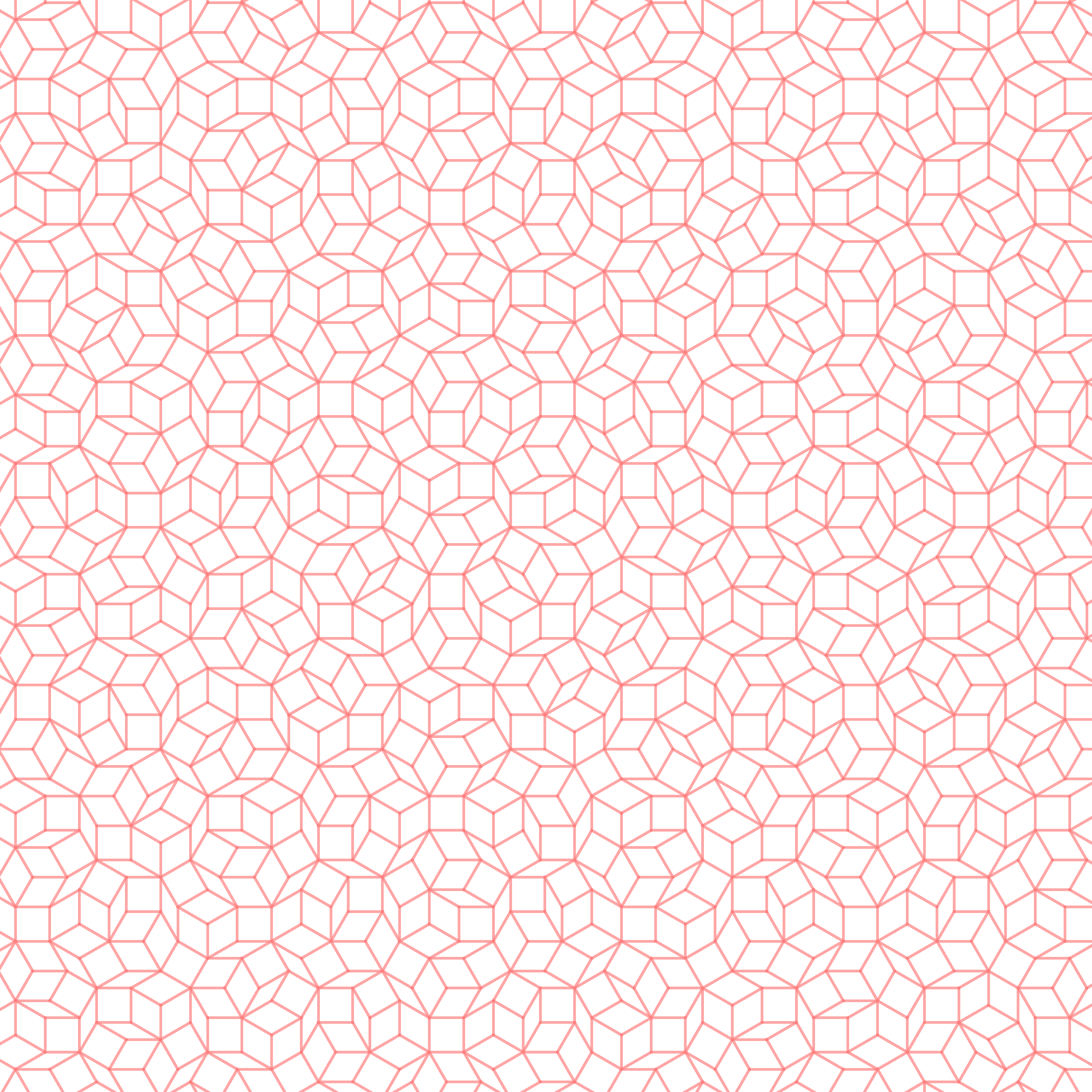}
  \captionof{figure}{The dual 2D 12-fold A1 tiling.}
  \label{2D_12foldA1_PurePenrose}
\end{center}

\begin{center}
  \includegraphics[width=2.4in]{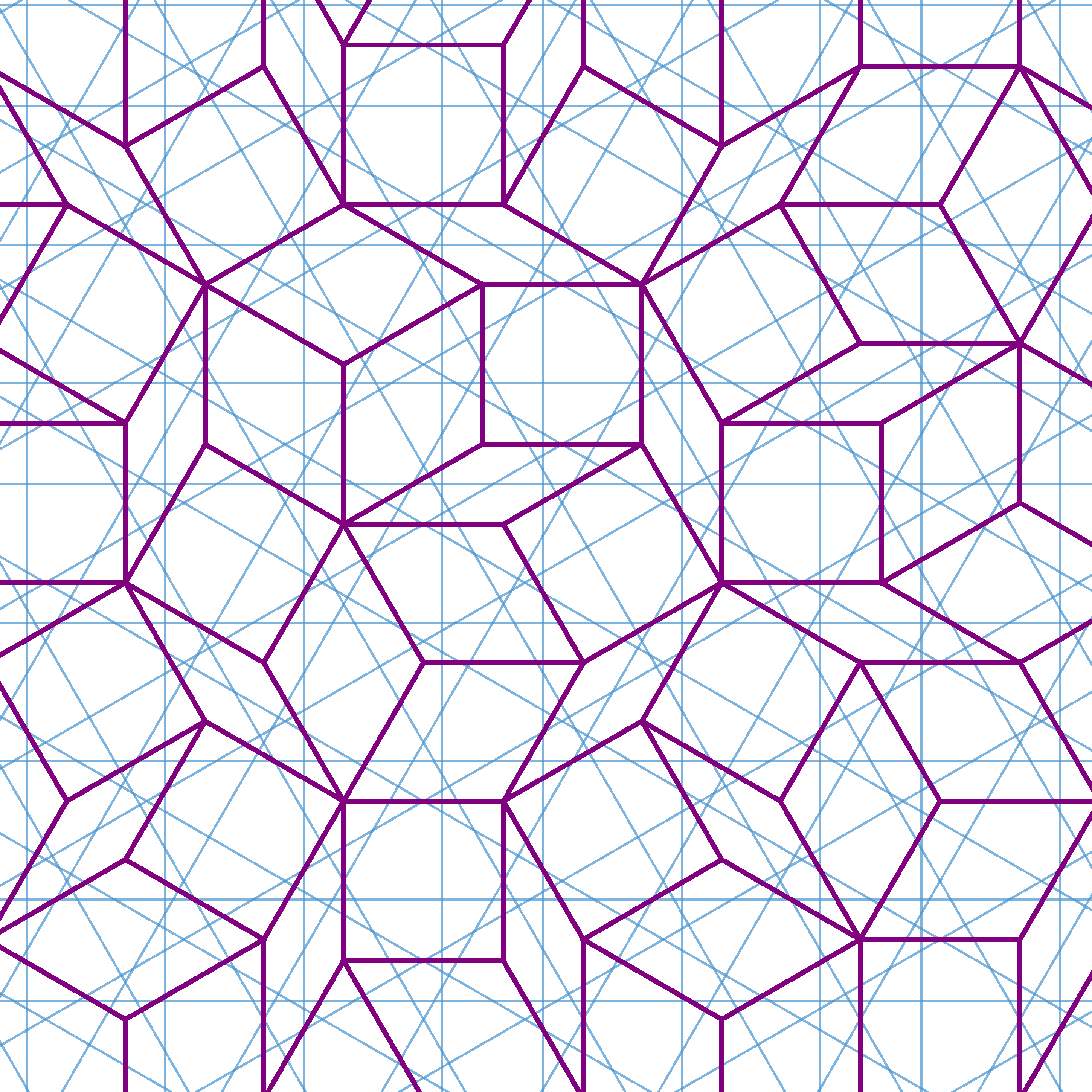}
  \captionof{figure}{The 2D 12-fold A1 tiling (thick, purple), with Ammann lines (thin, blue).}
  \label{2D_12foldA1_AmmannLines}
  \vspace{10mm}
  \includegraphics[width=2.4in]{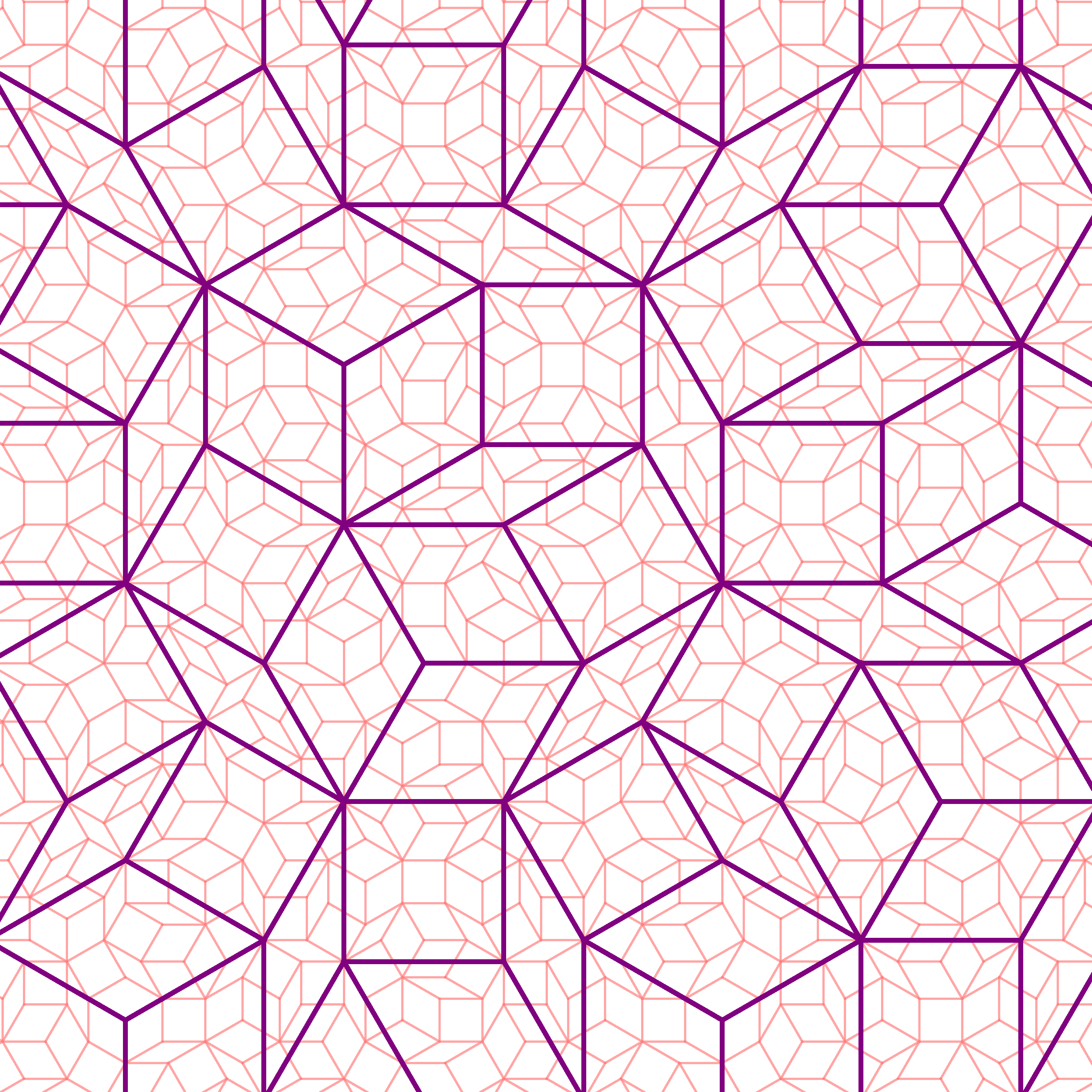}
  \captionof{figure}{The 2D 12-fold A1 tiling (thick, purple), and its inflation (thin, pink).}
   \label{2D_12foldA1_Inflation}
\end{center}

\begin{center}
  \includegraphics[width=2.4in]{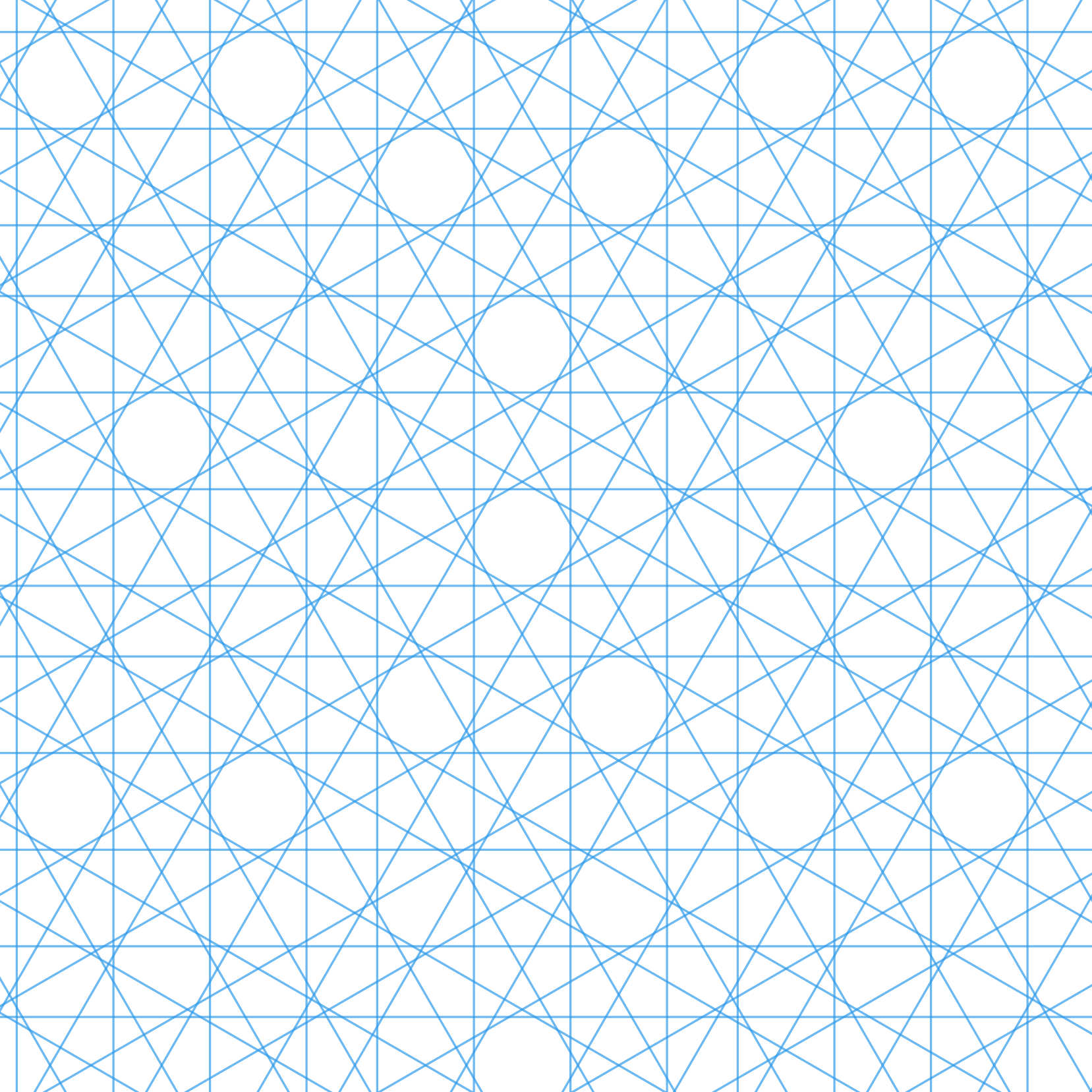}
  \captionof{figure}{The 2D 12-fold A2 Ammann pattern.}
  \label{2D_12foldA2_PureAmmann}
  \vspace{10mm}
  \includegraphics[width=2.4in]{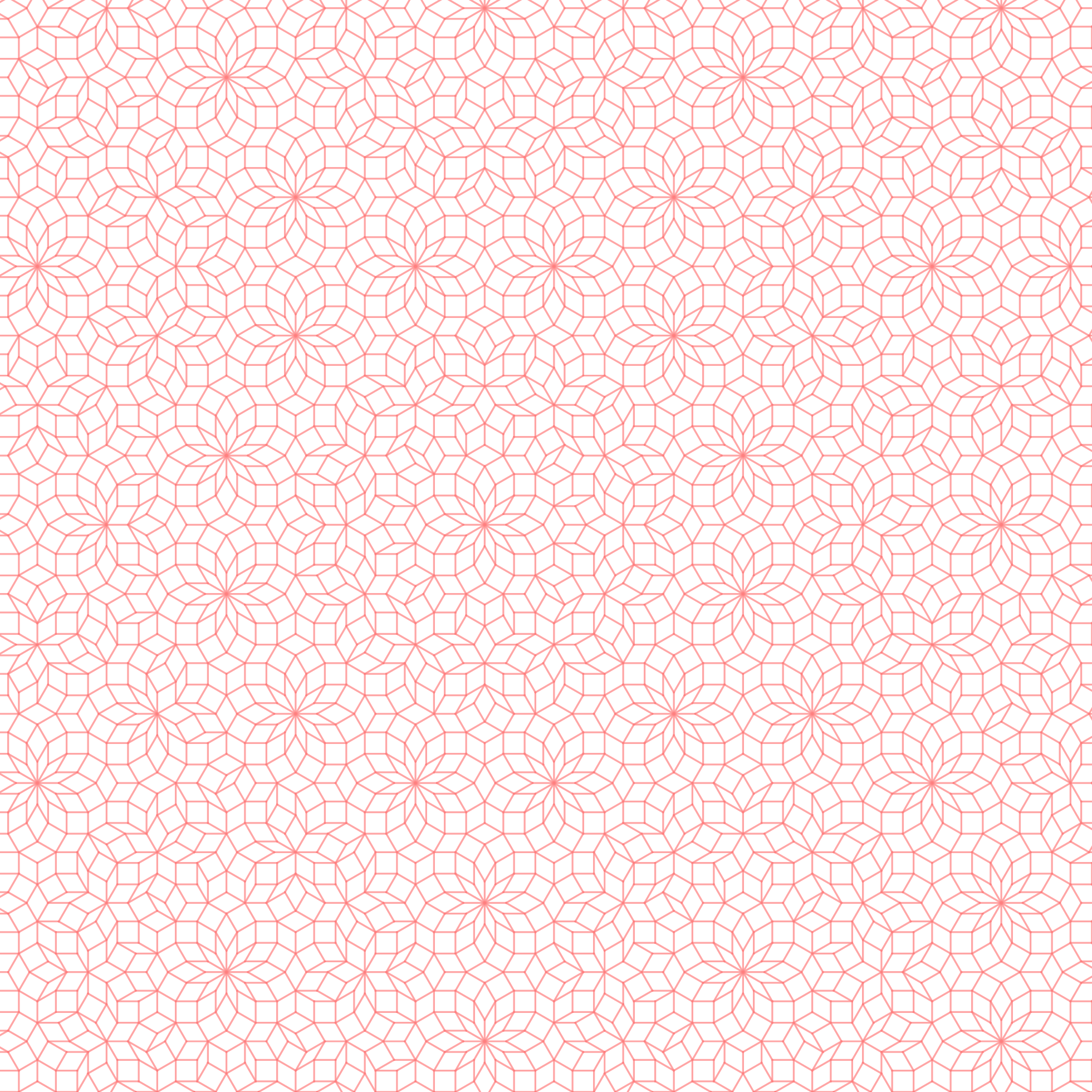}
  \captionof{figure}{The dual 2D 12-fold A2 tiling.}
  \label{2D_12foldA2_PurePenrose}
\end{center}

\begin{center}
  \includegraphics[width=2.4in]{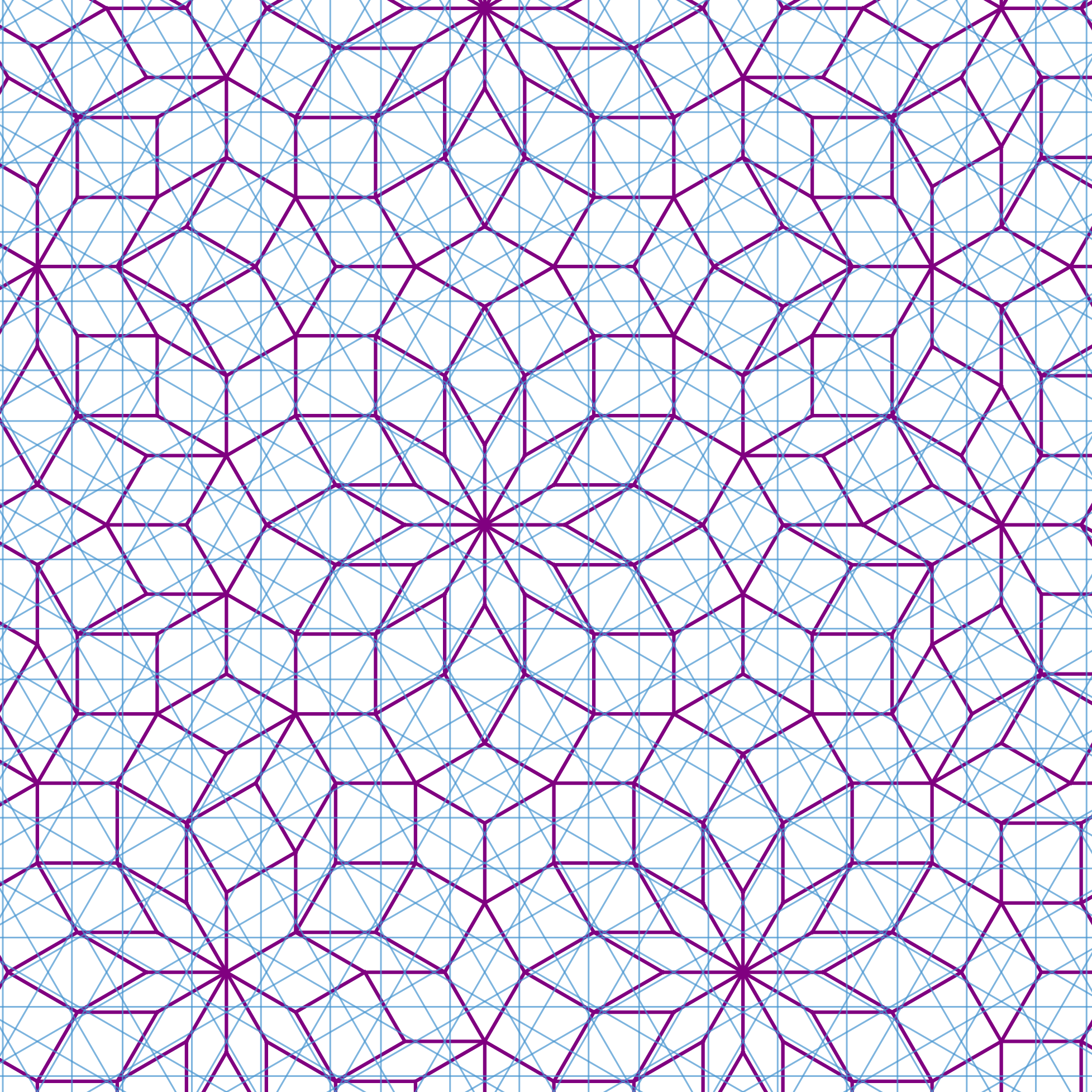}
  \captionof{figure}{The 2D 12-fold A2 tiling (thick, purple), with Ammann lines (thin, blue).}
  \label{2D_12foldA2_AmmannLines}
  \vspace{10mm}
  \includegraphics[width=2.4in]{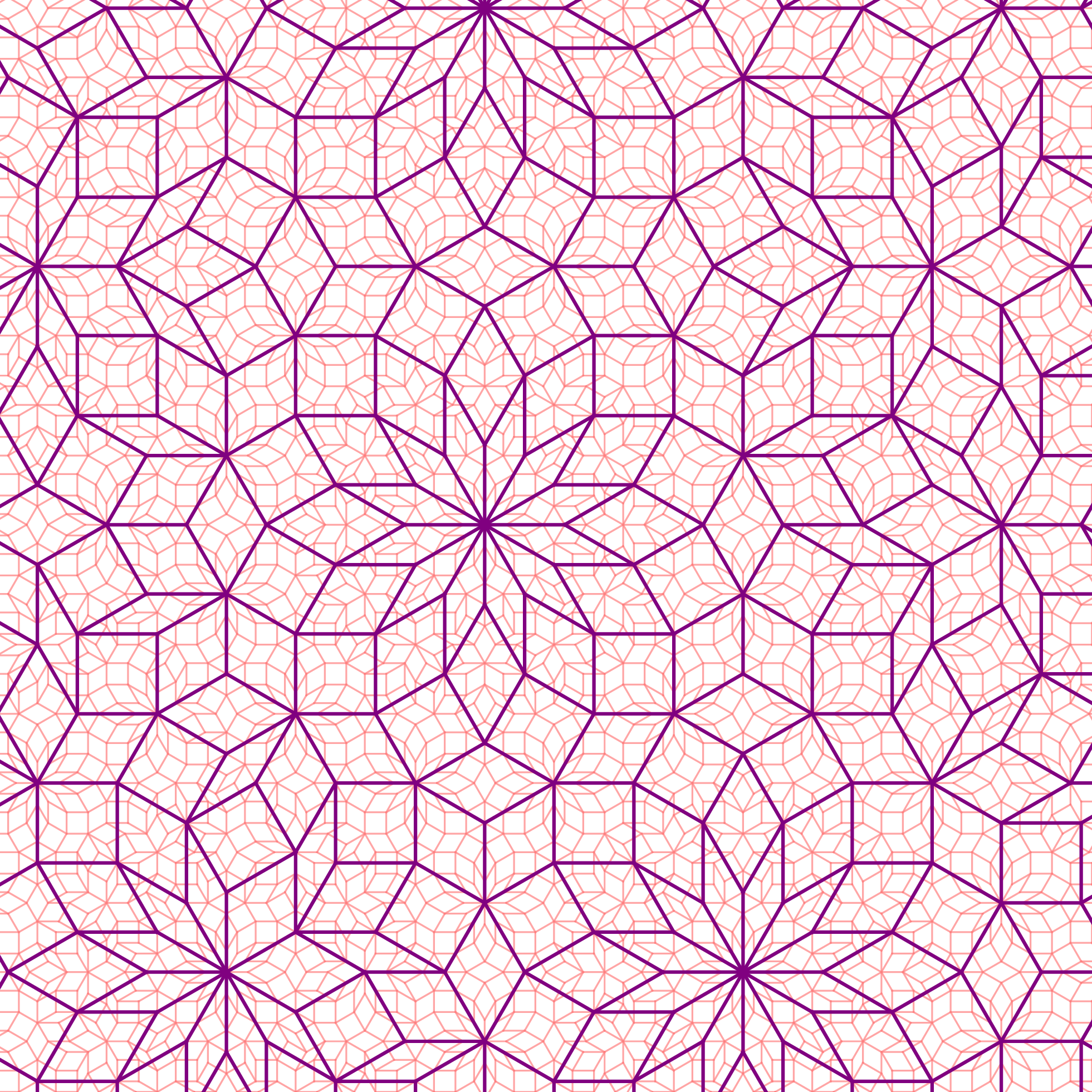}
  \captionof{figure}{The 2D 12-fold A2 tiling (thick, purple), and its inflation (thin, pink).}
   \label{2D_12foldA2_Inflation}
\end{center}

\begin{center}
  \includegraphics[width=2.4in]{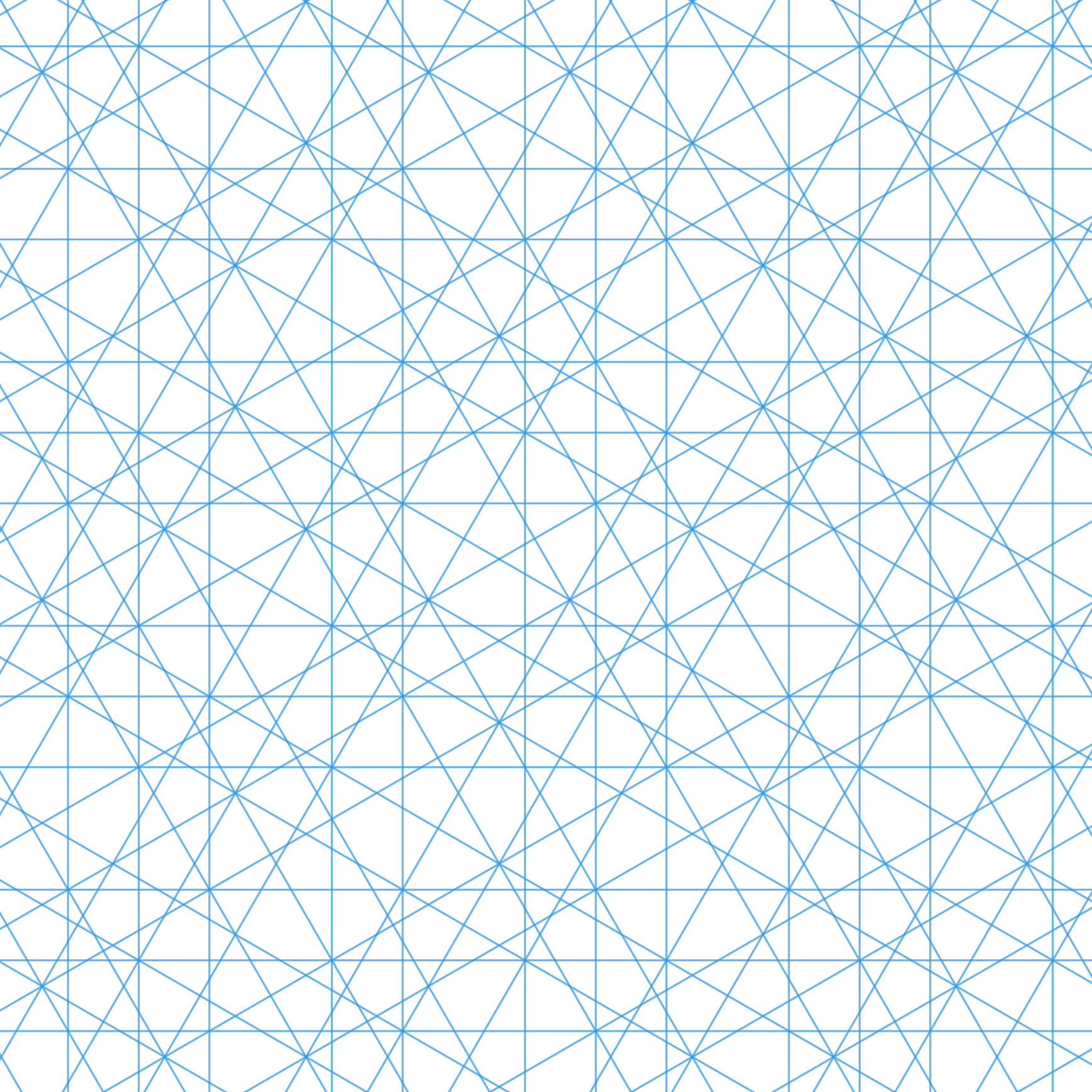}
  \captionof{figure}{The 2D 12-fold B1 Ammann pattern.}
  \label{2D_12foldB1_PureAmmann}
  \vspace{10mm}
  \includegraphics[width=2.4in]{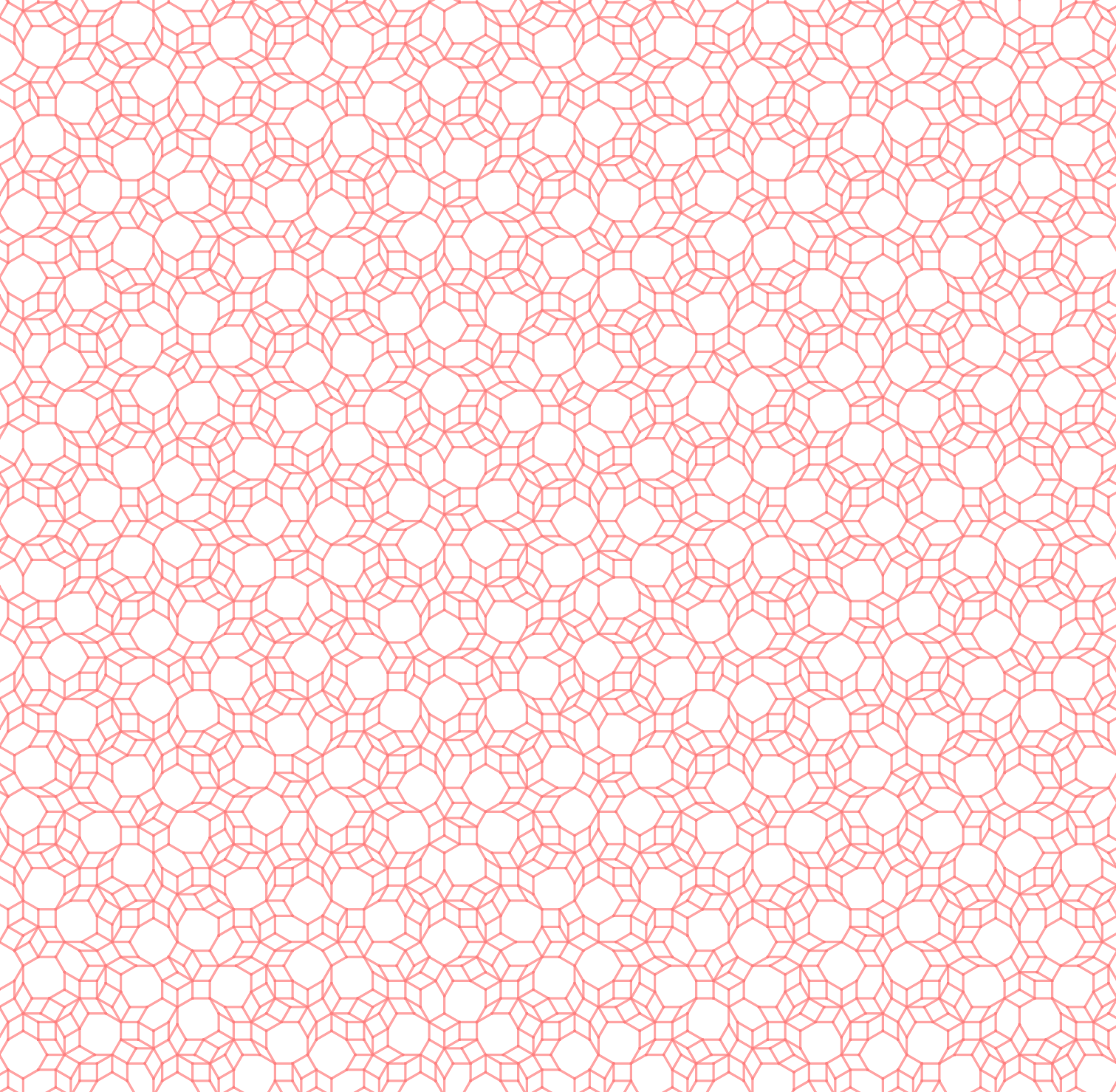}
  \captionof{figure}{The dual 2D 12-fold B1 tiling.}
  \label{2D_12foldB1_PurePenrose}
\end{center}

\begin{center}
  \includegraphics[width=2.4in]{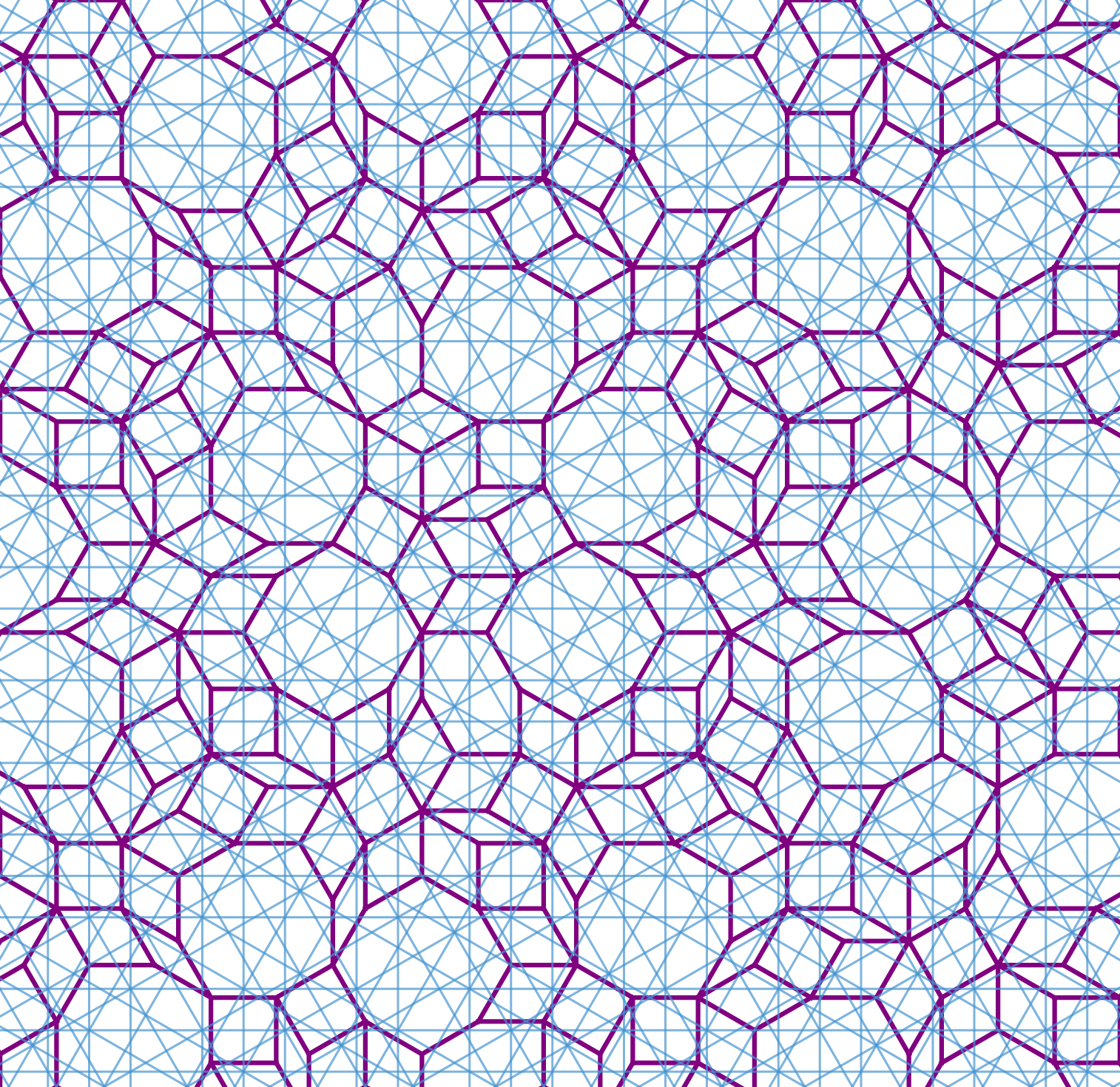}
  \captionof{figure}{The 2D 12-fold B1 tiling (thick, purple), with Ammann lines (thin, blue).}
  \label{2D_12foldB1_AmmannLines}
  \vspace{10mm}
  \includegraphics[width=2.4in]{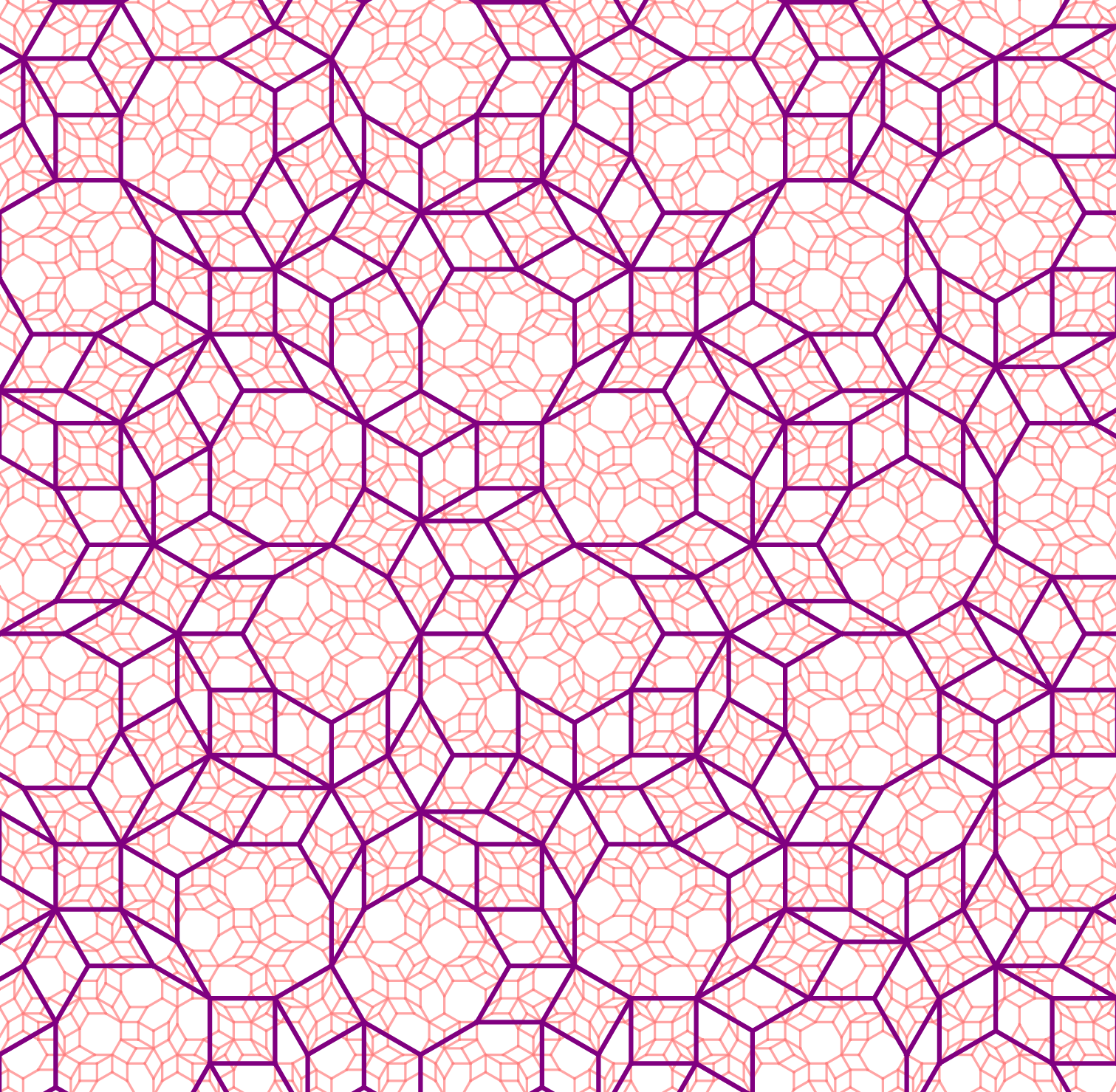}
  \captionof{figure}{The 2D 12-fold B1 tiling (thick, purple), and its inflation (thin, pink).}
  \label{2D_12foldB1_Inflation}
\end{center}

\begin{center}
  \includegraphics[width=2.4in]{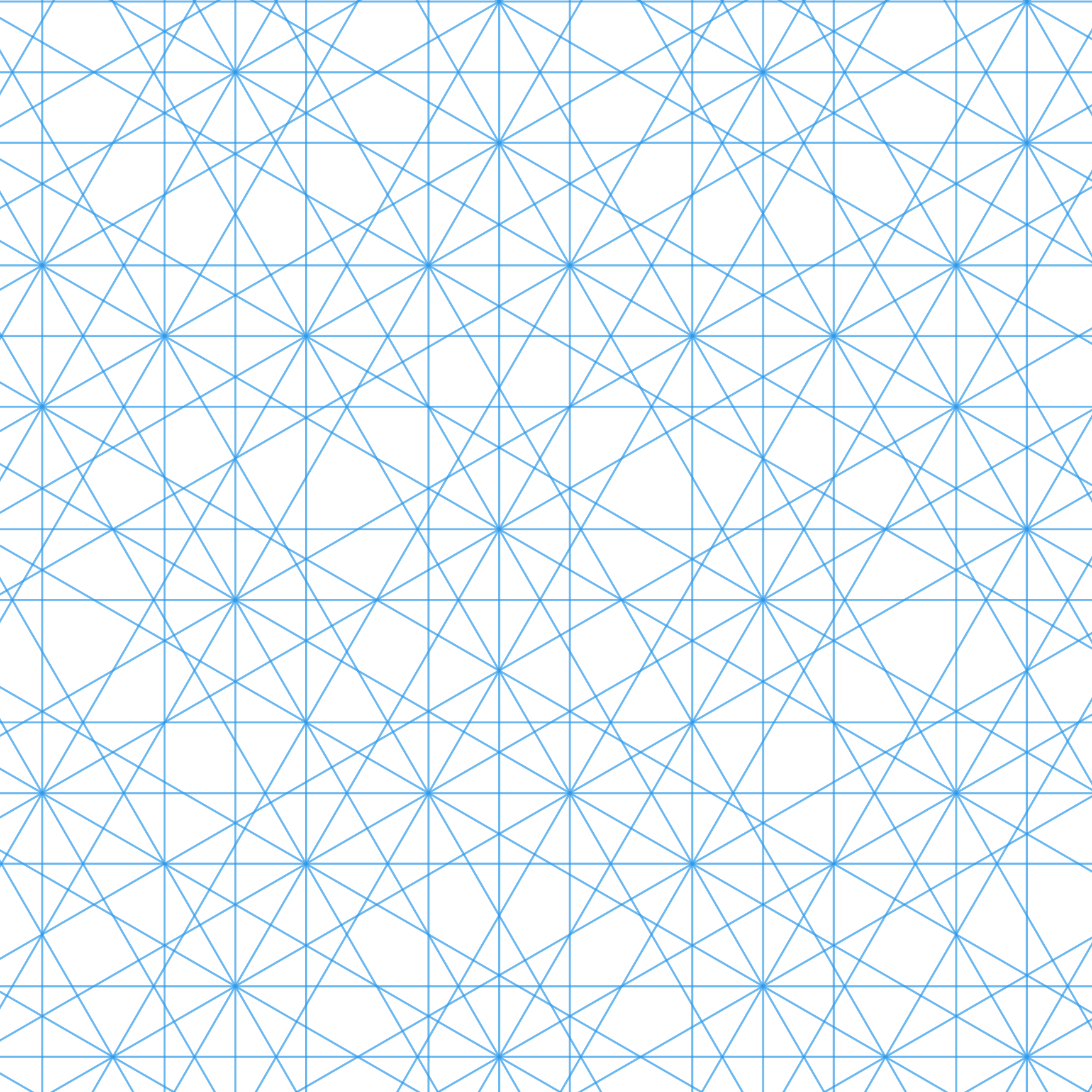}
  \captionof{figure}{The 2D 12-fold B2 Ammann pattern.}
  \label{2D_12foldB2_PureAmmann}
  \vspace{10mm}
  \includegraphics[width=2.4in]{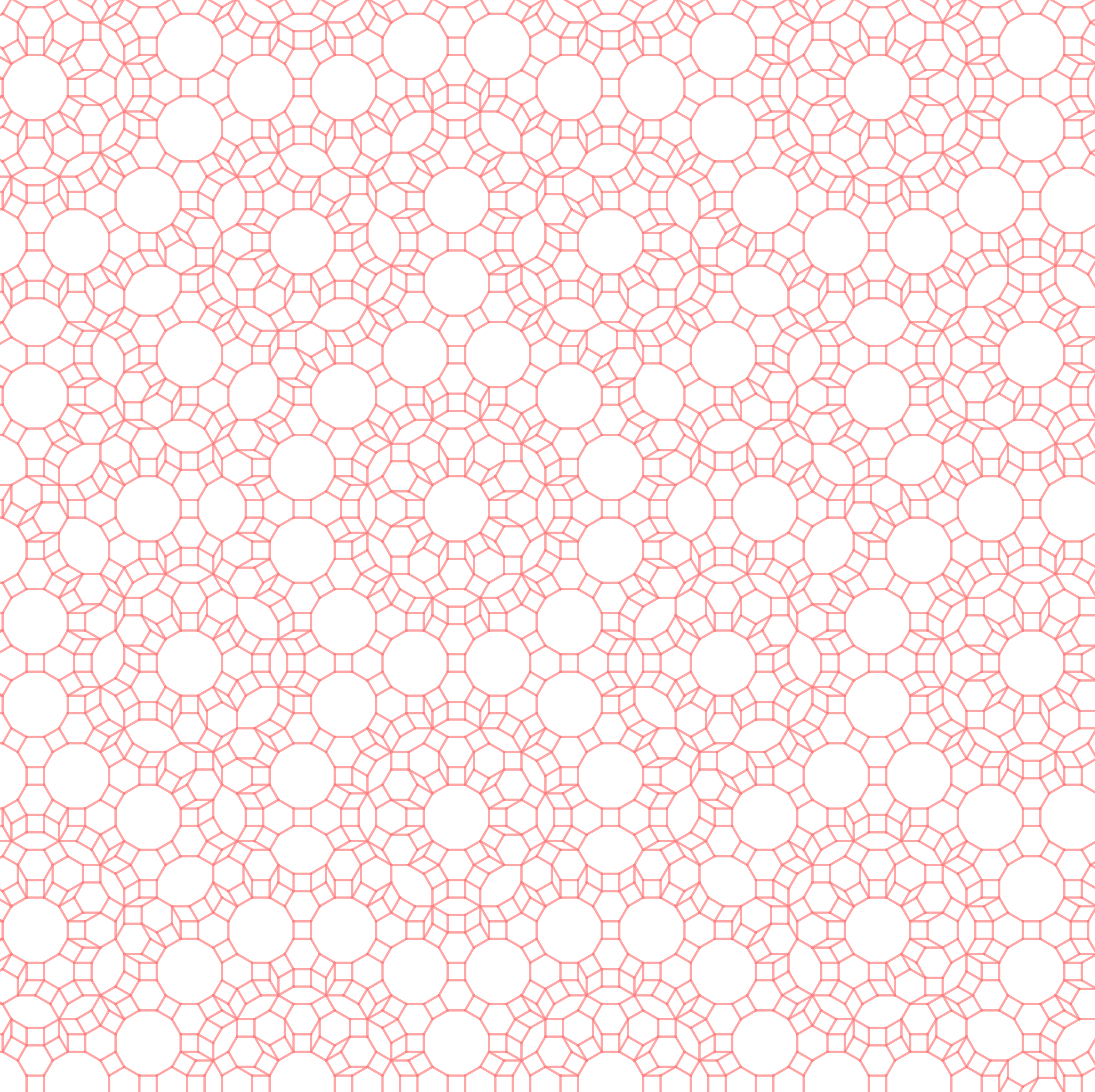}
  \captionof{figure}{The dual 2D 12-fold B2 tiling.}
  \label{2D_12foldB2_PurePenrose}
\end{center}

\begin{center}
  \includegraphics[width=2.4in]{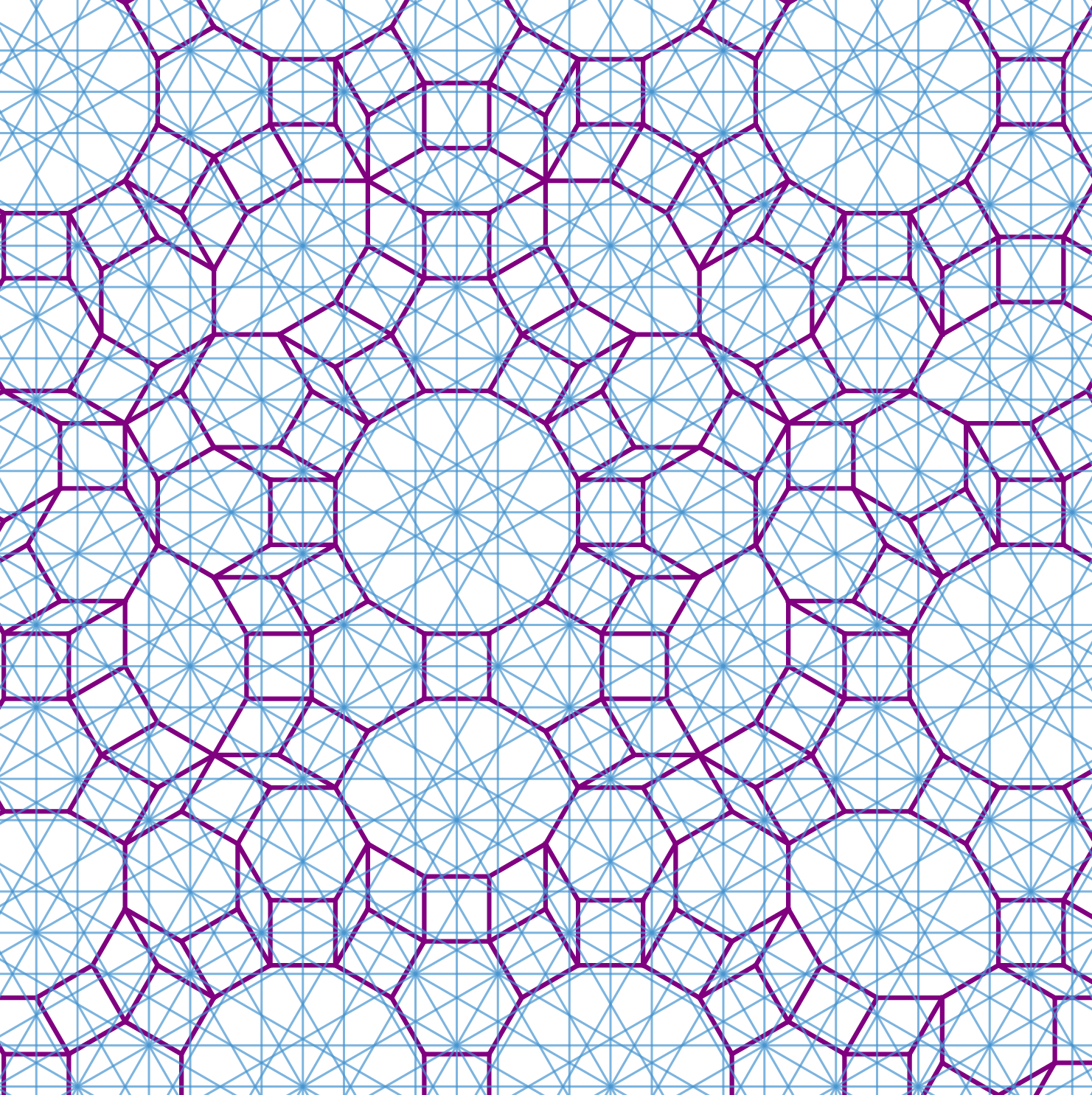}
  \captionof{figure}{The 2D 12-fold B2 tiling (thick, purple), with Ammann lines (thin, blue).}
  \label{2D_12foldB2_AmmannLines}
  \vspace{10mm}
  \includegraphics[width=2.4in]{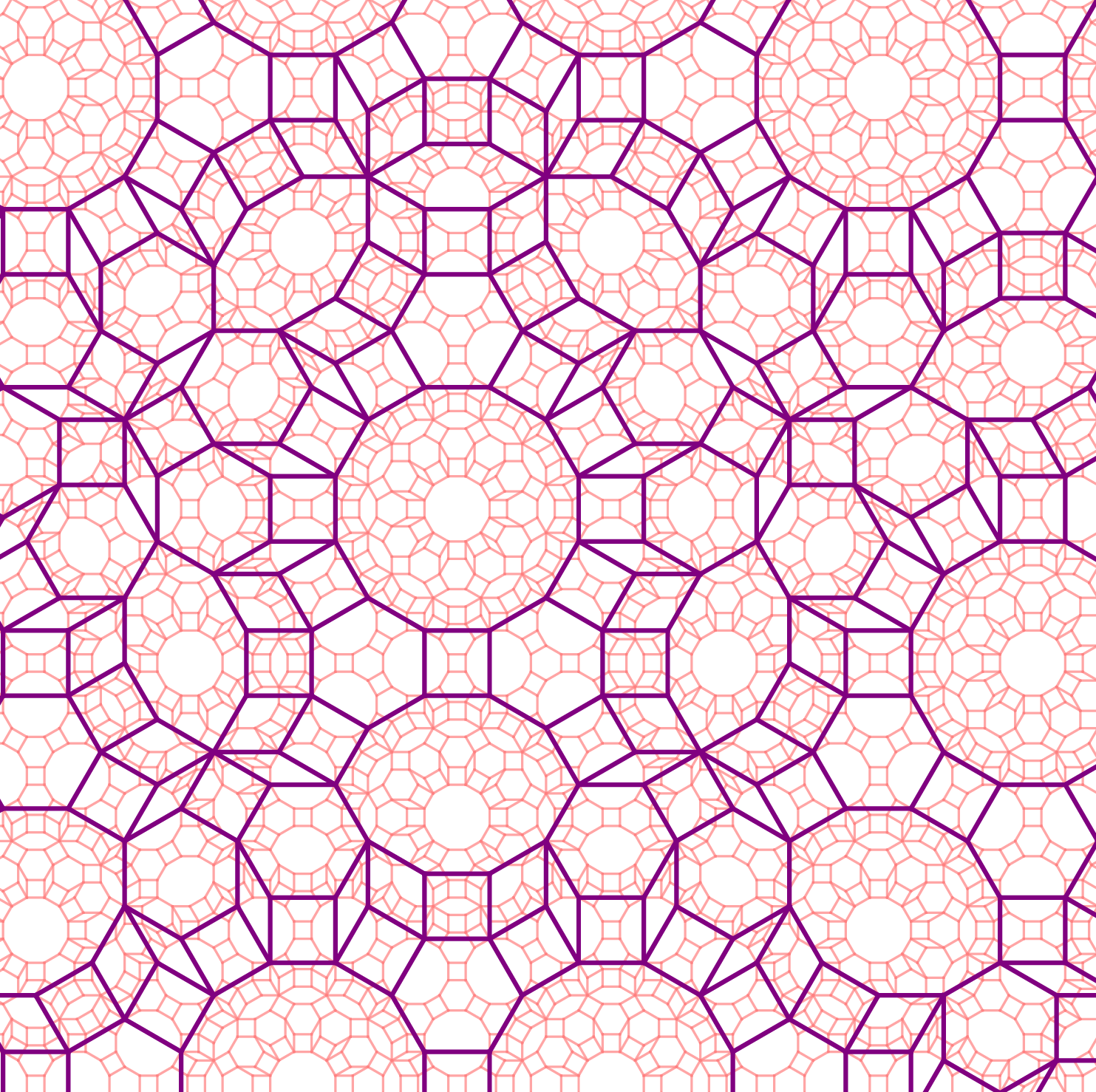}
  \captionof{figure}{The 2D 12-fold B2 tiling (thick, purple), and its inflation (thin, pink).}
  \label{2D_12foldB2_Inflation}
\end{center}

\begin{center}
  \includegraphics[width=2.4in]{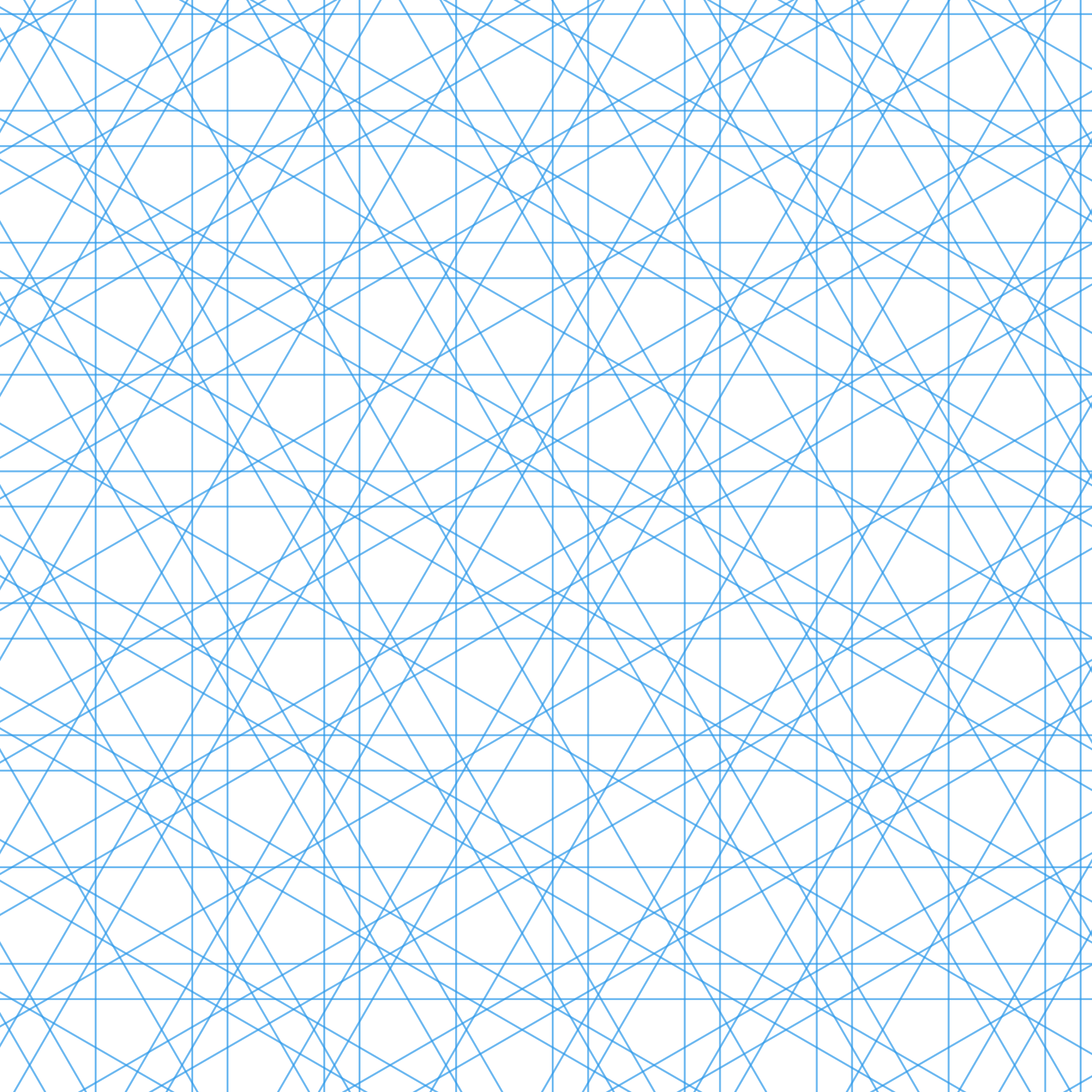}
  \captionof{figure}{The 2D 12-fold C1 Ammann pattern.}
  \label{2D_12foldC1_PureAmmann}
  \vspace{10mm}
  \includegraphics[width=2.4in]{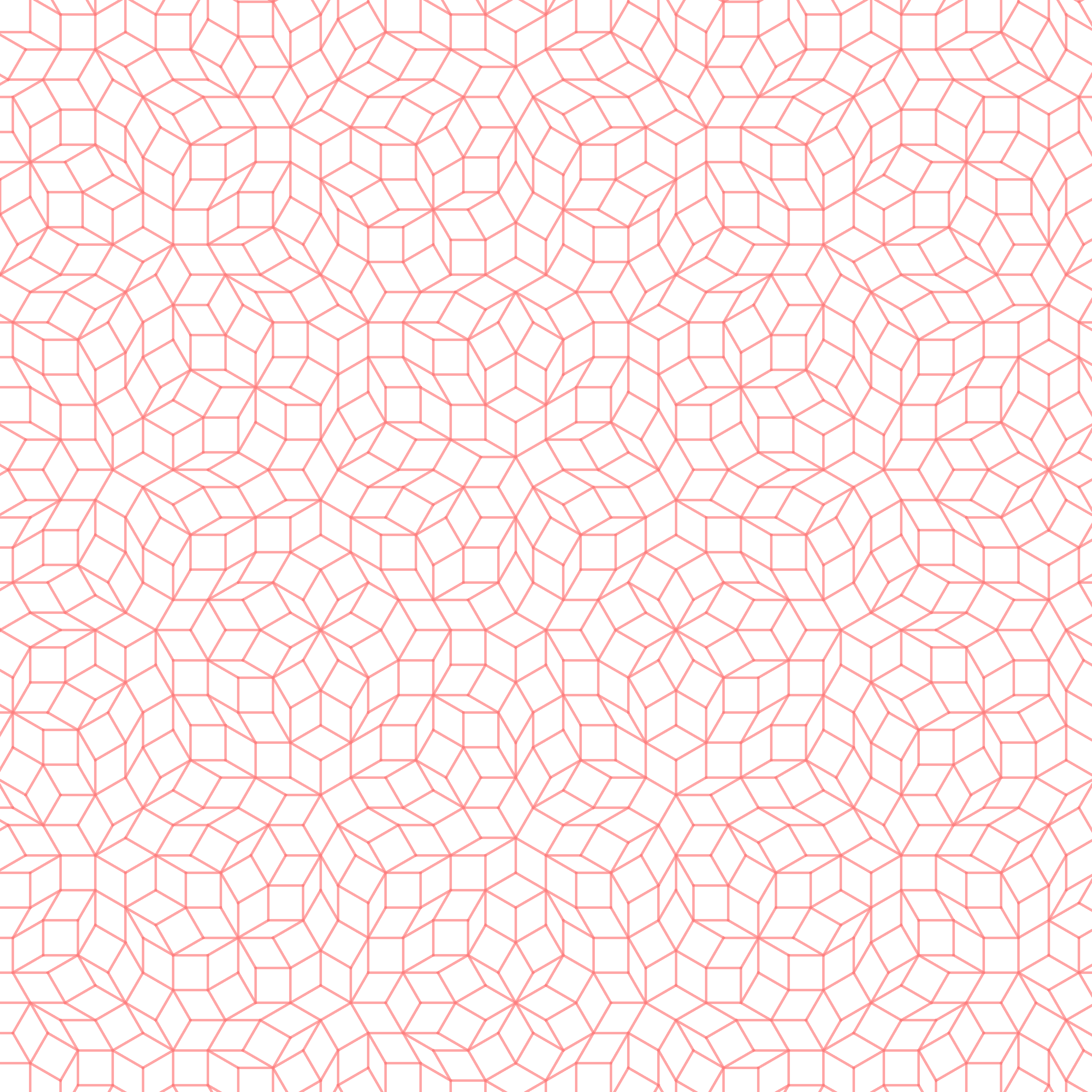}
  \captionof{figure}{The dual 2D 12-fold C1 tiling.}
  \label{2D_12foldC1_PurePenrose}
\end{center}

\begin{center}
  \includegraphics[width=2.4in]{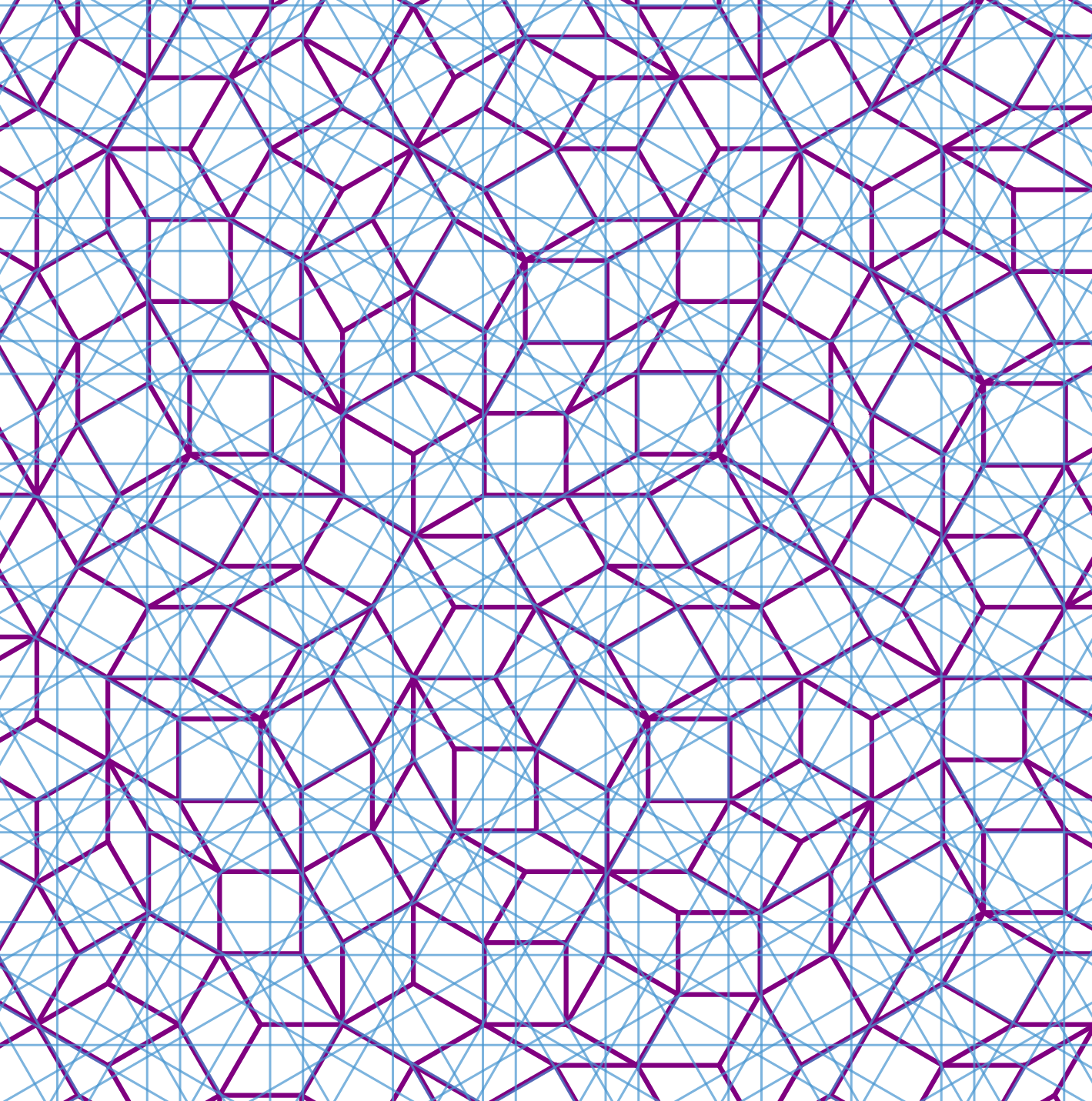}
  \captionof{figure}{The 2D 12-fold C1 tiling (thick, purple), with Ammann lines (thin, blue).}
  \label{2D_12foldC1_AmmannLines}
  \vspace{10mm}
  \includegraphics[width=2.4in]{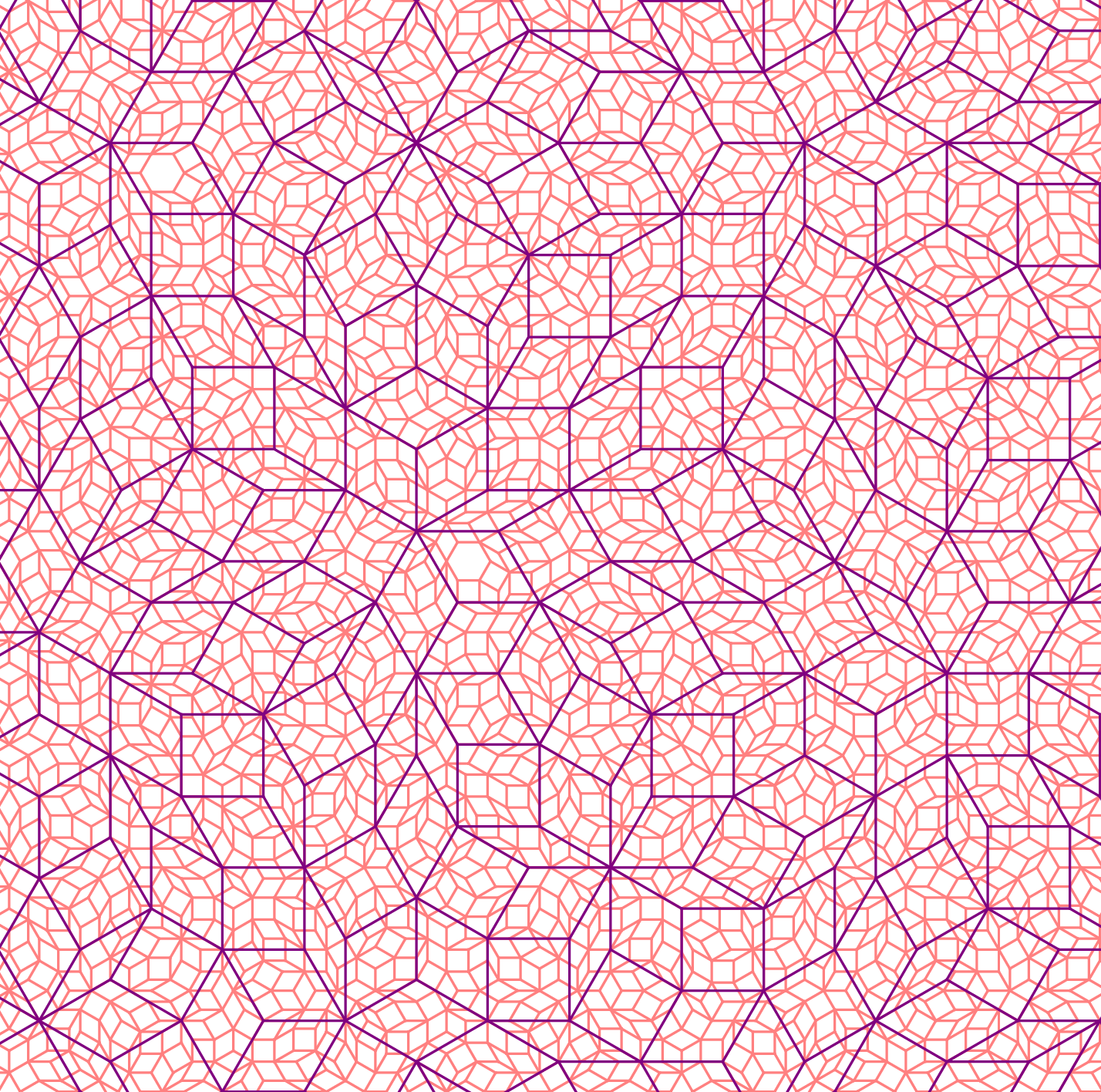}
  \captionof{figure}{The 2D 12-fold C1 tiling (thick, purple), and its inflation (thin, pink).}
  \label{2D_12foldC1_Inflation}
\end{center}

\begin{center}
  \includegraphics[width=2.4in]{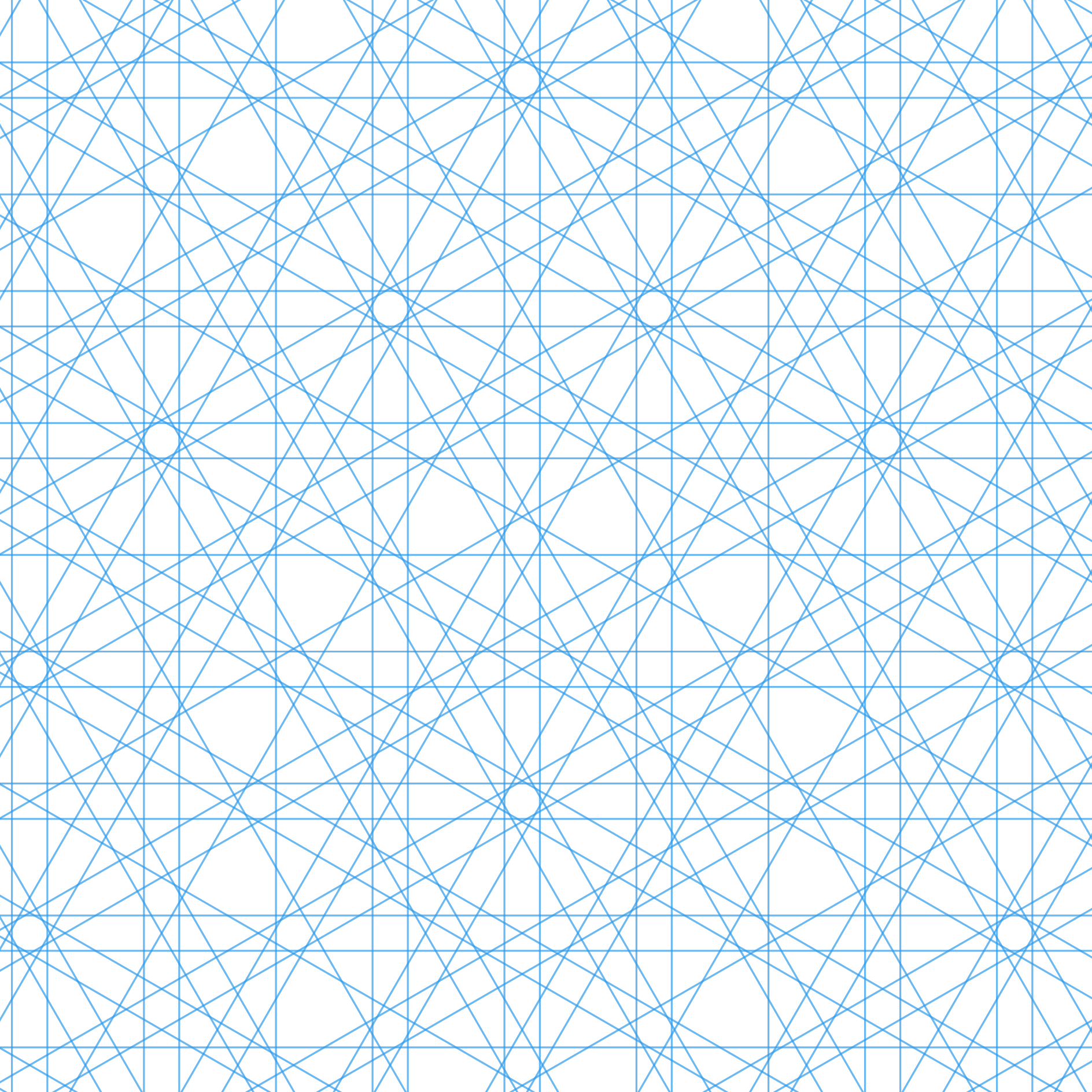}
  \captionof{figure}{The 2D 12-fold C2 Ammann pattern.}
  \label{2D_12foldC2_PureAmmann}
  \vspace{10mm}
  \includegraphics[width=2.4in]{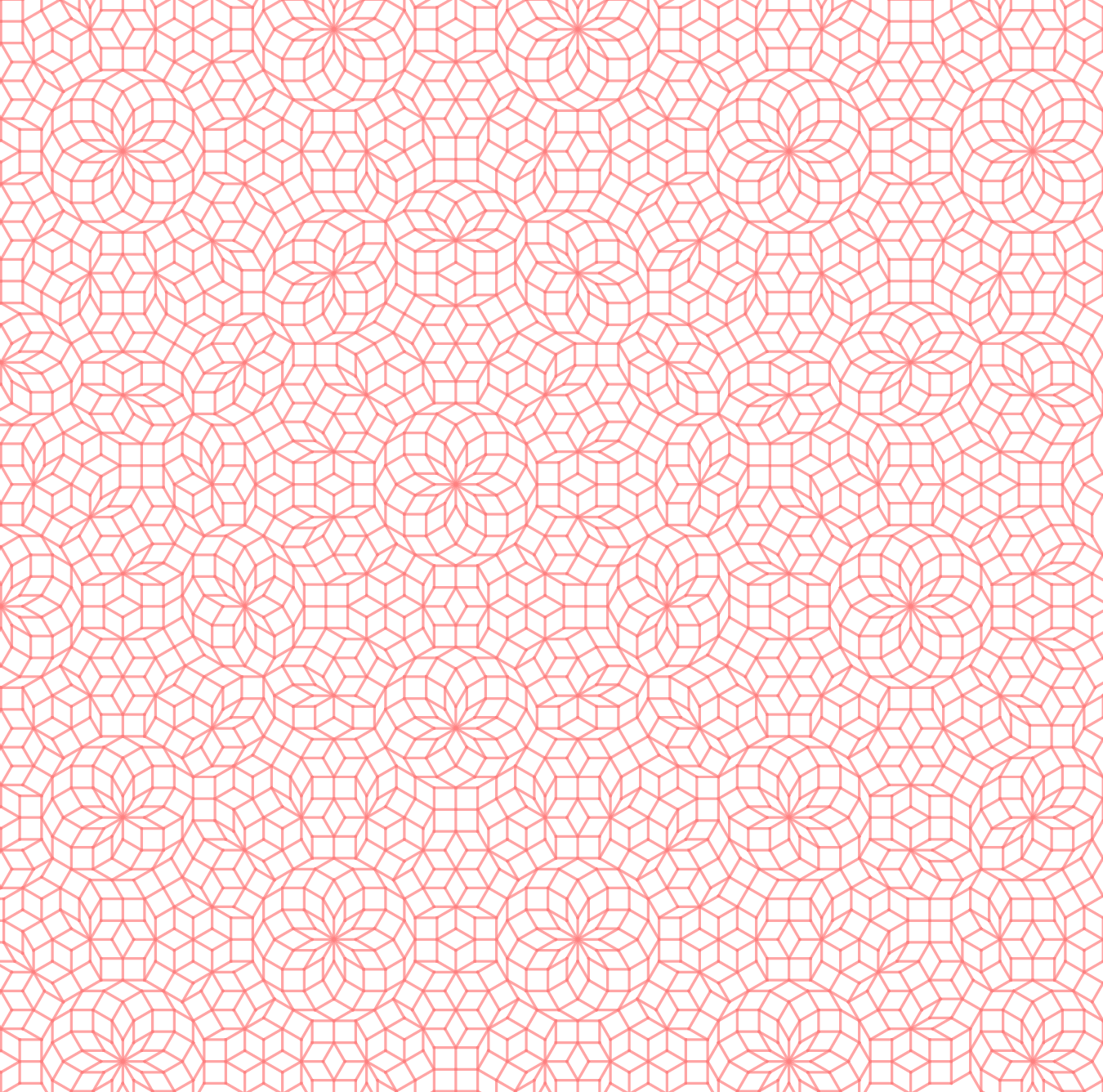}
  \captionof{figure}{The dual 2D 12-fold C2 Penrose-like tiling.}
  \label{2D_12foldC2_PurePenrose}
\end{center}

\begin{center}
  \includegraphics[width=2.4in]{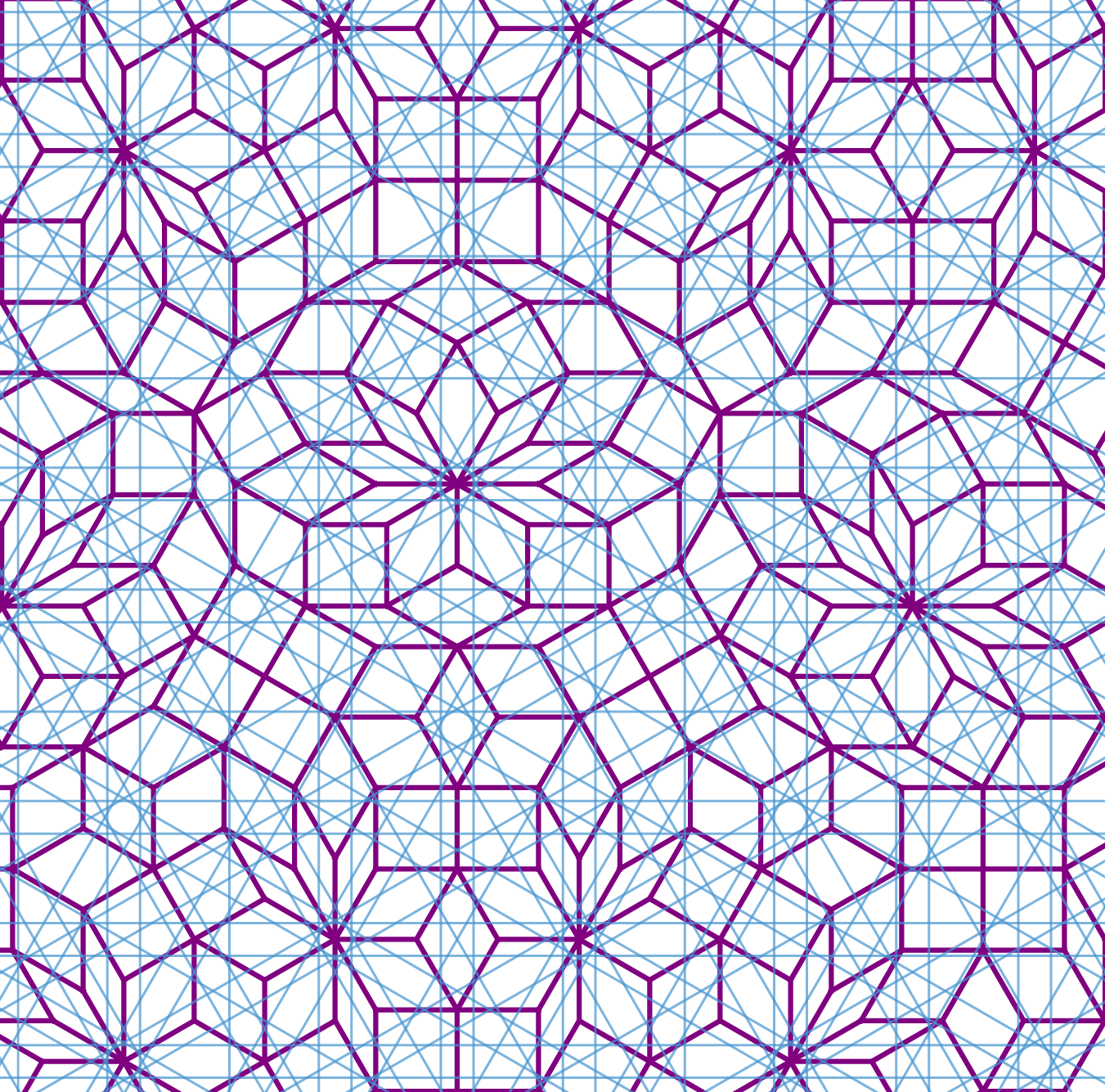}
  \captionof{figure}{The 2D 12-fold C2 tiling (thick, purple), with Ammann lines (thin, blue).}
  \label{2D_12foldC2_AmmannLines}
  \vspace{10mm}
  \includegraphics[width=2.4in]{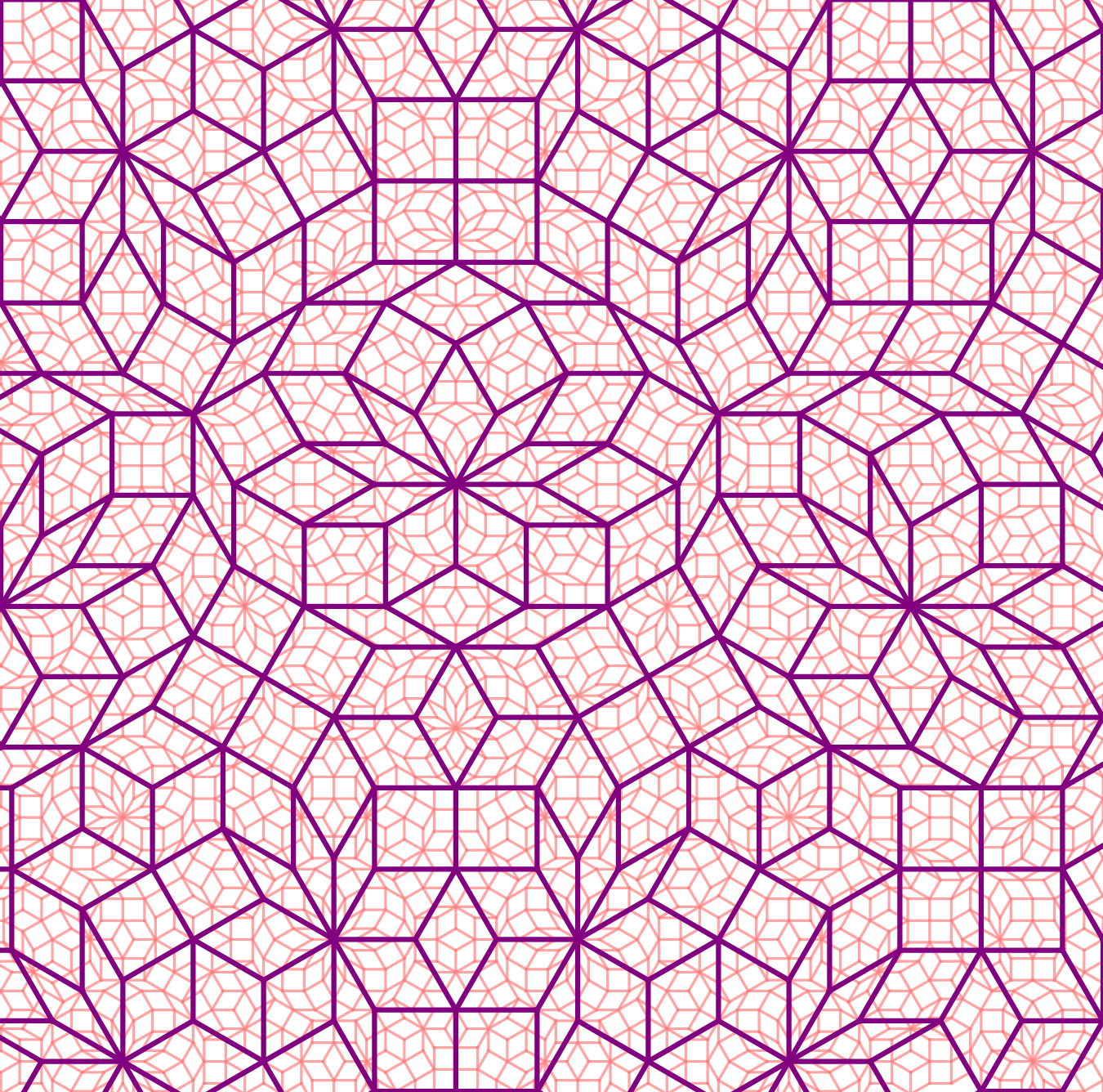}
  \captionof{figure}{The 2D 12-fold C2 tiling (thick, purple), and its inflation (thin, pink).}
  \label{2D_12foldC2_Inflation}
\end{center}

\begin{center}
  \includegraphics[width=2.4in]{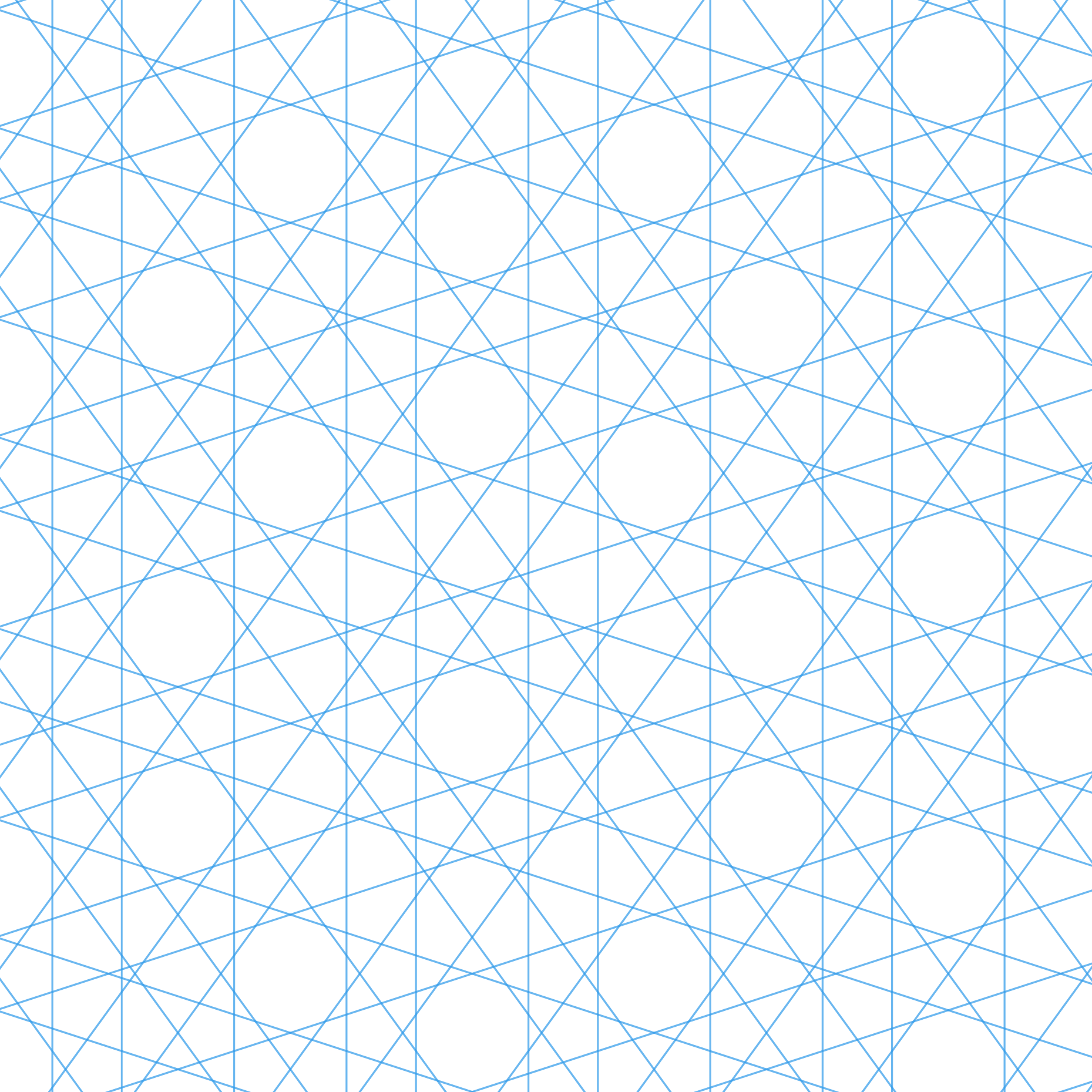}
  \captionof{figure}{The 2D 10-fold Ammann trio: pattern (i).}
  \label{2D_10foldTrio1_PureAmmann}
  \vspace{10mm}
  \includegraphics[width=2.4in]{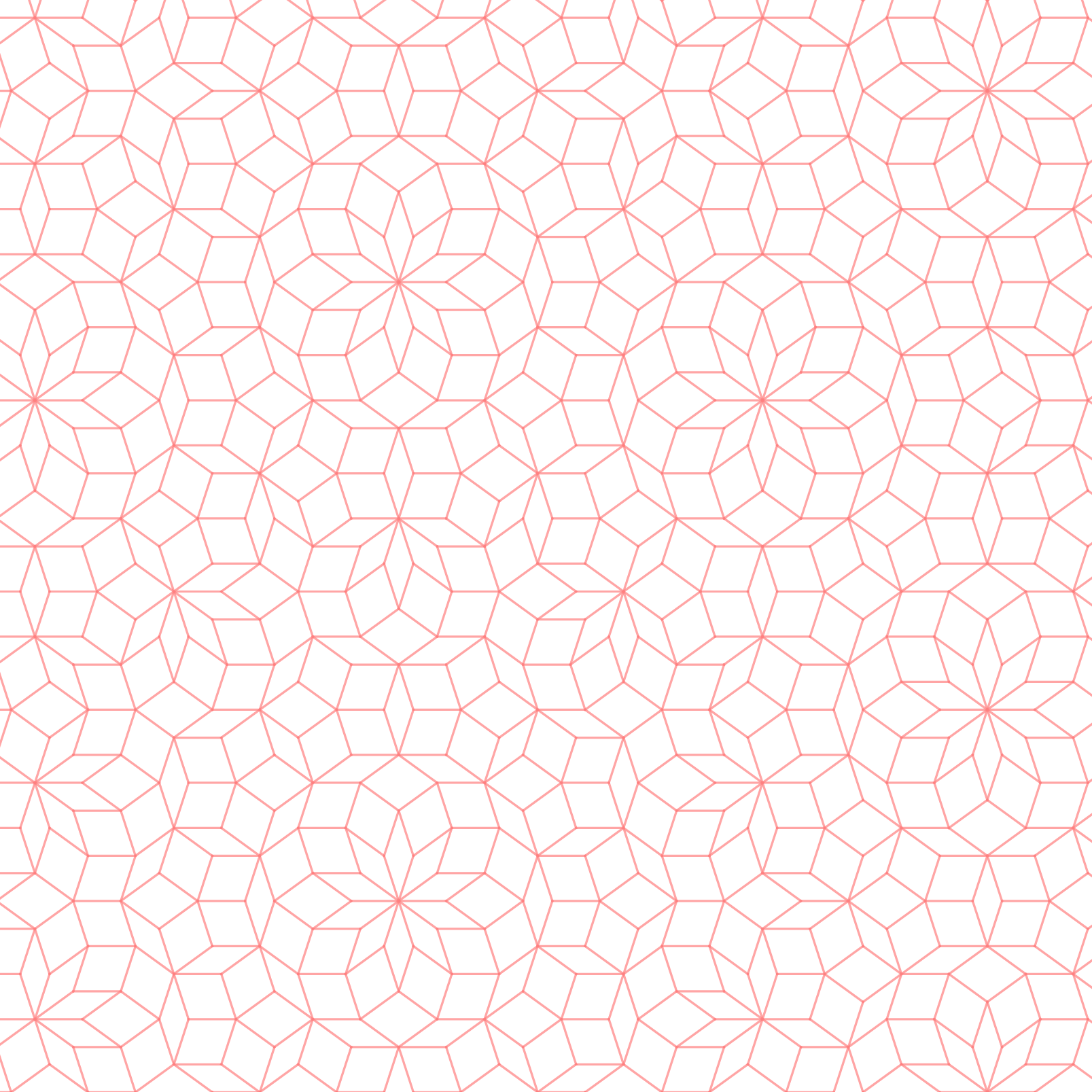}
  \captionof{figure}{The 2D 10-fold Ammann trio: dual tiling (i).}
   \label{2D_10foldTrio1_PurePenrose}
\end{center}

\begin{center}
  \includegraphics[width=2.4in]{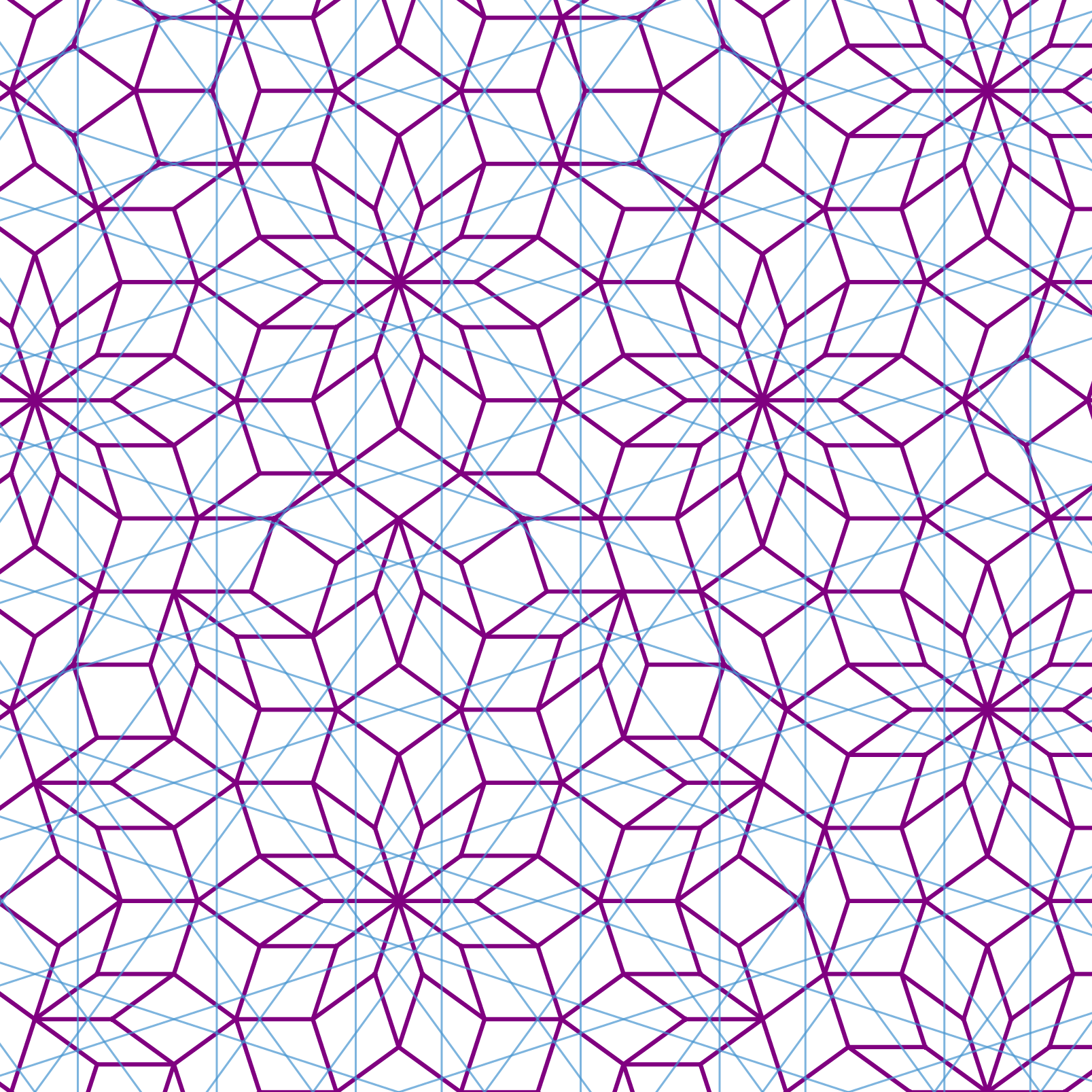}
  \captionof{figure}{2D 10-fold Ammann trio: 2D 10-fold tiling (i) (thick, purple), with Ammann pattern (i) (thin, blue).}
  \label{2D_10foldTrio1_AmmannLines}
  \vspace{10mm}
  \includegraphics[width=2.4in]{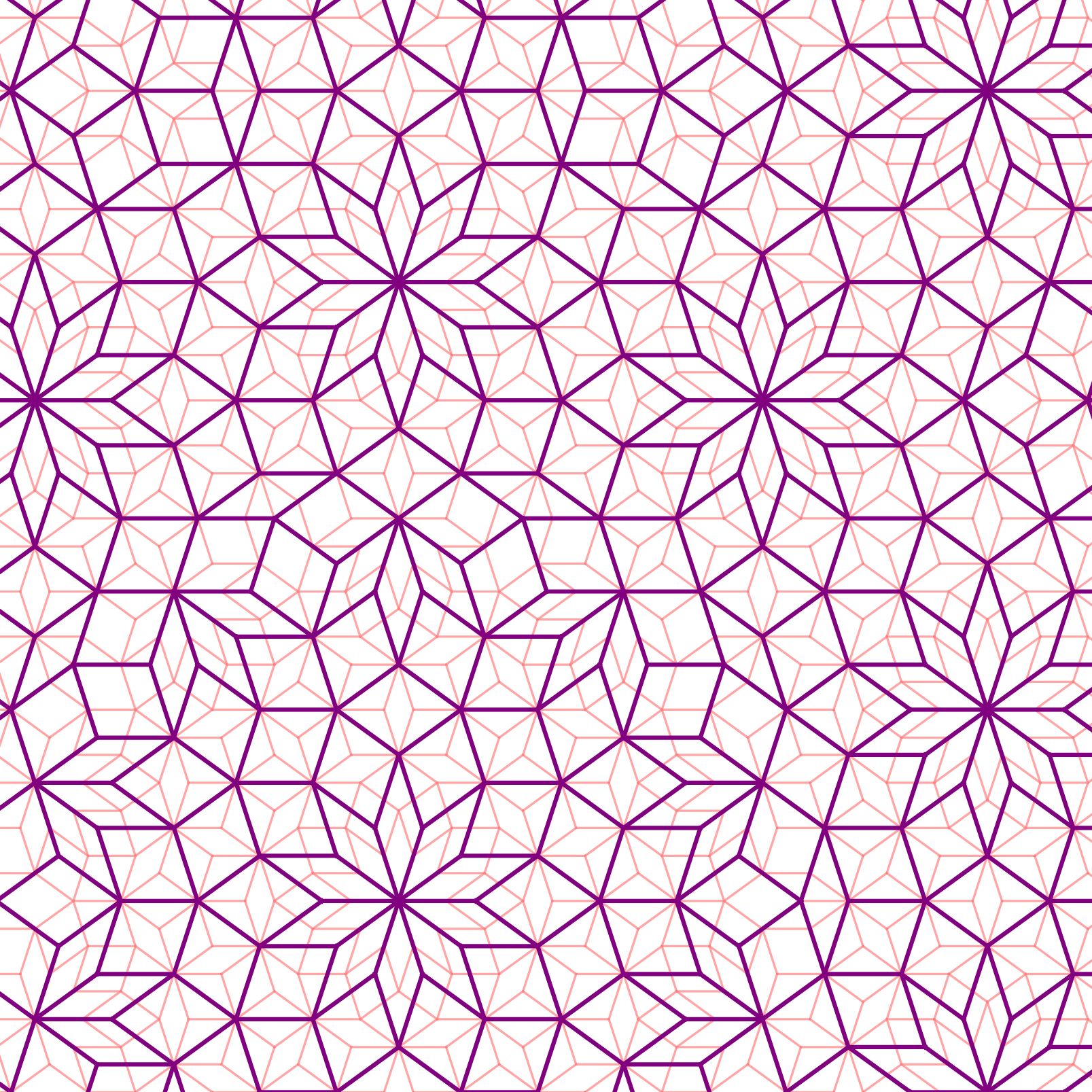}
  \captionof{figure}{2D 10-fold Ammann trio: 2D 10-fold tiling (i) (thick, purple), and its inflation (thin, pink).}
   \label{2D_10foldTrio1_Inflation}
\end{center}

\begin{center}
  \includegraphics[width=2.4in]{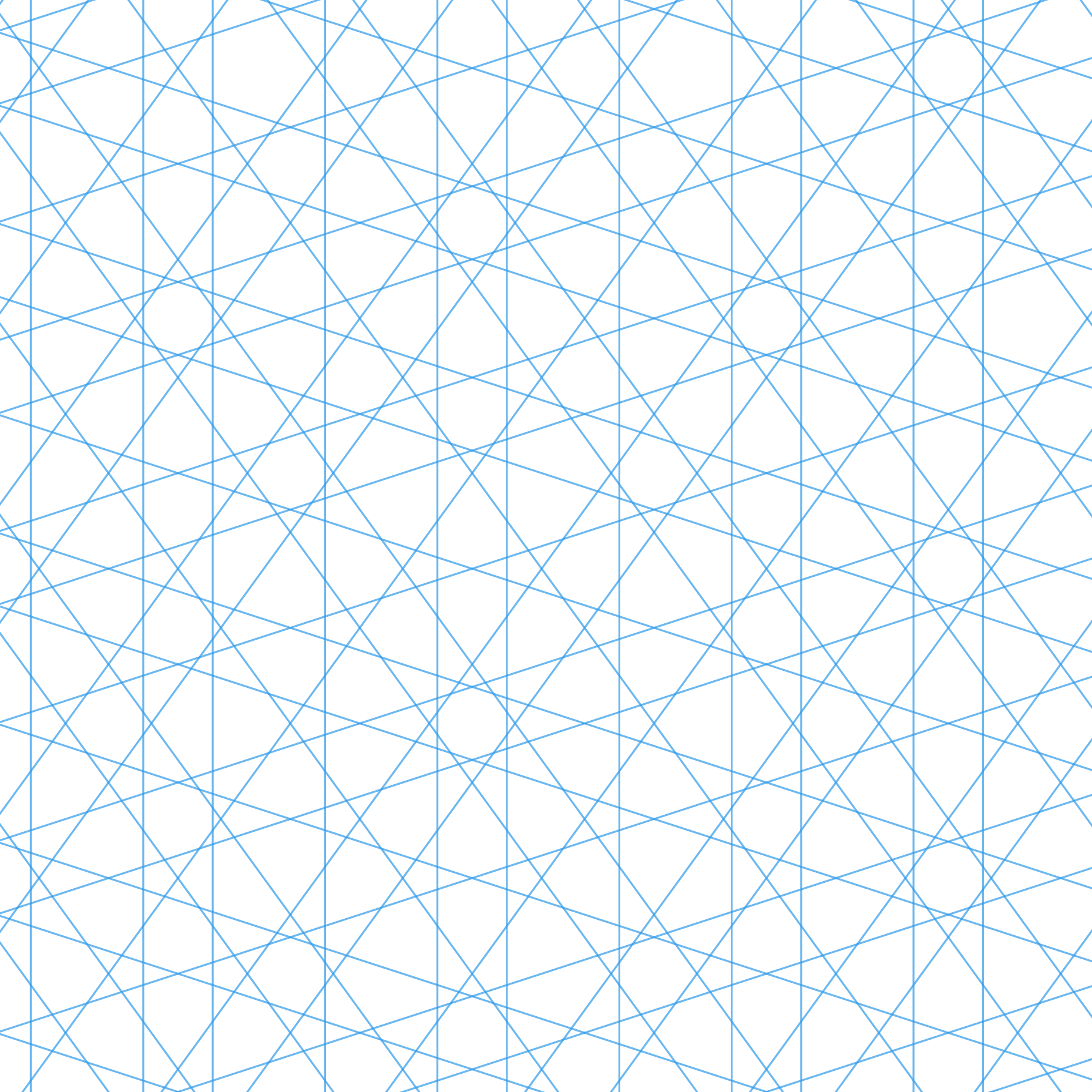}
  \captionof{figure}{The 2D 10-fold Ammann trio: pattern (ii).}
  \label{2D_10foldTrio2_PureAmmann}
  \vspace{10mm}
  \includegraphics[width=2.4in]{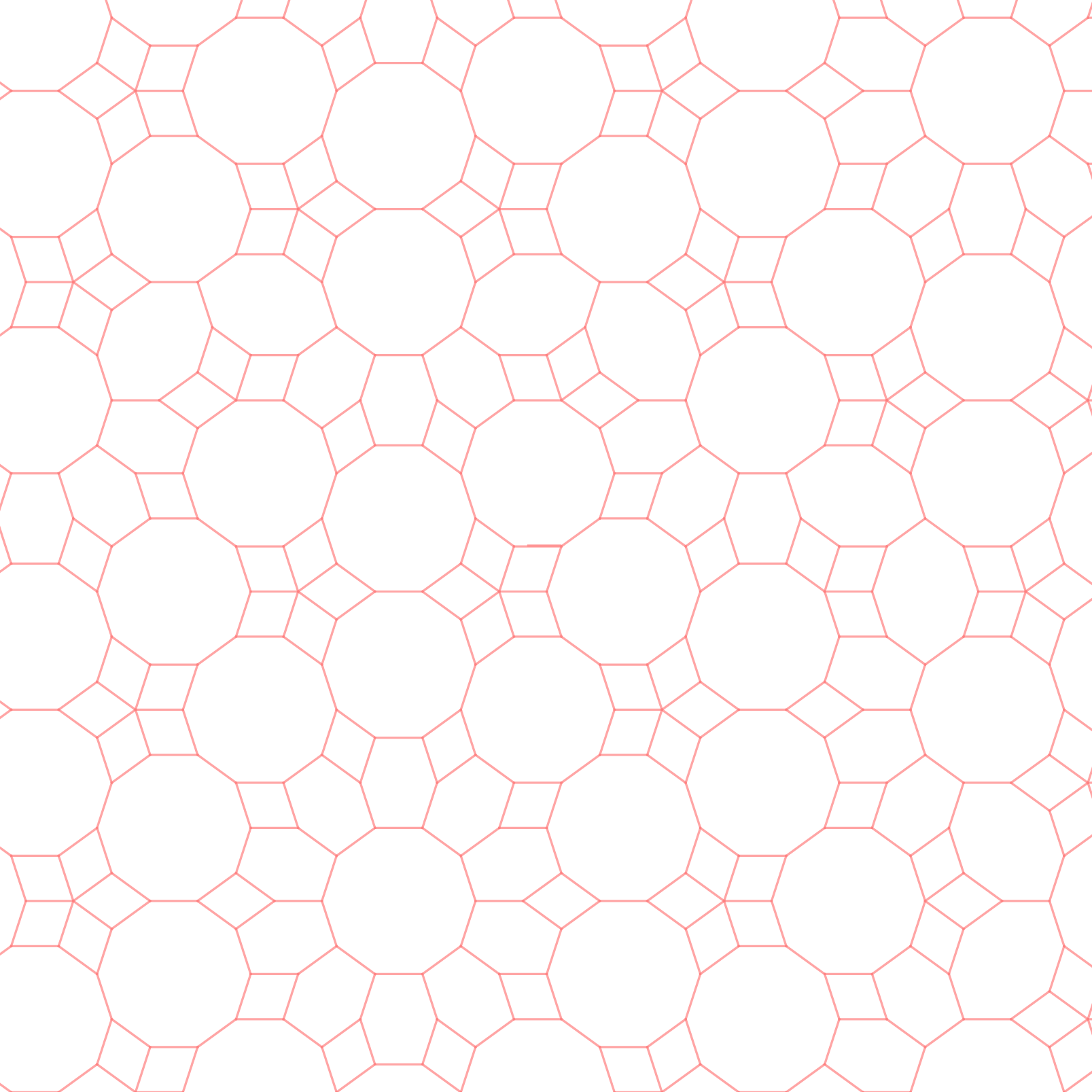}
  \captionof{figure}{The 2D 10-fold Ammann trio: dual tiling (ii).}
   \label{2D_10foldTrio2_PurePenrose}
\end{center}

\begin{center}
  \includegraphics[width=2.4in]{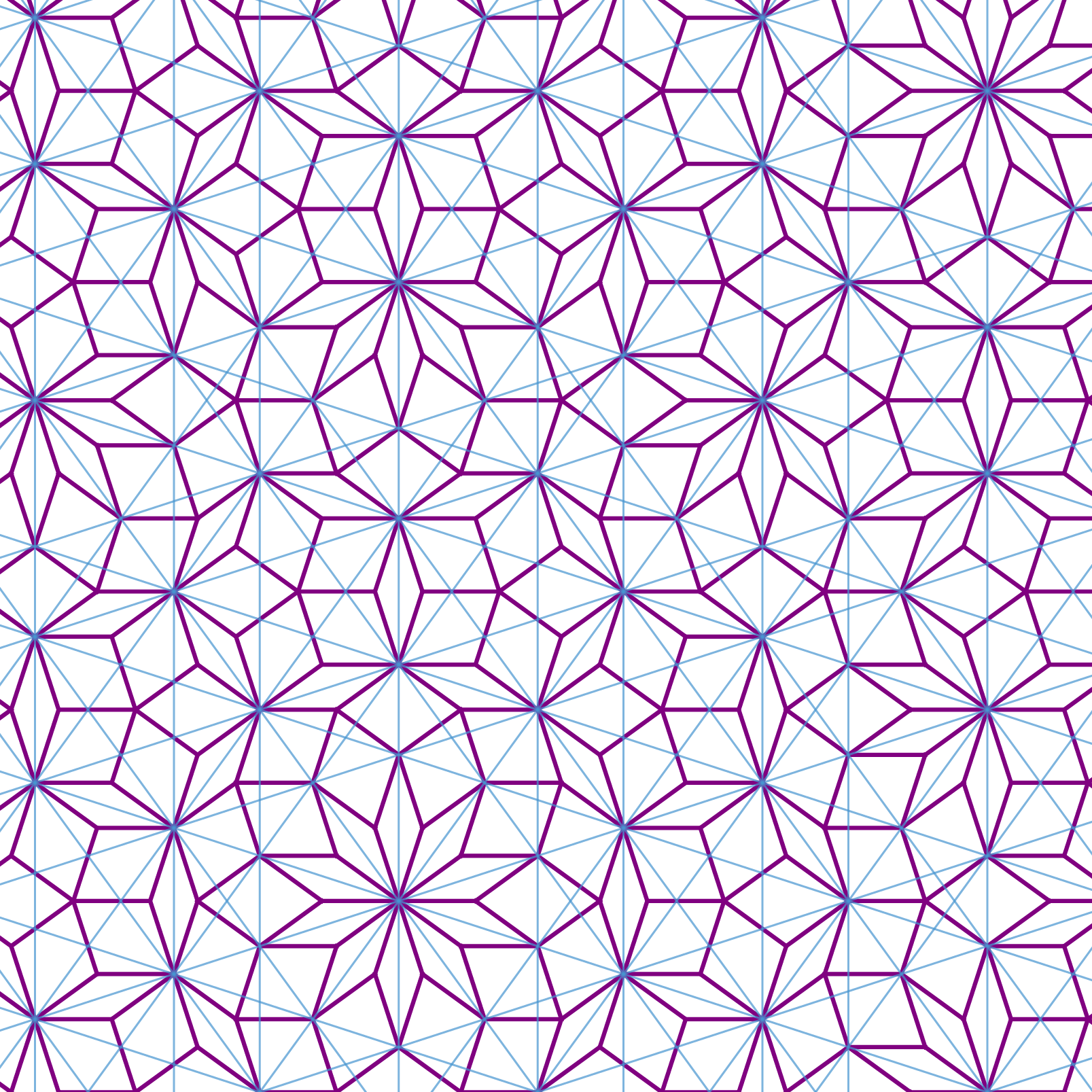}
  \captionof{figure}{2D 10-fold Ammann trio: 2D 10-fold tiling (ii) (thick, purple), with Ammann pattern (ii) (thin, blue).}
  \label{2D_10foldTrio2_AmmannLines}
  \vspace{10mm}
  \includegraphics[width=2.4in]{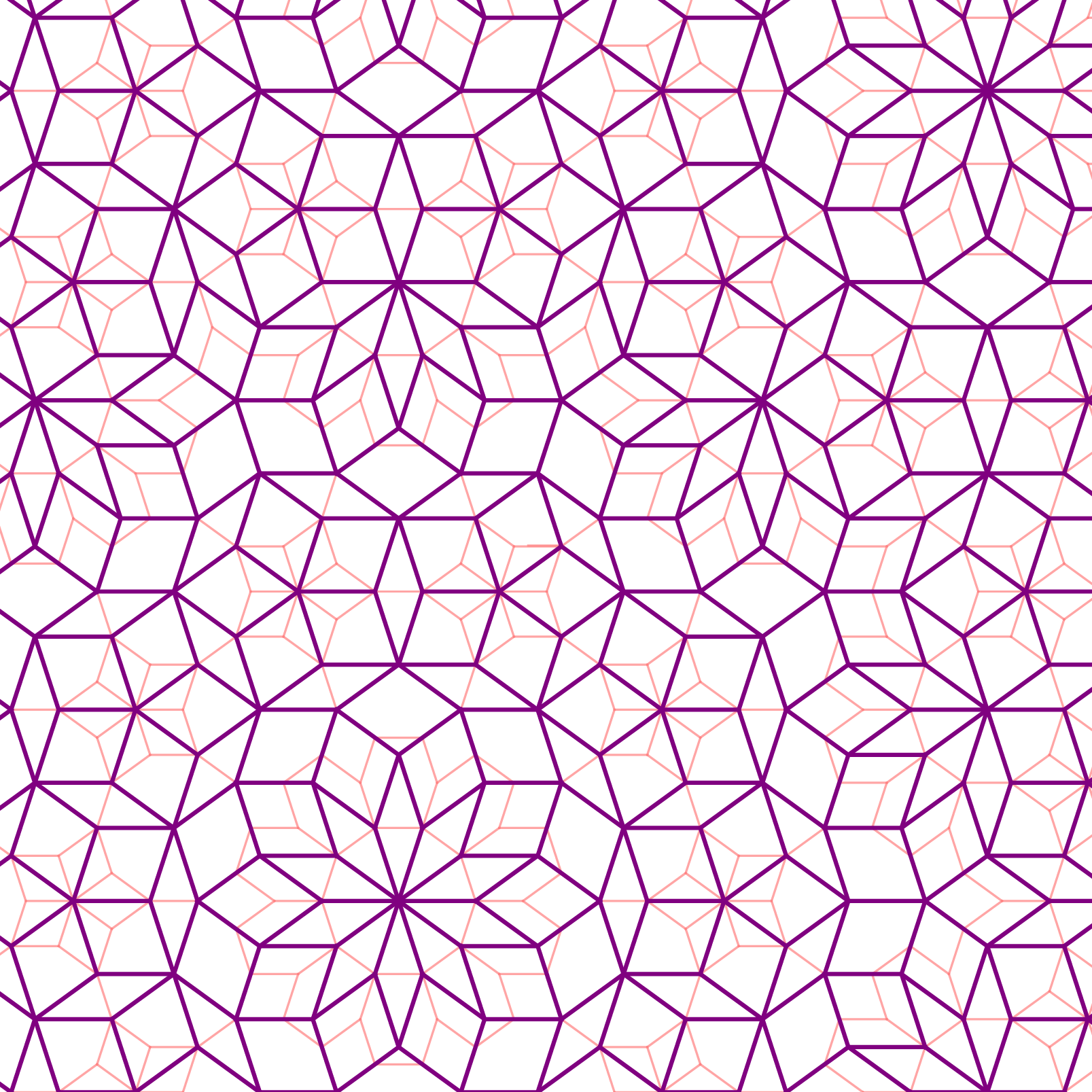}
  \captionof{figure}{2D 10-fold Ammann trio: 2D 10-fold tiling (ii) (thick, purple), and its inflation (thin, pink).}
   \label{2D_10foldTrio2_Inflation}
\end{center}

\begin{center}
  \includegraphics[width=2.4in]{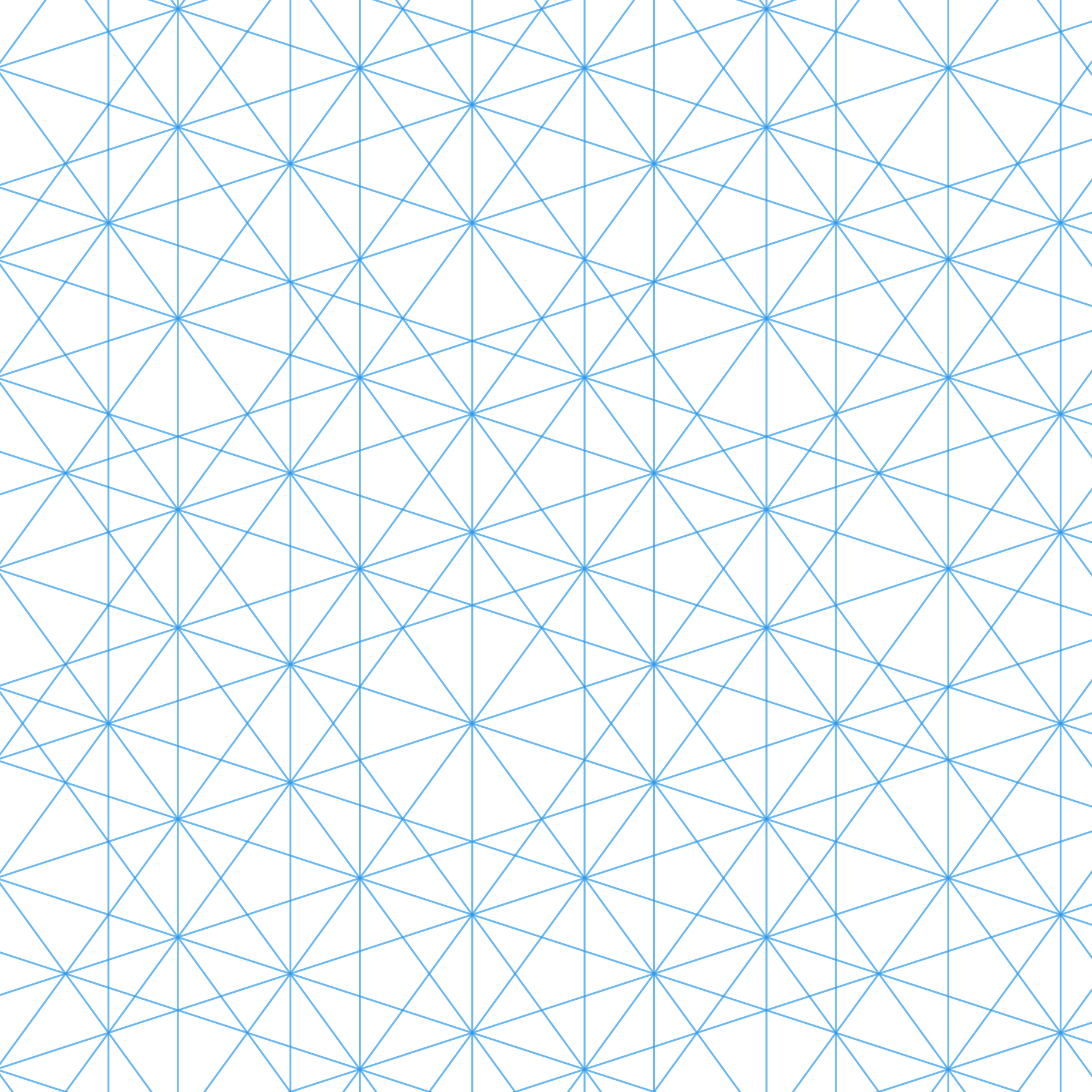}
  \captionof{figure}{The 2D 10-fold Ammann trio: pattern (iii).}
  \label{2D_10foldTrio3_PureAmmann}
  \vspace{10mm}
  \includegraphics[width=2.4in]{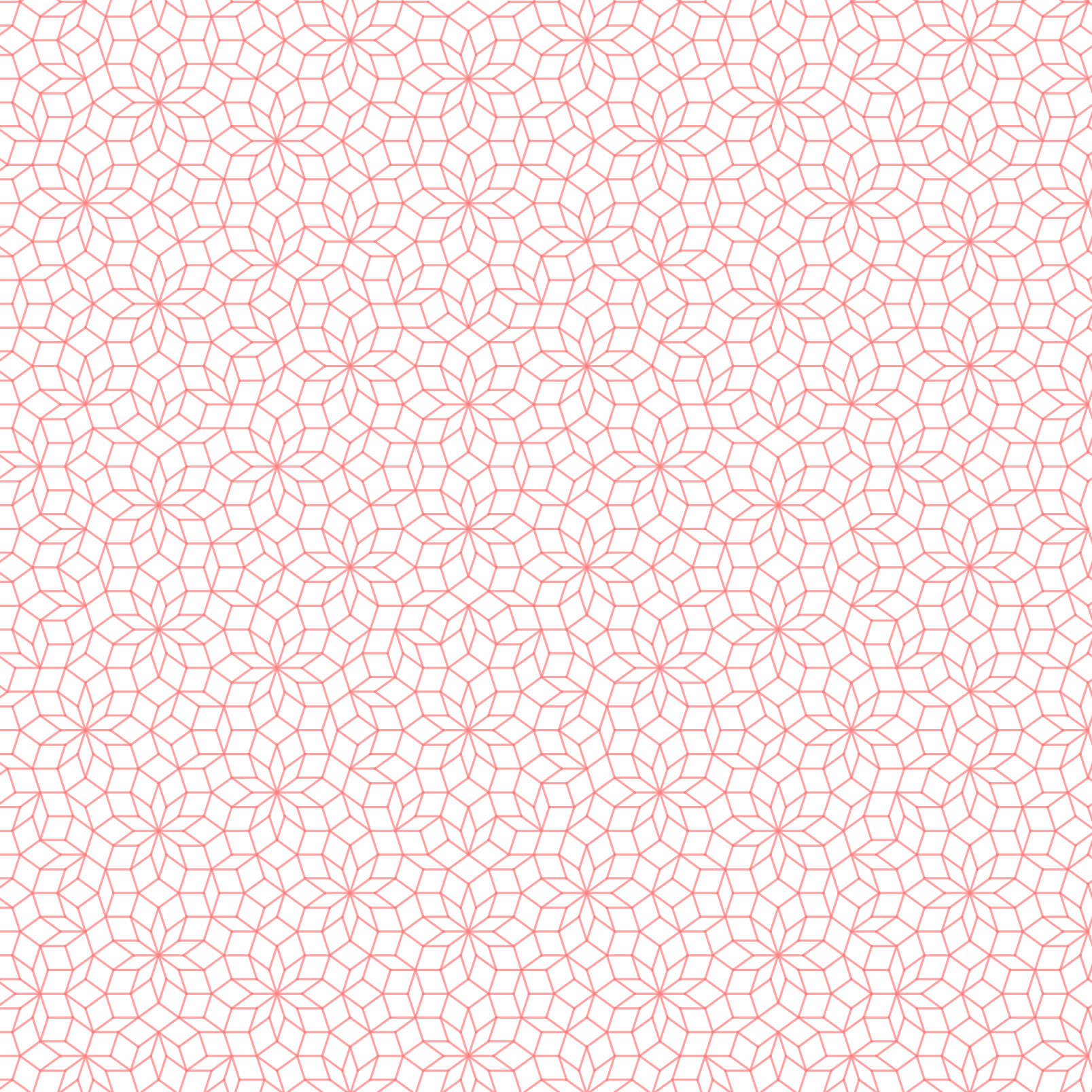}
  \captionof{figure}{The 2D 10-fold Ammann trio: dual tiling (iii).}
   \label{2D_10foldTrio3_PurePenrose}
\end{center}

\begin{center}
  \includegraphics[width=2.4in]{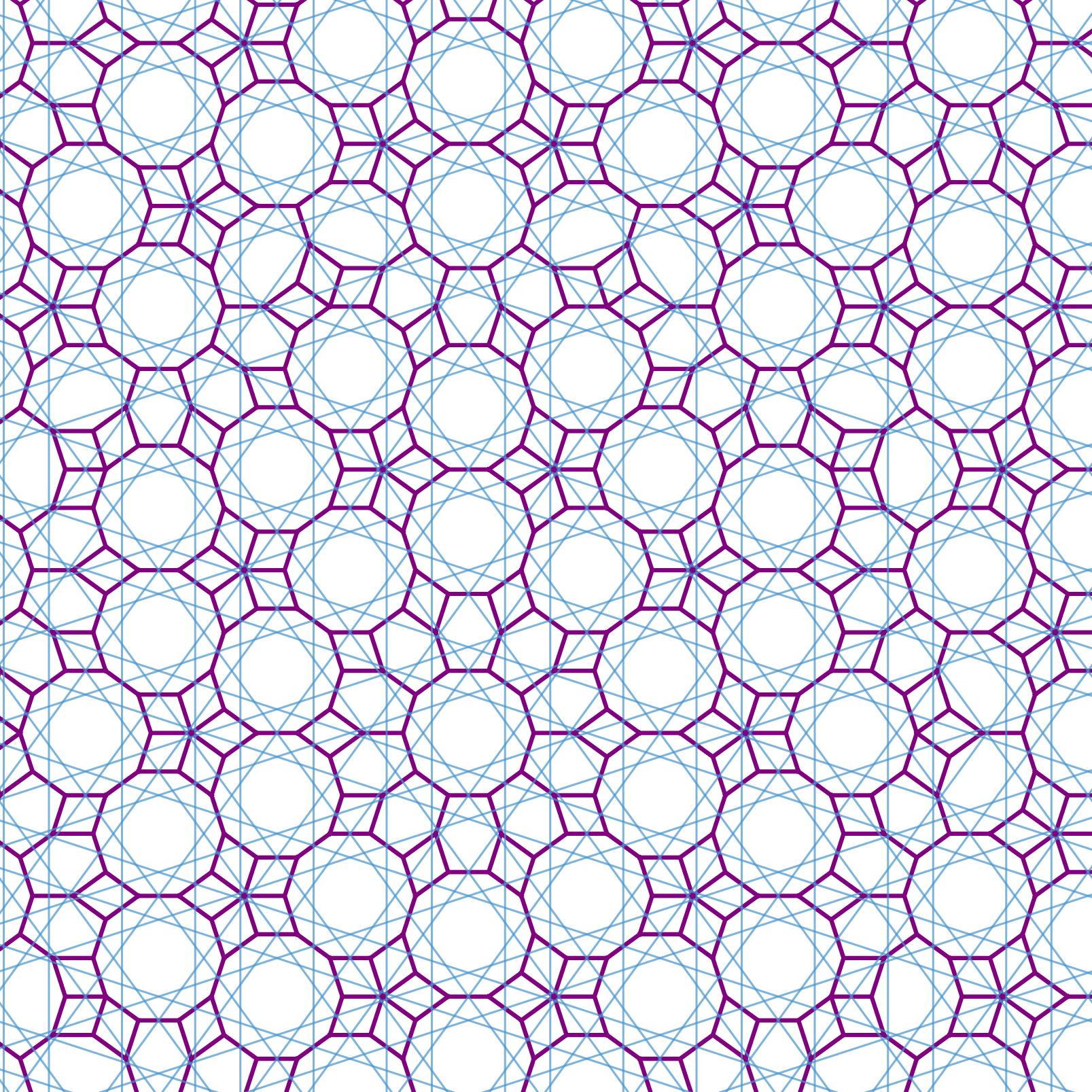}
  \captionof{figure}{2D 10-fold Ammann trio: 2D 10-fold tiling (iii) (thick, purple), with Ammann pattern (iii) (thin, blue).}
  \label{2D_10foldTrio3_AmmannLines}
  \vspace{10mm}
  \includegraphics[width=2.4in]{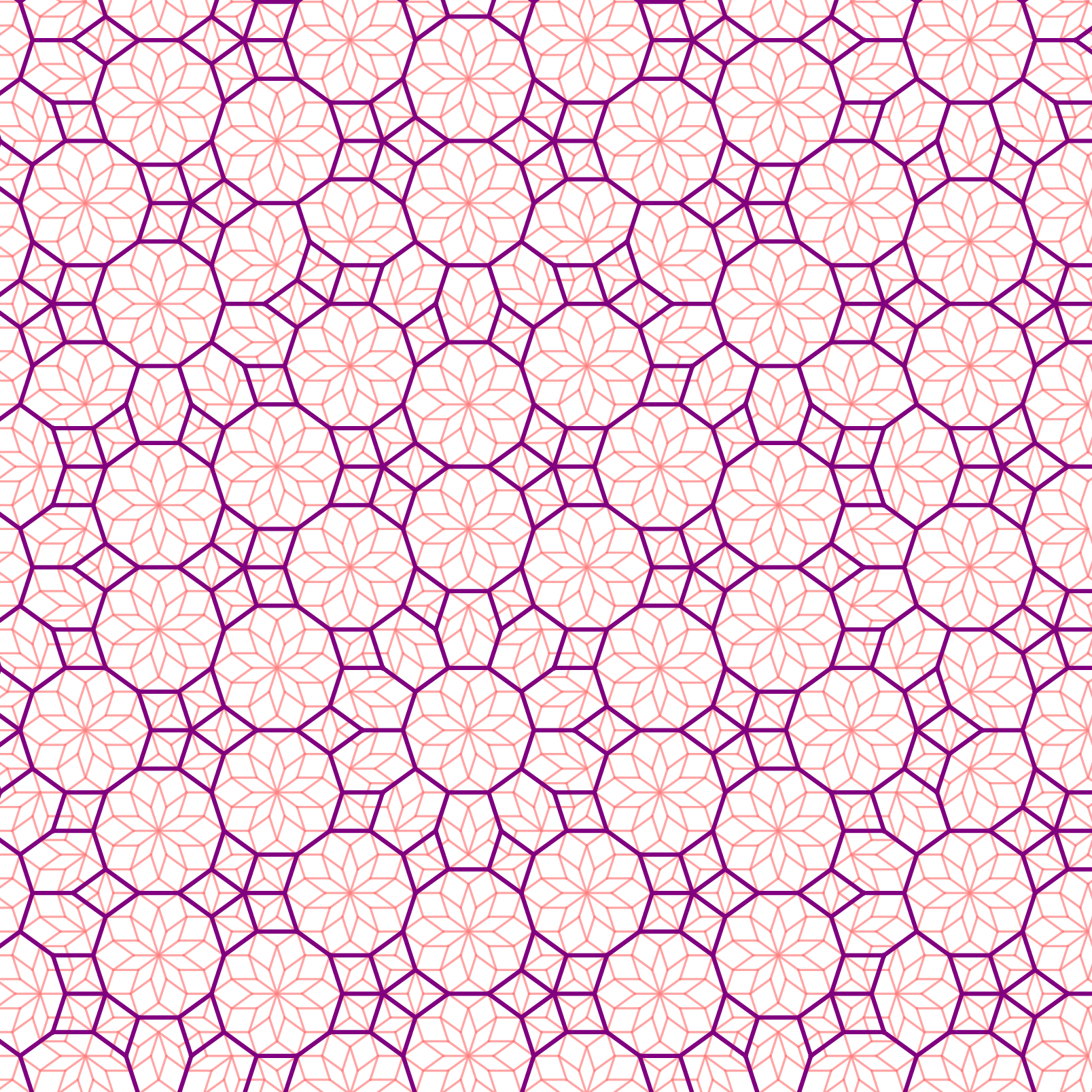}
  \captionof{figure}{2D 10-fold Ammann trio: 2D 10-fold tiling (iii) (thick, purple), and its inflation (thin, pink).}
   \label{2D_10foldTrio3_Inflation}
\end{center}

\begin{center}
  \includegraphics[width=2.4in]{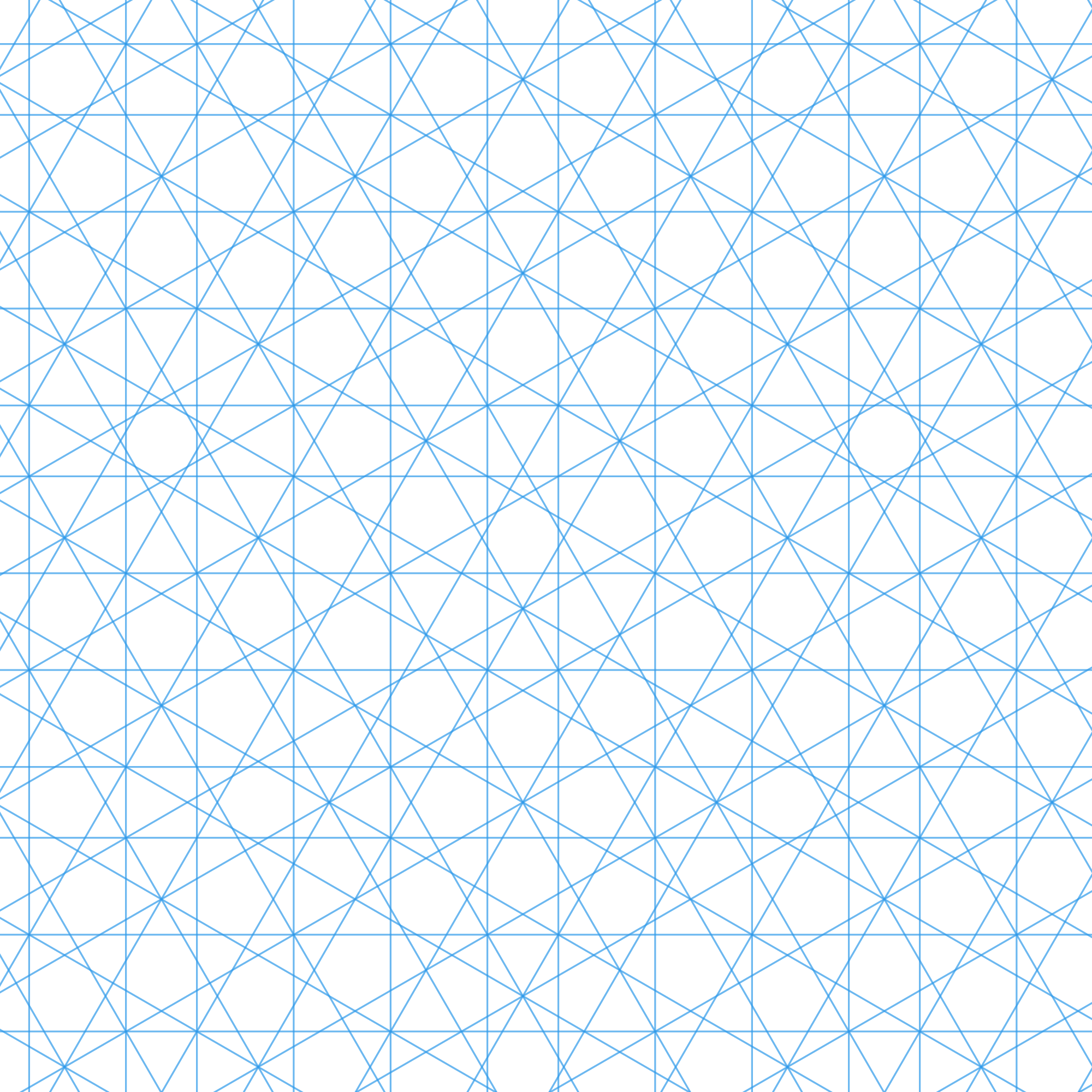}
  \captionof{figure}{The 2D 12-fold Ammann duo A: pattern (i).}
  \label{2D_12foldA_AmmannDuo_1}
  \vspace{10mm}
  \includegraphics[width=2.4in]{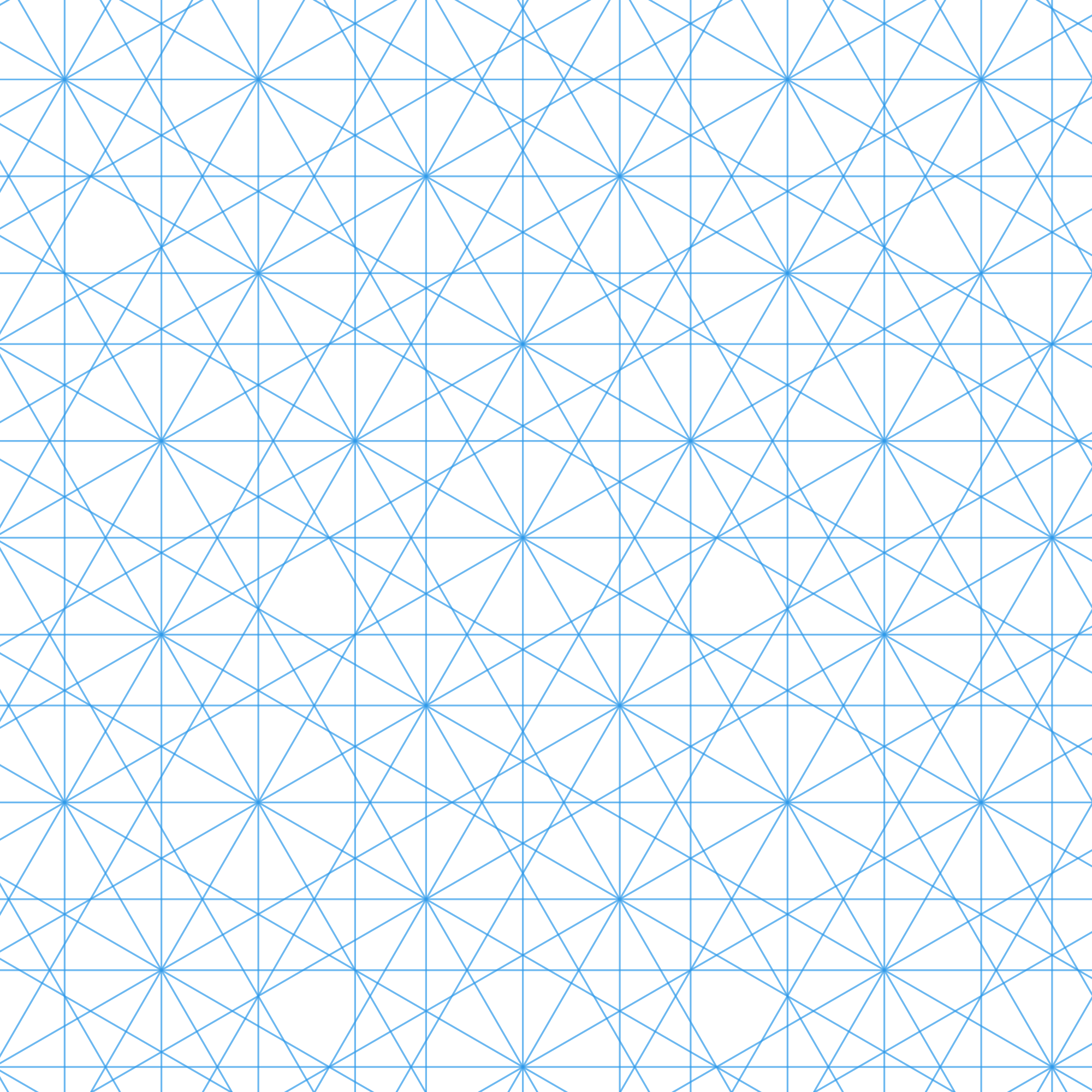}
  \captionof{figure}{The 2D 12-fold Ammann duo A: pattern (ii).}
   \label{2D_12foldA_AmmannDuo_2}
\end{center}

\begin{center}
  \includegraphics[width=2.4in]{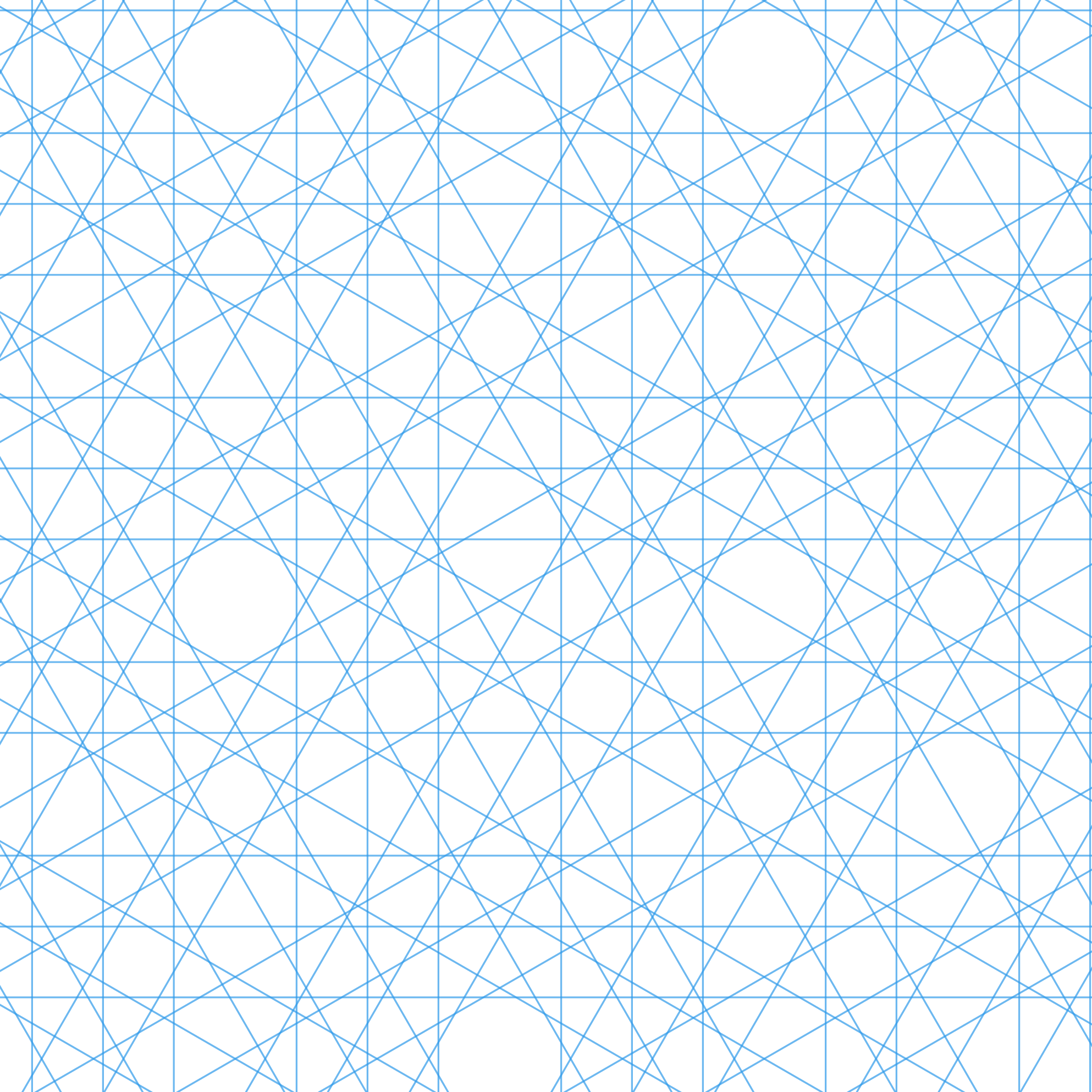}
  \captionof{figure}{The 2D 12-fold Ammann duo B: pattern (i).}
  \label{2D_12foldB_AmmannDuo_1}
  \vspace{10mm}
  \includegraphics[width=2.4in]{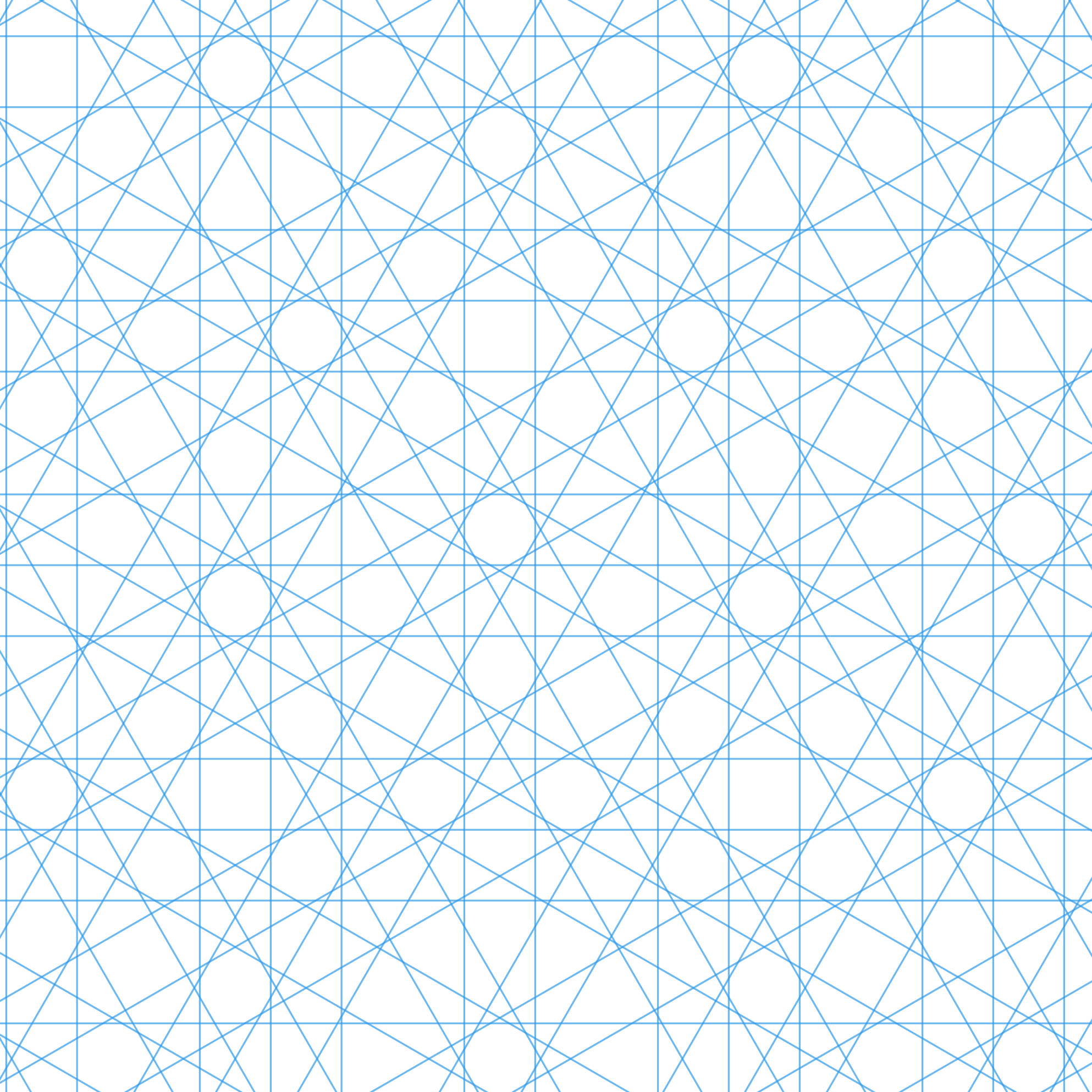}
  \captionof{figure}{The 2D 12-fold Ammann duo B: pattern (ii).}
   \label{2D_12foldB_AmmannDuo_2}
\end{center}

\begin{center}
  \includegraphics[width=2.4in]{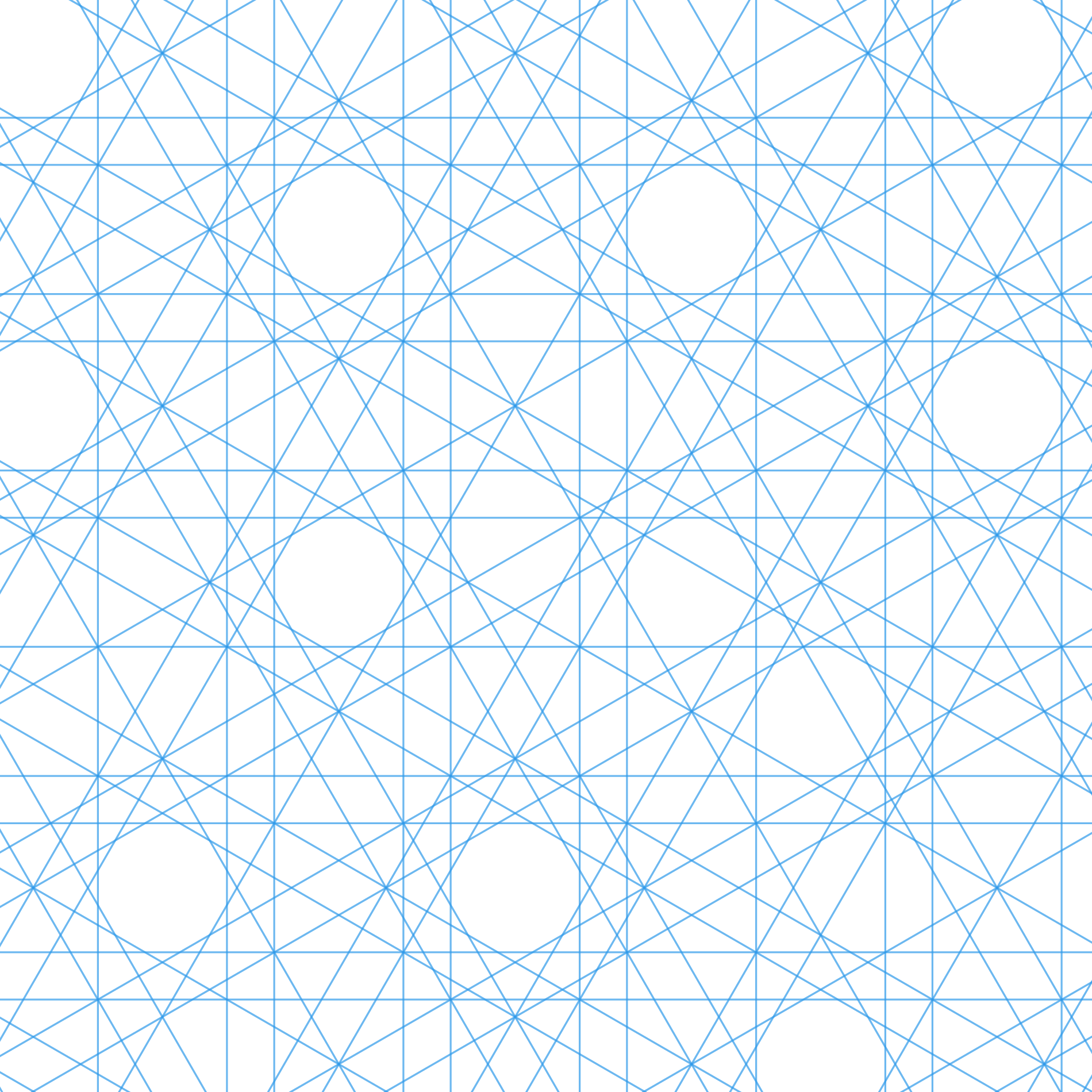}
  \captionof{figure}{The 2D 12-fold Ammann duo C: pattern (i).}
  \label{2D_12foldC_AmmannDuo_1}
  \vspace{10mm}
  \includegraphics[width=2.4in]{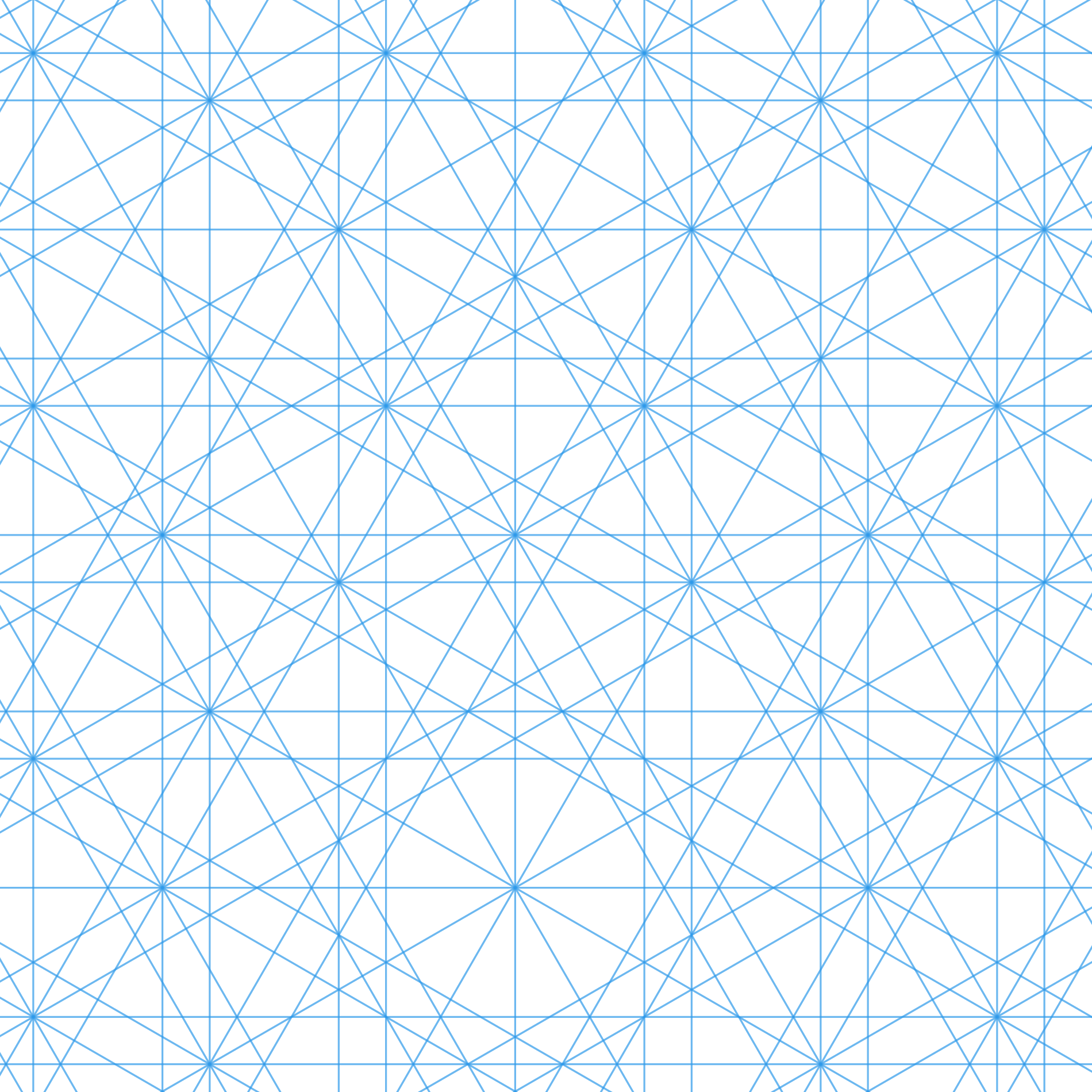}
  \captionof{figure}{The 2D 12-fold Ammann duo C: pattern (ii).}
   \label{2D_12foldC_AmmannDuo_2}
\end{center}

\section{Matching Rules}
\label{Matching}

In this section, we turn to the issue of matching rules.

We emphasize that the Ammann pattern and its dual Penrose-like tiling can {\it both} be thought of as quasicrystalline tilings, with perfect local matching rules that we can systematically construct.  To see this, first note that the Ammann pattern and its dual Penrose-like tiling are both examples of substitution tilings -- {\it i.e.}\ they are both equipped with inflation rules and, by repeated iteration of those rules, arbitrarily large patches of the tiling can be generated from an initial seed.   If a substitution tiling is in the class covered by the Goodman-Strauss theorem \cite{GoodmanStrauss}, then one can systematically construct perfect local matching rules for the tiling ({\it i.e.}\ local rules that constrain how nearby tiles may join, so that any two legal tilings are forced to be locally indistinguishable from one another).  In particular, let us call an inflation rule ``wall-to-wall" if, when we apply the inflation rule to each tile, the new (smaller) tiles are entirely contained within the original (larger) tile (so that the smaller tiles never cross the boundaries of the larger tiles).  If we have a substitution tiling whose inflation rule is not wall-to-wall, but we can slice the set of proto-tiles up into a finite set of smaller proto-tiles that inherit an inflation rule that {\it is} wall-to-wall, then we can apply Goodman-Strauss's result to infer (and systematically derive) perfect local matching rules.

At first glance, finding the right way to slice up a given tiling (to make its inflation rule wall-to-wall) looks like it might be difficult; but we will now explain that, for any Ammann pattern, it is actually easily done.  To see why, first consider the ten special 1D inflation rules collected in Table 1 and Figure 5 of Ref.~\cite{BoyleSteinhardt1D}: as inflation rules for the two-tile set $\{L,S\}$, four of these these ten cases (2b, 3b, 4b and 4d) are automatically wall-to-wall, while the remaining six cases (1, 2a, 3a, 3c, 4a, 4c) are not.  But it is easy to check that, in the remaining six cases, if we cut each tile in half ({\it i.e.}\ we cut $S$ into left and right halves, $S_{{\rm left}}$ and $S_{{\rm right}}$, and we cut $L$ into left and right halves, $L_{{\rm left}}$ and $L_{{\rm right}}$) the new four-tile set $\{S_{{\rm left}},S_{{\rm right}},L_{{\rm left}},L_{{\rm right}}\}$ {\it is} wall-to-wall.  

\begin{figure}
  \begin{center}
    \includegraphics[width=2.4in]{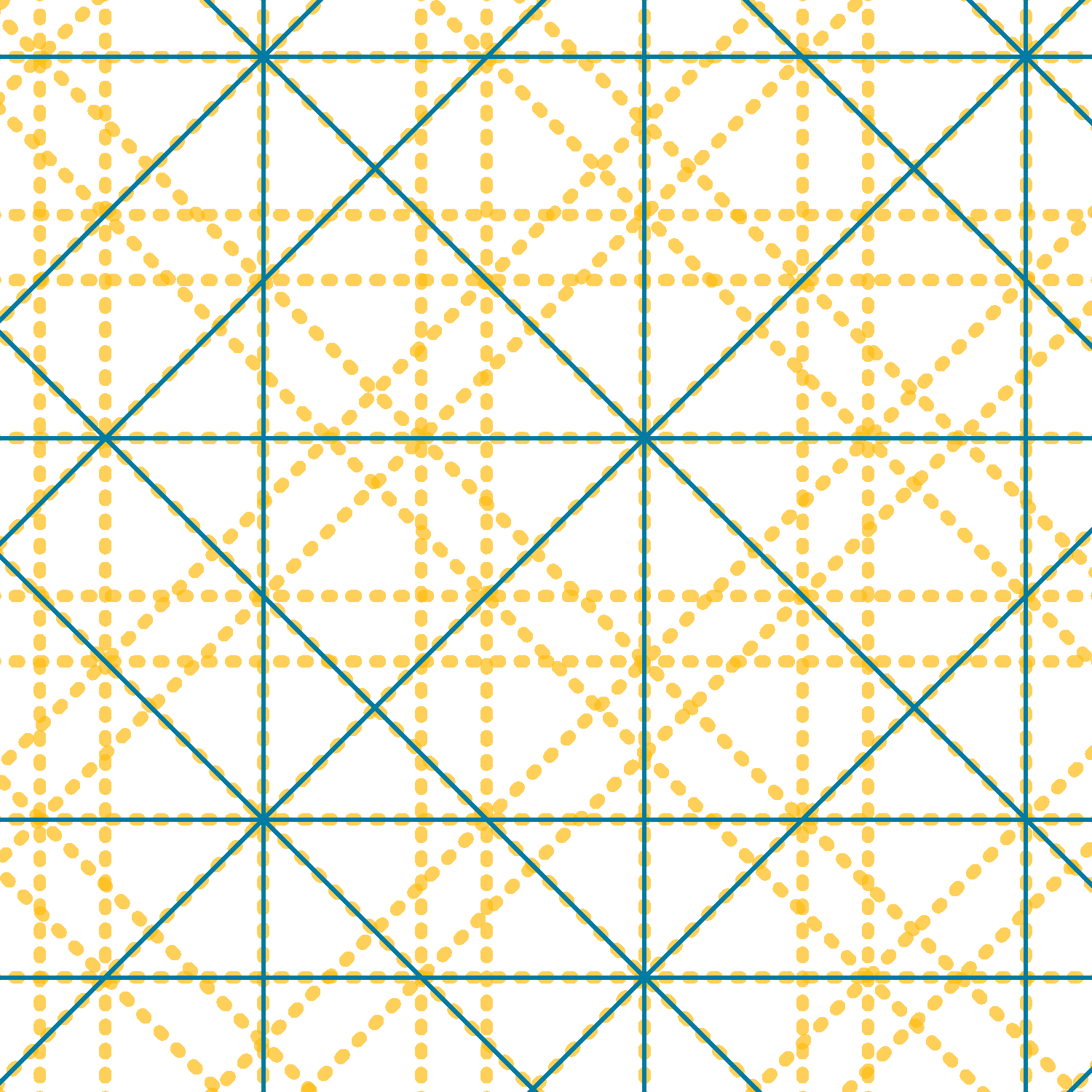}
  \end{center}
  \caption{An 8-fold ``B" Ammann pattern, superposed on its inflation, illustrates how the larger tiles (with thin solid green edges) are decomposed into ``smaller"
  tiles (with thick dotted yellow edges) in wall-to-wall fashion.  That is, for any tile created by the dotted yellow lines, each segment of its boundary lies within or 
  right along the boundary of a larger tile created by the solid green lines; or equivalently, no tile created by the dotted yellow lines crosses a solid green line.}  
  \label{AmmannInflation8B}
\end{figure}

Now we turn from the 1D quasilattices to the Ammann patterns, and proceed in two steps.  Step 1: First consider the simpler case where the Ammann pattern is constructed from 1D tiles that inflate wall-to-wall.  In this case, since the boundaries of the Ammann tiles are directly given by the Ammann lines/planes/hyperplanes, the Ammann tiles will also inflate wall-to-wall (see Fig.~\ref{AmmannInflation8B} for an illustration of this point).  Step 2: Next consider the seemingly-more-complicated case where the 1D tiles (and hence the Ammann tiles) do {\it not} inflate wall-to-wall.  We want to find a way to slice up the Ammann tiles so that the pieces {\it do} inflate wall-to-wall.  However, if we try to work with the Ammann tiles directly, the problem looks complicated.  Fortunately, the Ammann pattern's 1D decomposition comes to the rescue.  If starting with the 1D tiles, then it is clear how to split them up: we just split them in half, as explained above.  But that, in turn, determines how to slice up the Ammann tiles: go back to the original Ammann pattern and, halfway between every pair of parallel lines/planes/hyperplanes,  add another parallel line/plane/hyperplane, and in this way, the original Ammann tiles are sliced into smaller tiles that inflate wall-to-wall (for the same reason that the Ammann tiles inflated wall-to-wall in step 1).  Thus, we can apply Goodman-Strauss's result to infer (and systematically construct) local matching rules for an Ammann pattern (regarded as a tiling).  And then, since the Ammann pattern and its corresponding Penrose-like tiling are dual to one another, and mutually locally derivable, it follows that the dual Penrose-like tiling also has perfect local matching rules \cite{BaakeGrimme}.  {\it Hence, matching rules are another feature of the Ammann patterns (viewed as tesselations) and Penrose-like tilings generated by our procedure.  Since this applies to all the irreducible Ammann patterns, of which there were previously only four, we have uncovered by our procedure an infinite number of new tilings with matching rules.}

So far we have argued (using \cite{GoodmanStrauss}) that local matching rules exist and can be systematically derived.  Now let us turn to the question of how those rules can be best implemented.  In the original (2D, 10-fold) Penrose tiling, something nice happens: if we begin with an Ammann pattern and use it to derive the Penrose proto-tiles and their Ammann decorations (as described in Section 4), we find that the Ammann decorations of the tiles, combined with the requirement that they join together across tile boundaries to form infinite straight lines, precisely provide the perfect local matching rule we want!  By contrast, if we repeat this procedure in the 8-fold and 12-fold cases, the resulting Ammann decorations are not strong enough to give a perfect matching rule.  In fact, they are compatible with a periodic arrangement of the tiles.  Is there some natural way to adjust the Ammann decoration so that it provides a perfect matching rule in all cases?  

Our Coxeter-pair formalism suggests a beautiful answer.  First, let us distinguish between the Ammann pattern that is used to {\it generate} the tiling (by dualization) and the Ammann pattern that is used to {\it decorate} the tiling.  Suppose that, to {\it generate} the tiling, we use the same minimal Ammann pattern as before.  But then, 
in constructing the Ammann pattern that will {\it decorate} the tiling, instead of taking the initial star of $d^{\parallel}$-dimensional vectors to be a {\it minimal} star ({\it i.e.}\ with the minimal set of vectors compatible with the Ammann pattern's desired orientational symmetry), we take it to be the $\theta^{\parallel}$ root system itself (where $\theta^{\parallel}$ is the lower-dimensional non-crystallographic member of the associated Coxeter pair).  Note that in 2D, where $\theta^{\parallel}=I_{2}^{n}$, these two possibilities (the minimal star vs the root star) yield exactly the {\it same} Ammann pattern when $n$ is odd.  But when $n$ is even, the root star yields an Ammann pattern that is the superposition of {\it two} minimal Ammann patterns (rotated by $\pi/n$ relative to one another).  This neatly explains why in the 10-fold ($I_{2}^{5}$) case, the minimal Ammann decoration already yields a perfect local matching rule, while in the 8-fold ($I_{2}^{8}$) and 12-fold ($I_{2}^{12}$) cases, it does not; and, moreover, why Socolar found \cite{Socolar89} that a perfect local matching rule could be obtained in the 8-fold and 12-fold cases by adding a second minimal Ammann decoration rotated by $\pi/n$ relative to the first.  In fact, the Coxeter pair perspective goes further.   The Ammann pattern based on the (lower-dimensional) $\theta^{\parallel}$ root system is naturally obtained from a ``Coxeter slice" of the higher-dimensional honeycomb based on the $\theta$ roots system, via the approach of Section 3. This, in turn, means that when $n$ is even, the $I_{2}^{n}$ Ammann decoration is actually a superposition of two minimal Ammann grids built from two {\it different} 1D quasilattices (based on two {\it different} rows in Table 1 of Ref.~\cite{BoyleSteinhardt1D}, corresponding to the same scale factor).  We plan to flesh out this idea in future work.

\section{Discussion}
\label{Discussion}

As we already introduced, motivated and outlined our results at length in Section I, here we will just provide a brief recapitulation.

In Section \ref{Coxeter_Pairs}, we introduced the notion of a Coxeter pair: a natural pairing between a lower-rank non-crystallographic reflection group and a higher-rank crystallographic partner.  We found all such ``Coxeter pairs" and collected them in Table \ref{CoxeterPairs}.  They form two infinite families plus four exceptional cases.  

In Section \ref{general_construction}, we define what we mean by an irreducible Ammann pattern, we explain how to construct all such patterns using a geometric construction based on Coxeter pairs, and we obtain a useful closed form analytic expression that describes all such Ammann patterns in a simple and unified way.

The tilings obtained by dualizing these Ammann patterns share many of the key properties of the Penrose tiling, so we call them Penrose-like.  In Section \ref{Penrose_Tilings},  we derived a dualization formula that allows us to convert any irreducible Ammann pattern into a dual Penrose-like tiling, and to obtain the Ammann decorations and inflation rules for this tiling.

Using these results, in Section \ref{Minimal_Penrose_Tilings} we construct all the {\it minimal} Ammann patterns ({\it i.e.}\ Ammann patterns with an irreducible non-crystallographic symmetry, built from the minimal set of 1D quasilattices compatible with that symmetry), and all the corresponding dual Penrose-like tilings, along with their Ammann decorations and inflation rules.  There are 11 cases in 2D, 9 in 3D, one in 4D, and none in higher dimensions, and the 2D results are shown explicitly in Figs. \ref{A4rootfig}-\ref{2D_12foldC2_Inflation}.  The same procedure also yields the four minimal ``Ammann cycles" (one 10-fold-symmetric ``Ammann trio" and three 12-fold-symmetric ``Ammann duos") which are shown in Figs. \ref{2D_10foldTrio1_PureAmmann}-\ref{2D_12foldC_AmmannDuo_2}.

In Section \ref{Discussion}, we discuss the matching rules for our tilings.  We first explain how the Ammann perspective allows us to see that the Goodman-Strauss theorem \cite{GoodmanStrauss} applies to our tilings, and guaruntees that they have perfect local matching rules.  Then we point out that the Coxeter-pair perspective suggests a particular concrete and elegant implementation of the local matching rules: we leave the detailed fleshing out of this final point as a topic for future work.

Finally, note that for the 2D Ammann patterns and the 10-fold-symmetric Ammann trio, we have explicitly constructed their dual Penrose-like tilings, their Ammann decorations, and inflations.  By contrast, although we have also constructed the three 12-fold-symmetric Ammann duos shown in Figs.~\ref{2D_12foldA_AmmannDuo_1}-\ref{2D_12foldC_AmmannDuo_2}, we have not yet constructed their dual tilings, Ammann decorations, or inflations.  This is another point we leave for future work.

\acknowledgments 
We thank Joshua Socolar for a helpful private communication first drawing our attention to the 12-fold ``A2" Ammann pattern and Penrose tiling, as well as the 10-fold Ammann trio.   Research at Perimeter Institute is supported by the Government of Canada, through Innovation, Science and Economic Development, Canada and by the Province of Ontario through the Ministry of Research, Innovation and Science.  PJS was supported in part by the Princeton University Innovation Fund for New Ideas in the Natural Sciences.

\appendix

\section{Finding all Coxeter pairs}
\label{Finding_all_Coxeter_pairs}

In order to find the complete list of Coxeter pairs, we begin by reviewing three notions: Coxeter element, Coxeter number and Coxeter plane.  Corresponding to each $d$-dimensional fundamental root ${\bf f}_{j}\in\theta$ we have a reflection matrix $F_{j}$ that acts on the $d$-dimensional coordinate ${\bf x}$ as follows:
\begin{equation}
  F_{j}:{\bf x}\to {\bf x}-2\frac{{\bf x}\cdot {\bf f}_{j}}{{\bf f}_{j}\cdot{\bf f}_{j}}{\bf f}_{j}.
\end{equation}
The product of all $d$ of these reflection matrices is a ``Coxeter element" of $\theta$:
\begin{equation}
  C=F_{1}\ldots F_{d}.
\end{equation}
The ``Coxeter number" of $\theta$ is the smallest positive integer $h$ for which 
\begin{equation}
  C^{h}=1,
\end{equation}
and it is given by a simple formula: the total number of roots in $\theta$ divided by the rank $d$.  (In Table \ref{CoxeterNumbers}, we list the number of roots and the Coxeter number for each crystallographic root system.)  Thus, the eigenvalues of $C$ are $h$th roots of unity ${\rm exp}[\pm2\pi i k/h]$.  In particular, two of the eigenvalues are ${\rm exp}[\pm 2\pi i/h]$, and the two corresponding eigenvectors ${\bf u}_{\pm}$ span a two-dimensional plane called the ``Coxeter plane".   The orthogonal projection of the roots of $\theta$ onto the Coxeter plane is the the two-dimensional projection of maximal symmetry.  If $\theta$ has Coxeter number $h$, then its roots project onto the Coxeter plane to form an $I_{2}^{h}$-like pattern (as Figs.~\ref{A4rootfig} and  
\ref{B4rootfig} illustrate in case where $h$ is odd or even, respectively).

\begin{table}
\begin{center}
\begin{tabular}{c|c|c}
crystallographic root system & $\#$ of roots & Coxeter number $(h)$ \\
\hline
$A_{n}\;(n\geq1)$ & $n(n+1)$ & $n+1$ \\
$B_{n}\;(n\geq2)$ & $2n^{2}$ & $2n$ \\
$C_{n}\;(n\geq3)$ & $2n^{2}$ & $2n$ \\
$D_{n}\;(n\geq4)$ & $2n(n-1)$ & $2(n-1)$ \\
\hline
$G_{2}$ & $12$ & $6$ \\
$F_{4}$ & $48$ & $12$ \\
$E_{6}$ & $72$ & $12$ \\
$E_{7}$ & $126$ & $18$ \\
$E_{8}$ & $240$ & $30$ \\
\end{tabular}
\end{center}
\caption{For the various crystallographic root systems, we list the total number of roots, and the Coxeter number $h$.}
\label{CoxeterNumbers}
\end{table} 

Now we turn to our basic question: given an irreducible non-crystallographic root system $\theta^{\parallel}$, when does it have a corresponding irreducible crystallographic partner $\theta$ such that: (i) $\theta^{\parallel}$ and $\theta$ have the same rational rank $d$; and (ii) the maximally-symmetric projection of $\theta$ onto $d^{\parallel}$ dimensions yields $N$ copies of $\theta^{\parallel}$?  Let us answer this question by considering the various possible values of $d^{\parallel}$ in turn.  In the $d^{\parallel}=2$ case, the question becomes the following: for which $n$ does the non-crystallographic root system $I_{2}^{n}$ have a crystallographic partner of rank $\phi(n)$ and Coxeter number $h=n$?  Consider the various possibilities in Table \ref{CoxeterNumbers}:
\begin{itemize}
\item If $\theta$ is $A_{\phi(n)}$, the $h$ requirement becomes $\phi(n)+1=n$: this is precisely the requirement that $n$ is prime.  
\item If $\theta$ is $B_{\phi(n)}$ or $C_{\phi(n)}$, the $h$ requirement becomes $2\phi(n)=n$; from Euler's product formula $\phi(n)=\prod_{p|n}[1-(1/p)]$ (where the product is over all distinct prime factors $p$ of $n$), we see that this is precisely the requirement that $n$ is a power of $2$.
\item If $\theta$ is $D_{\phi(n)}$, the $h$ requirement becomes $2[\phi(n)-1]=n$.  This equation has no solution since, on the one hand, it says $n$ is even but, on the other hand, if $n$ is even, then $\phi(n)\leq n/2$.
\item Finally, among the exceptional cases $\theta=\{G_{2},F_{4},E_{6},E_{7},E_{8}\}$, we easily check that the only two that work are 
$F_{4}$ ($n=12$) and $E_{8}$ ($n=30$).
\end{itemize}
Next consider the $d^{\parallel}=3$ case: then $\theta^{\parallel}=H_{3}$, which has rational rank $6$.  From Table \ref{CoxeterNumbers} we see that the only candidate crystallographic partner $\theta$ of rank $d=6$ is $D_{6}$, since this is the only rank-six root system whose total number of roots ($60$) is a multiple of the total number of $H_{3}$ roots ($30$); and, as is well known (and as we see in Subsection \ref{3Dfrom6D}), $D_{6}$ does indeed have a projection onto $d^{\parallel}=3$ dimensions with $G(H_{3})$ (icosahedral) symmetry, where the projected roots split into two copies of $H_{3}$.  Finally consider the $d^{\parallel}=4$ case: then $\theta^{\parallel}=H_{4}$, which has rational rank $8$.  From Table \ref{CoxeterNumbers} we see that the only candidate crystallographic partner $\theta$ of rank $d=8$ is $E_{8}$, since this is the only rank eight root system whose total number of roots ($240$) is a multiple of the total number of $H_{4}$ roots ($120$); and, as is well known (an as we see in Subsection \ref{4Dfrom8D}), $E_{8}$ does indeed have a projection onto $d^{\parallel}=4$ dimensions with $G(H_{4})$ (hyper-icosahedral) symmetry, where the projected roots split into two copies of $H_{4}$.  Since there are no more irreducible non-crystallographic Coxeter groups in $d^{\parallel}>4$, this completes the enumeration of all Coxeter pairs: they come in two infinite families and four exceptional cases, as summarized in Table \ref{CoxeterPairs}.

\section{Uniqueness of the $H_{4}$ space group}
\label{H4_space_group}

As mentioned above, the finite irreducible non-crystallographic reflection groups were classified by Coxeter \cite{Cox1, Cox2, Cox3, RegularPolytopes, CoxeterMoser}: they are $I_{2}^{n}$ (the symmetries of an equilateral $n$-gon in 2D), $H_{3}$ (the symmetries of the icosahedron in 3D), and $H_{4}$ (the symmetries of the 600-cell in 4D).  

In this Appendix we complete the enumeration of the space groups associated with these point groups.  In particular, since the 2D ($I_{2}^{n}$) and 3D ($H_{3}$) space groups were already treated in Refs.~\cite{Mermin2D1988, Mermin3D1988, Mermin1992}, it remains for us to analyze the 4D ($H_{4}$) case: we will prove that there is a unique 4D space group associated to $H_{4}$ (namely, the symmorphic space group).

We follow the formalism developed in Section 2 of Ref.~\cite{Mermin1992} (see also Ref.~\cite{Mermin2D1988}) for working out all the space groups corresponding to a given point group.  In that formalism, given a point group $G$, the first preliminary step is to determine the relevant set of lattices corresponding to $G$ (see Section 2F in Ref.~\cite{Mermin1992}).  Since the point groups of interest to us are the irreducible reflection groups, the relevant lattices are precisely the reflection (quasi)lattices defined in Ref.~\cite{ReflectionQuasilattices}.  In particular, the unique reflection quasilattice corresponding to $H_{4}$ is the $H_{4}$ root quasilattice (integer linear combinations of the $H_{4}$ roots).  

\begin{figure}
  \begin{center}
    \includegraphics[width=1.5in]{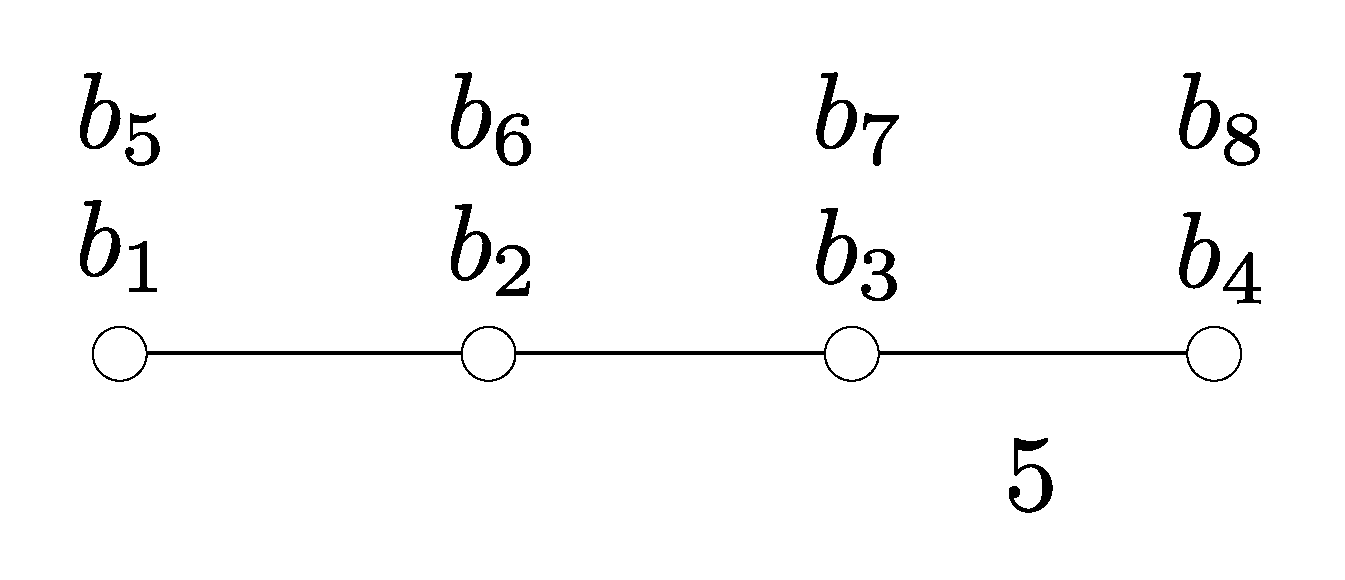}
  \end{center}
  \caption{The labelling of the $H_{4}$ roots.}
 \label{H4CoxeterDynkinDiagram}
\end{figure}

The second preliminary step is to choose a set of primitive generating vectors for the $H_{4}$ reflection quasilattice $\Lambda$ -- {\it i.e.}\ a set of vectors $\{{\bf b}_{k}\}$ such that $\Lambda$ precisely consists of all integer linear combinations $\sum_{k}n_{k}{\bf b}_{k}$ (again, see Section 2F in \cite{Mermin1992}).  Let $\{{\bf b}_{1}, {\bf b}_{2}, {\bf b}_{3}, {\bf b}_{4}\}$ be the simple roots of $H_{4}$, labelled as in Fig.~\ref{H4CoxeterDynkinDiagram}; {\it e.g.}\ the roots could concretely be given by
${\bf b}_{1}=\{1,0,0,0\}$, ${\bf b}_{2}=-\frac{1}{2}\{1,1,1,1\}$, ${\bf b}_{3}=\frac{1}{2}\{0,1,\sigma,\tau\}$, ${\bf b}_{4}=\frac{1}{2}\{0,\sigma,\tau,-1\}$, where $\tau=(1/2)(1+\sqrt{5})$ is the golden ratio and $\sigma=(1/2)(1-\sqrt{5})$ is its algebraic conjugate.  And let $\{{\bf b}_{5}, {\bf b}_{6}, {\bf b}_{7}, {\bf b}_{8}\}=\tau\{{\bf b}_{1}, {\bf b}_{2}, {\bf b}_{3}, {\bf b}_{4}\}$ be another copy of the simple roots, multiplied by the golden ratio.    Then the eight vectors $\{{\bf b}_{1},\ldots,{\bf b}_{8}\}$ form a convenient set of primitive generating vectors for $H_{4}$.

The third preliminary step is to choose a set of generators and relations for $G(H_{4}$), the reflection group corresponding to $H_{4}$.  We will take the standard set of generators and relations for a reflection group: in other words, the generators are $\{R_{1},R_{2},R_{3},R_{4}\}$ (the four reflections corresponding to the four simple roots $\{{\bf b}_{1},{\bf b}_{2},{\bf b}_{3},{\bf b}_{4}\}$, respectively).  The corresponding relations are then neatly summarized by the $H_{4}$ Coxeter-Dynkin diagram in Fig.~\ref{H4CoxeterDynkinDiagram}; explicitly, they are:
\begin{subequations}
  \label{relations}
  \begin{eqnarray}
    \label{relations1}
    &R_{1}^{2}=R_{2}^{2}=R_{3}^{2}=R_{4}^{2}=1,& \\
    \label{relations2}
    &(R_{1}R_{2})^{3}=(R_{2}R_{3})^{3}=(R_{3}R_{4})^{5}=1,& \\
    \label{relations3}
    &(R_{1}R_{3})^{2}=(R_{1}R_{4})^{2}=(R_{2}R_{4})^{2}=1.&
  \end{eqnarray}
\end{subequations}

Now, as explained in Section 2B of \cite{Mermin1992}, if a system has point group $G$ then, in Fourier space, for any element $g\in G$, the density $\rho({\bf k})$ transforms as
\begin{equation}
  \rho(g{\bf k})={\rm e}^{2\pi i \Phi_{g}({\bf k})}\rho({\bf k}).
\end{equation}
Classifying the possible space groups boils down to constraining and classifying the gauge-inequivalent phase functions $\Phi_{g}({\bf k})$ that can consistently appear in this equation, where the phase functions $\Phi_{g}({\bf k})$ and $\Phi_{g}'({\bf k})$ are said to be gauge equivalent if there is a function $\chi({\bf k})$ such that 
\begin{equation}
  \label{gauge_transformation}
  \Phi_{g}'({\bf k})\equiv\Phi_{g}({\bf k})+\chi(g{\bf k})-\chi({\bf k})
\end{equation}
for all $g\in G$ and for all ${\bf k}\in\Lambda$ (see Section 2C in \cite{Mermin1992}).  As in \cite{Mermin1992}, we use the symbol ``$\equiv$" to denote equality up to an additive integer.  

In constraining and classifying the possible phase functions $\Phi_{g}({\bf k})$, we don't need to consider each $g\in G$ separately.  Instead, since each $g\in G$ may be written as a product of the four generators $\{R_{1},R_{2},R_{3},R_{4}\}$, it suffices to consider the four corresponding phase functions $\{\Phi_{1}({\bf k})$, $\Phi_{2}({\bf k})$, $\Phi_{3}({\bf k})$, $\Phi_{4}({\bf k})\}$; any arbitrary phase function $\Phi_{g}({\bf k})$ may then be expressed in terms of these four by successive application of Eq.~(2.11) in \cite{Mermin1992}.  Furthermore, we don't need to consider each wavevector ${\bf k}\in\Lambda$ separately: instead, it is enough to consider the cases where ${\bf k}$ is one of the eight primitive generating vectors $\{{\bf b}_{1},\ldots,{\bf b}_{8}\}$.  Thus our goal here is to constrain the $4\times 8 = 32$ numbers, $\Phi_{i}({\bf b}_{j})$.  From each relation in Eq.~(\ref{relations}) we obtain constraints on these 32 numbers, and we will show that, when taken together, these constraints imply that the 32 values $\Phi_{i}({\bf b}_{j})$ all vanish (in an appropriate gauge).  Thus there is a unique 4D space group corresponding to $H_{4}$: the symmorphic one.

Before we begin analyzing the various constraints on $\Phi_{i}({\bf b}_{j})$, it will be useful to have the matrix expressions for the four generators $\{R_{1},R_{2},R_{3},R_{4}\}$.  Note that the generator $R_{i}$ maps the primitive generating vector ${\bf b}_{j}$ to an integer linear combination of the eight ${\bf b}_{k}$'s so that, in the $\{{\bf b}_{1},\ldots,{\bf b}_{8}\}$ basis, it is an $8\times8$ integer matrix.
\begin{subequations}
  \begin{equation}
    \label{R1}
    R_{1}=\left(\begin{array}{cccccccc}
    -1 & +1 & 0 & 0 & 0 & 0 & 0 & 0 \\
    0 & +1 & 0 & 0 & 0 & 0 & 0 & 0 \\
    0 & 0 & +1 & 0 & 0 & 0 & 0 & 0 \\
    0 & 0 & 0 & +1 & 0 & 0 & 0 & 0 \\
    0 & 0 & 0 & 0 & -1 & +1 & 0 & 0 \\
    0 & 0 & 0 & 0 & 0 & +1 & 0 & 0 \\
    0 & 0 & 0 & 0 & 0 & 0 & +1 & 0 \\
    0 & 0 & 0 & 0 & 0 & 0 & 0 & +1
    \end{array}\right), 
  \end{equation}
  \begin{equation}
    \label{R2}
    R_{2}=\left(\begin{array}{cccccccc}
    +1 & 0 & 0 & 0 & 0 & 0 & 0 & 0 \\
    +1 & -1 & +1 & 0 & 0 & 0 & 0 & 0 \\
    0 & 0 & +1 & 0 & 0 & 0 & 0 & 0 \\
    0 & 0 & 0 & +1 & 0 & 0 & 0 & 0 \\
    0 & 0 & 0 & 0 & +1 & 0 & 0 & 0 \\
    0 & 0 & 0 & 0 & +1 & -1 & +1 & 0 \\
    0 & 0 & 0 & 0 & 0 & 0 & +1 & 0 \\
    0 & 0 & 0 & 0 & 0 & 0 & 0 & +1
    \end{array}\right),
  \end{equation}
  \begin{equation}
    \label{R3}
    R_{3}=\left(\begin{array}{cccccccc}
    +1 & 0 & 0 & 0 & 0 & 0 & 0 & 0 \\
    0 & +1 & 0 & 0 & 0 & 0 & 0 & 0 \\
    0 & +1 & -1 & 0 & 0 & 0 & 0 & +1 \\
    0 & 0 & 0 & +1 & 0 & 0 & 0 & 0 \\
    0 & 0 & 0 & 0 & +1 & 0 & 0 & 0 \\
    0 & 0 & 0 & 0 & 0 & +1 & 0 & 0 \\
    0 & 0 & 0 & +1 & 0 & +1 & -1 & +1 \\
    0 & 0 & 0 & 0 & 0 & 0 & 0 & +1
    \end{array}\right),
  \end{equation}
  \begin{equation}
    \label{R4}
    R_{4}=\left(\begin{array}{cccccccc}
    +1 & 0 & 0 & 0 & 0 & 0 & 0 & 0 \\
    0 & +1 & 0 & 0 & 0 & 0 & 0 & 0 \\
    0 & 0 & +1 & 0 & 0 & 0 & 0 & 0 \\
    0 & 0 & 0 & -1 & 0 & 0 & +1 & 0 \\
    0 & 0 & 0 & 0 & +1 & 0 & 0 & 0 \\
    0 & 0 & 0 & 0 & 0 & +1 & 0 & 0 \\
    0 & 0 & 0 & 0 & 0 & 0 & +1 & 0 \\
    0 & 0 & +1 & 0 & 0 & 0 & +1 & -1
    \end{array}\right).
  \end{equation}
\end{subequations}
Also note that each relation in Eq.~(\ref{relations}) is of the form $1=g^{n}$, which translates into the constraint $0\equiv\Phi_{g^{n}}({\bf k})$.  Then, by repeated application of Eq.~(2.11) in \cite{Mermin1992}, this may be re-written in the more convenient (but equivalent) form
\begin{equation}
  \label{power_law_constraint}
  0\equiv\Phi_{g}([1+g^{1}+\ldots+g^{n-1}]{\bf k}).
\end{equation}

Now we proceed to analyze the constraints on $\Phi_{i}({\bf b}_{j})$ coming from each relation in (\ref{relations}).  The overview is as follows: we first analyze the four relations in (\ref{relations1}) and, in the process, completely fix the gauge freedom; we then proceed to analyze the other relations (\ref{relations2}, \ref{relations3}) and see that they force the remaining gauge-fixed phase functions to vanish.

We proceed to consider the first constraint in (\ref{relations1}); we will describe the analysis of this first relation in detail, to illustrate how the calculation proceeds.  The first relation in (\ref{relations1}) is $R_{1}^{2}=1$.  From Eq.~(\ref{power_law_constraint}), we see that this translates into the constraint
$0\equiv\Phi_{1}([1+R_{1}]{\bf k})$; and we can use Eq.~(\ref{R1}) to calculate:
\begin{equation}
  1+R_{1}=\left(\begin{array}{cccccccc}
    0 & +1 & 0 & 0 & 0 & 0 & 0 & 0 \\
    0 & +2 & 0 & 0 & 0 & 0 & 0 & 0 \\
    0 & 0 & +2 & 0 & 0 & 0 & 0 & 0 \\
    0 & 0 & 0 & +2 & 0 & 0 & 0 & 0 \\
    0 & 0 & 0 & 0 & 0 & +1 & 0 & 0 \\
    0 & 0 & 0 & 0 & 0 & +2 & 0 & 0 \\
    0 & 0 & 0 & 0 & 0 & 0 & +2 & 0 \\
    \;0\; & 0 & 0 & 0 & \;0\; & 0 & 0 & +2
  \end{array}\right).
\end{equation}
From the six non-vanishing columns of this matrix we read off the six non-trivial constraints:
\begin{subequations}
  \begin{eqnarray}
    0&\equiv&\Phi_{1}({\bf b}_{1})+2\Phi_{1}({\bf b}_{2}), \\
    0&\equiv&2\Phi_{1}({\bf b}_{3}), \\
    0&\equiv&2\Phi_{1}({\bf b}_{4}), \\
    0&\equiv&\Phi_{1}({\bf b}_{5})+2\Phi_{1}({\bf b}_{6}), \\
    0&\equiv&2\Phi_{1}({\bf b}_{7}), \\
    0&\equiv&2\Phi_{1}({\bf b}_{8}).
  \end{eqnarray}
\end{subequations}
Now, from (\ref{gauge_transformation}), we see that by choosing $\chi({\bf b}_{1})=-\Phi_{1}({\bf b}_{2})$ and $\chi({\bf b}_{5})=-\Phi_{1}({\bf b}_{6})$ we can do a gauge transformation to set $\Phi_{1}'({\bf b}_{1})\equiv\Phi_{1}'({\bf b}_{2})\equiv\Phi_{1}'({\bf b}_{5})\equiv\Phi_{1}'({\bf b}_{6})\equiv0$, while the other phases $\{\Phi_{1}'({\bf b}_{3})$, $\Phi_{1}'({\bf b}_{4})$, $\Phi_{1}'({\bf b}_{7})$, $\Phi_{1}'({\bf b}_{8})\}$ are constrained to be $0$ or $1/2$.  

The remaining relations in Eq.~(\ref{relations1}) may be analyzed in a completely analogous way, so we will be brief:

For the second relation in (\ref{relations1}): $R_{2}^{2}=1 \Rightarrow 0\equiv\Phi_{2}([1+R_{2}]{\bf k})$.  From the six non-zero columns of the matrix $1+R_{2}$ we read off the six constraints
\begin{subequations}
  \begin{eqnarray}
    0&\equiv&2\Phi_{2}({\bf b}_{1})+\Phi_{2}({\bf b}_{2}), \\
    0&\equiv&\Phi_{2}({\bf b}_{2})+2\Phi_{2}({\bf b}_{3}), \\
    0&\equiv&2\Phi_{2}({\bf b}_{4}), \\
    0&\equiv&2\Phi_{2}({\bf b}_{5})+\Phi_{2}({\bf b}_{6}), \\
    0&\equiv&\Phi_{2}({\bf b}_{6})+2\Phi_{2}({\bf b}_{7}), \\
    0&\equiv&2\Phi_{2}({\bf b}_{8}).
  \end{eqnarray}
\end{subequations}
Then by choosing $\chi({\bf b}_{2})=-\Phi_{2}({\bf b}_{1})$ and $\chi({\bf b}_{6})=-\Phi_{2}({\bf b}_{5})$ we can set $\Phi_{2}'({\bf b}_{1})\equiv\Phi_{2}'({\bf b}_{2})\equiv\Phi_{2}'({\bf b}_{5})\equiv\Phi_{2}'({\bf b}_{6})\equiv0$, while the other phases $\{\Phi_{2}'({\bf b}_{3})$, $\Phi_{2}'({\bf b}_{4})$, $\Phi_{2}'({\bf b}_{7})$, $\Phi_{2}'({\bf b}_{8})\}$ are constrained to be $0$ or $1/2$.

For the third relation in (\ref{relations1}): $R_{3}^{2}=1 \Rightarrow 0\equiv\Phi_{3}([1+R_{3}]{\bf k})$.  From the six non-zero columns of the matrix $1+R_{3}$ we read off the six constraints
\begin{subequations}
  \begin{eqnarray}
    0&\equiv&2\Phi_{3}({\bf b}_{1}), \\
    0&\equiv&2\Phi_{3}({\bf b}_{2})+\Phi_{3}({\bf b}_{3}), \\
    0&\equiv&2\Phi_{3}({\bf b}_{4})+\Phi_{3}({\bf b}_{7}), \\
    0&\equiv&2\Phi_{3}({\bf b}_{5}), \\
    0&\equiv&2\Phi_{3}({\bf b}_{6})+\Phi_{3}({\bf b}_{7}), \\
    0&\equiv&\Phi_{3}({\bf b}_{3})+\Phi_{3}({\bf b}_{7})+2\Phi_{3}({\bf b}_{8}).
  \end{eqnarray}
\end{subequations}
Then by choosing $\chi({\bf b}_{3})=\Phi_{3}({\bf b}_{4})-\Phi_{3}({\bf b}_{8})$ and $\chi({\bf b}_{7})=-\Phi_{3}({\bf b}_{4})$ we can set 
$\Phi_{3}'({\bf b}_{3})\equiv\Phi_{3}'({\bf b}_{4})\equiv\Phi_{3}'({\bf b}_{7})\equiv\Phi_{3}'({\bf b}_{8})\equiv0$, while the other phases $\{\Phi_{3}'({\bf b}_{1})$, $\Phi_{3}'({\bf b}_{2})$, $\Phi_{3}'({\bf b}_{5})$, $\Phi_{3}'({\bf b}_{6})\}$ are constrained to be $0$ or $1/2$.

For the fourth relation in (\ref{relations1}): $R_{4}^{2}=1 \Rightarrow 0\equiv\Phi_{4}([1+R_{4}]{\bf k})$.  From the six non-zero columns of the matrix $1+R_{4}$ we read off the six constraints
\begin{subequations}
  \begin{eqnarray}
    0&\equiv&2\Phi_{4}({\bf b}_{1}), \\
    0&\equiv&2\Phi_{4}({\bf b}_{2}), \\
    0&\equiv&2\Phi_{4}({\bf b}_{3})+\Phi_{4}({\bf b}_{8}), \\
    0&\equiv&2\Phi_{4}({\bf b}_{5}), \\
    0&\equiv&2\Phi_{4}({\bf b}_{6}), \\
    0&\equiv&\Phi_{4}({\bf b}_{4})+2\Phi_{4}({\bf b}_{7})+\Phi_{4}({\bf b}_{8}).
  \end{eqnarray}
\end{subequations}
Then by choosing $\chi({\bf b}_{4})=\Phi_{4}({\bf b}_{3})-\Phi_{4}({\bf b}_{7})$ and $\chi({\bf b}_{8})=-\Phi_{4}({\bf b}_{3})$ we can set 
$\Phi_{4}'({\bf b}_{3})\equiv\Phi_{4}'({\bf b}_{4})\equiv\Phi_{4}'({\bf b}_{7})\equiv\Phi_{4}'({\bf b}_{8})\equiv0$, while the other phases
$\{\Phi_{4}'({\bf b}_{1})$, $\Phi_{4}'({\bf b}_{2})$, $\Phi_{4}'({\bf b}_{5})$, $\Phi_{4}'({\bf b}_{6})\}$ are constrained to be $0$ or $1/2$.

Let us pause to summarize the situation thus far: we have analyzed the four relations in (\ref{relations1}).  In the process, we have chosen values for $\{\chi({\bf b}_{1}),\ldots,\chi({\bf b}_{8})\}$ -- in this way we completely fix the gauge and set 16 of the 32 quantities $\Phi_{i}({\bf b}_{j})$ to zero, while the remaining 16 quantities have been constrained to be $0$ or $1/2$.  

We will now show that the remaining relations -- those in Eqs.~(\ref{relations2}) and (\ref{relations3}) -- actually constrain these remaining 16 quantities to all be $0$.  Note that each of the remaining constraints has the form $1=(R_{i}R_{j})^{n}$.  Using (\ref{power_law_constraint}), this may be rewritten in the form
$0\equiv\Phi_{R_{i}R_{j}}([1+R_{i}R_{j}+\ldots+(R_{i}R_{j})^{n-1}]{\bf k})$ and then, by a further application of Eq.~(2.11) in \cite{Mermin1992}, in the more convenient form 
\begin{eqnarray}
  \label{product_power_law}
  0&\equiv&\Phi_{i}\big(R_{j}[1+R_{i}R_{j}+\ldots+(R_{i}R_{j})^{n-1}]{\bf k}\big) \nonumber \\
  &+&\Phi_{j}\big([1+R_{i}R_{j}+\ldots+(R_{i}R_{j})^{n-1}]{\bf k}\big).
\end{eqnarray}

Let us apply this to the first constraint in (\ref{relations2}): $(R_{1}R_{2})^{3}=1$.  From Eq.~(\ref{product_power_law}), this becomes 
$0\equiv\Phi_{1}(R_{2}[1+R_{1}R_{2}+(R_{1}R_{2})^{2}]{\bf k})+\Phi_{2}([1+R_{1}R_{2}+(R_{1}R_{2})^{2}]{\bf k})$.  From Eqs.~(\ref{R1}, \ref{R2}) we find that $1+R_{1}R_{2}+(R_{1}R_{2})^{2}$ and $R_{2}[1+R_{1}R_{2}+(R_{1}R_{2})^{2}]$ are both
\begin{equation}
  \left(\begin{array}{cccccccc}
    \;0\; & \;0\; & \;1\; & \;0\; & \;0\; & \;0\; & \;0\; & \;0\; \\
    0 & 0 & 2 & 0 & 0 & 0 & 0 & 0 \\
    0 & 0 & 3 & 0 & 0 & 0 & 0 & 0 \\
    0 & 0 & 0 & 3 & 0 & 0 & 0 & 0 \\
    0 & 0 & 0 & 0 & 0 & 0 & 1 & 0 \\
    0 & 0 & 0 & 0 & 0 & 0 & 2 & 0 \\
    0 & 0 & 0 & 0 & 0 & 0 & 3 & 0 \\
    0 & 0 & 0 & 0 & 0 & 0 & 0 & 3
  \end{array}\right).
\end{equation}
Now, if we take into account the 16 quantities that have already been previously set to zero ($\equiv0$), along with the fact that other 16 quantities are either integer or half-integer ($\equiv0$ or $\equiv1/2$) so that any even multiple of such a quantity is $\equiv0$, we find that the four non-zero columns of this matrix yield the four constraints:
\begin{subequations}
  \begin{eqnarray}
    \Phi_{1}({\bf b}_{3})&\equiv&\Phi_{2}({\bf b}_{3}) \\
    \Phi_{1}({\bf b}_{4})&\equiv&\Phi_{2}({\bf b}_{4}) \\
    \Phi_{1}({\bf b}_{7})&\equiv&\Phi_{2}({\bf b}_{7}) \\
    \Phi_{1}({\bf b}_{8})&\equiv&\Phi_{2}({\bf b}_{8}).
  \end{eqnarray}
\end{subequations}

Again, the remaining relations in Eq.~(\ref{relations2}, \ref{relations3}) may be analyzed in a completely analogous way, so in brief:

From the second relation in Eq.~(\ref{relations2}), $(R_{2}R_{3})^{3}=1$:
\begin{subequations}
  \begin{eqnarray}
    \Phi_{2}({\bf b}_{3})&\equiv&\Phi_{3}({\bf b}_{1}) \\
    \Phi_{2}({\bf b}_{4})&\equiv&\Phi_{3}({\bf b}_{6}) \\
    \Phi_{2}({\bf b}_{7})&\equiv&\Phi_{3}({\bf b}_{5}) \\
    \Phi_{2}({\bf b}_{8})&\equiv&\Phi_{3}({\bf b}_{2})+\Phi_{3}({\bf b}_{6})
  \end{eqnarray}
\end{subequations}
From the third relation in Eq.~(\ref{relations2}), $(R_{3}R_{4})^{5}=1$:
\begin{subequations}
  \begin{eqnarray}
    \Phi_{3}({\bf b}_{1})&\equiv&\Phi_{4}({\bf b}_{1}) \\
    \Phi_{3}({\bf b}_{2})&\equiv&\Phi_{4}({\bf b}_{2}) \\
    \Phi_{3}({\bf b}_{5})&\equiv&\Phi_{4}({\bf b}_{5}) \\
    \Phi_{3}({\bf b}_{6})&\equiv&\Phi_{4}({\bf b}_{6})
  \end{eqnarray}
\end{subequations}
From the first relation in Eq.~(\ref{relations3}), $(R_{1}R_{3})^{2}=1$:
\begin{subequations}
  \begin{eqnarray}
    \Phi_{1}({\bf b}_{4})&\equiv&\Phi_{3}({\bf b}_{1}) \\
    \Phi_{1}({\bf b}_{7})&\equiv&0 \\
    \Phi_{1}({\bf b}_{7})&\equiv&\Phi_{3}({\bf b}_{5}) \\
    \Phi_{1}({\bf b}_{7})&\equiv&\Phi_{1}({\bf b}_{3})
  \end{eqnarray}
\end{subequations}
From the second relation in Eq.~(\ref{relations3}), $(R_{1}R_{4})^{2}=1$:
\begin{subequations}
  \begin{eqnarray}
    \Phi_{4}({\bf b}_{1})&\equiv&0 \\
    \Phi_{1}({\bf b}_{8})&\equiv&0 \\
    \Phi_{4}({\bf b}_{5})&\equiv&0 \\
    \Phi_{1}({\bf b}_{4})&\equiv&0  
  \end{eqnarray}
\end{subequations}
From the third relation in Eq.~(\ref{relations3}), $(R_{2}R_{4})^{2}=1$:
\begin{subequations}
  \begin{eqnarray}
    \Phi_{4}({\bf b}_{2})&\equiv&0 \\
    \Phi_{2}({\bf b}_{8})&\equiv&0 \\
    \Phi_{4}({\bf b}_{6})&\equiv&0 \\
    \Phi_{2}({\bf b}_{4})&\equiv&0  
  \end{eqnarray}
\end{subequations}
It is now straightforward to check that the above constraints together imply that all 16 remaining quantities $\Phi_{i}({\bf b}_{j})$ vanish, so we have shown that it is possible to choose a gauge in which $\Phi_{1}({\bf k})\equiv\Phi_{2}({\bf k})\equiv\Phi_{3}({\bf k})\equiv\Phi_{4}({\bf k})\equiv0$ and hence $\Phi_{g}({\bf k})\equiv0$ for arbitrary $g\in G$ and ${\bf k}\in\Lambda$.  This completes the proof that there is a unique 4D space group corresponding to (the unique irreducible non-crystallographic roots system) $H_{4}$.

\end{document}